%% file: DR2HRD.tex
\documentclass[longauth]{aa_gaia}

\usepackage{graphicx}
\usepackage{txfonts}

\newcommand{\gaia}{\emph{Gaia}\xspace}

\newcommand{\gacs}[1]{{\footnotesize\texttt{#1}}}
\providecommand{\kms}{\,km\,s$^{-1}$}
\providecommand{\Vtot}{$V_\mathrm{tot}$}
\providecommand{\VT}{$V_\mathrm{T}$}
\providecommand{\Msol}{M$_\odot$}
\def\gmag{$G$}
\def\gbp{$G_{\rm BP}$}
\def\grp{$G_{\rm RP}$}
\def\grvs{$G_{\rm RVS}$}
\def\bpmrp{$G_{\rm BP}-G_{\rm RP}$}
\def\logg{$\log g$}
\usepackage{color}

\begin{document}

   \title{\gaia Data Release 2:\\ Observational Hertzsprung-Russell diagrams
\thanks{The full Table A.1 is only available at the CDS via anonymous ftp to cdsarc.u-strasbg.fr (130.79.128.5) or via http://cdsarc.u-strasbg.fr/viz-bin/qcat?J/A+A/616/A10}}

    \input{authors.tex}

   \date{Received ; accepted }

 
  \abstract
   {
   \gaia Data Release 2 provides high-precision astrometry and three-band photometry for about 1.3 billion sources over the full sky.
   The precision, accuracy, and homogeneity of both astrometry and photometry are unprecedented.}
   {We highlight the power of the \gaia DR2 in studying many fine structures of the Hertzsprung-Russell diagram (HRD). \gaia allows us to present many different HRDs, depending in particular on stellar population selections. We do not aim here for 
   completeness in terms of types of stars or stellar evolutionary aspects. Instead, we have chosen several illustrative examples. }
   {We describe some of the selections that can be made in \gaia DR2 to highlight the main structures of the \gaia HRDs. We select both 
   field and cluster (open and globular) stars, compare the observations with previous classifications and with stellar evolutionary 
   tracks, and we present variations of the \gaia HRD with age, metallicity, and kinematics. Late stages of stellar evolution such as hot subdwarfs, post-AGB stars, planetary
   nebulae, and white dwarfs are also analysed, as well as low-mass brown dwarf objects.}
   {The \gaia HRDs are unprecedented in both precision and coverage of the various Milky Way stellar populations and stellar evolutionary 
    phases. Many fine structures of the HRDs are presented. The clear split of the white dwarf sequence into hydrogen and helium white dwarfs is presented for the first time in an HRD.  The relation between kinematics and the HRD is nicely illustrated. Two different populations in a classical kinematic selection of the halo are unambiguously identified in the HRD. 
   Membership and mean parameters for a selected list of open clusters are provided. They allow drawing very detailed cluster sequences, highlighting fine structures, and providing extremely precise empirical isochrones that will lead to more insight in stellar physics.
   }
   {\gaia DR2 demonstrates the potential of combining precise astrometry and photometry for large samples for studies in stellar evolution and stellar population and opens an entire new area for HRD-based studies. }


   \maketitle
%

\section{Introduction}

The Hertzsprung-Russell diagram (HRD) is one of the most important tools in stellar studies. It illustrates empirically the relationship between stellar spectral type (or temperature or colour index) and luminosity (or absolute magnitude). The position of a star in the HRD is mainly given by its initial mass, chemical composition, and age, but effects such as rotation, stellar wind, magnetic field, detailed chemical abundance, over-shooting, and non-local
thermal equilibrium also play a role. 
Therefore, the detailed HRD features are important to constrain stellar structure and evolutionary studies as well as stellar atmosphere modelling. 
Up to now, a proper understanding of the physical process in the stellar interior and the exact 
contribution of each of the effects mentioned are missing because we lack large precise and homogeneous samples that cover the full HRD. 
Moreover, a precise HRD provides a great framework for exploring stellar populations and stellar systems. 

Up to now, the most complete solar neighbourhood empirical HRD could be obtained by combining the Hipparcos data \citep{1995A&A...304...69P} with nearby stellar catalogues to provide the faint end \cite[e.g.][]{1991adc..rept.....G,2015IAUGA..2253773H}. Clusters provide empirical HRDs for a range of ages and metal contents and are therefore widely used in stellar evolution studies. To be conclusive, they need homogeneous photometry for inter-comparisons and astrometry for good memberships. 

With its global census of the whole sky, homogeneous astrometry, and photometry of unprecedented accuracy, \gaia DR2 is setting a new major step in stellar, galactic, and extragalactic studies. 
It provides position, trigonometric parallax, and proper motion as well as three broad-band magnitudes (\gmag, \gbp, and \grp) for more than a billion objects brighter than \gmag$\sim$20, plus radial velocity for sources brighter than \grvs$\sim$12 mag and photometry for variable stars \citep{DR2-DPACP-36}. 
The amount, exquisite quality, and homogeneity of the data allows reaching a level of detail in the HRDs that has never been reached before. The number of open clusters with accurate parallax information is unprecedented, and new open clusters or associations will be discovered. \gaia~DR2 provides absolute parallax for faint red dwarfs and the faintest white dwarfs for
the first time. 

This paper is one of the papers accompanying the \gaia DR2 release. The following papers describe the data used here: \cite{DR2-DPACP-36} for an overview, \cite{DR2-DPACP-51} for the astrometry, \cite{DR2-DPACP-40} for the photometry, and \cite{DR2-DPACP-39} for the global validation. Someone interested in this HRD paper may also be interested in the variability in the HRD described in \cite{DR2-DPACP-35}, in the first attempt to derive an HRD using temperatures and luminosities from the \gaia DR2 data of \cite{DR2-DPACP-43}, in the kinematics of the globular clusters discussed in \cite{DR2-DPACP-34}, and in the field kinematics presented in \cite{DR2-DPACP-33}.

In this paper, Sect.~\ref{sec:building} presents a global description of how we built the \gaia HRDs of both field and cluster stars, the filters that we applied, and the handling of the extinction. In Sect.~\ref{sec:clustersdata} we present our selection of cluster data; the handling of the globular clusters is detailed in \cite{DR2-DPACP-34} and the handling of the open clusters is detailed in Appendix~\ref{sec:ocsolu}. Section~\ref{sec:HRDdetails} discusses the main structures of the \gaia DR2 HRD. The level of the details of the white dwarf sequence is so new that it leads to a more intense discussion,
which we present in a separate Sect.~\ref{sec:WDs}. In Sect.~\ref{sec:ociso} we compare clusters with a set of isochrones. In Sect.~\ref{sec:kine} we study the variation of the \gaia HRDs with kinematics. We finally conclude in Sect.~\ref{sec:conclu}.

\section{Building the \gaia HRDs\label{sec:building}}

This paper presents the power of the \gaia DR2 astrometry and photometry in studying fine structures of the HRD. For this, we selected the most precise data, without trying to reach completeness. In practice, this means selecting the most precise parallax and photometry, but also handling the extinction rigorously. This
can no longer be neglected with the depth of the \gaia precise data in this release.  

\subsection{Data filtering}
\label{sec:filters}

The \gaia DR2 is unprecedented in both the quality and the quantity of its astrometric and photometric data. Still, this is an intermediate data release without a full implementation of the complexity of the processing for an optimal usage of the data. A detailed description of the astrometric and photometric features is given in \cite{DR2-DPACP-51} and \cite{DR2-DPACP-40}, respectively, and \cite{DR2-DPACP-39} provides a global validation of them. Here we highlight the features that are important to be taken into account in building \gaia DR2 HRDs and present the filters we applied in this paper. 

Concerning the astrometric content \citep{DR2-DPACP-51}, the median uncertainty for the bright source (\gmag$<$14~mag) parallax is 0.03~mas. The systematics are lower than 0.1~mas, and the parallax zeropoint error is about 0.03~mas. Significant correlations at small spatial scale between the astrometric parameters are also observed. Concerning the photometric content \citep{DR2-DPACP-40}, the precision at \gmag=12 is around 1~mmag in the three passbands, with systematics at the level of 10~mmag. 

\cite{DR2-DPACP-51} described that a five-parameter solution is accepted only if at least six visibility periods are used (e.g. the number of groups of observations separated from other groups by a gap of at least four days, the parameter is named   \gacs{visibility\_periods\_used} in the \gaia archive). The observations need to be well spread out in time to provide reliable five-parameter solutions. Here we applied a stronger filter on this parameter: \gacs{visibility\_periods\_used$>$8}. This removes strong outliers, in particular at the faint end of the local HRD \citep{DR2-DPACP-39}. It also leads to more incompleteness, but this is not an issue for this paper. 

The astrometric excess noise is the extra noise  that must be postulated to explain the scatter of residuals in the astrometric solution. When it is high, it either means that the astrometric solution has failed and/or that the studied object is in a multiple system for which the single-star solution is not reliable. Without filtering on the astrometric excess noise, artefacts are present in particular between the white dwarf and the main sequence in the \gaia HRDs. Some of those stars are genuine binaries, but the majority are artefacts \citep{DR2-DPACP-39}. To still see the imprint of genuine binaries on the HRD while removing most of the artefacts, we adopted the filter proposed in Appendix C of \cite{DR2-DPACP-51}: $\sqrt{\chi^2/(\nu'-5)}<1.2 \max(1,\exp(-0.2 (G-19.5))$ with $\chi^2$ and $\nu'$ given as \gacs{astrometric\_chi2\_al} and \gacs{astrometric\_n\_good\_obs\_al,} respectively, in the \gaia archive. A similar clean-up of the HRD is obtained by the \gacs{astrometric\_excess\_noise$<$1} criterion, but this is less optimised for the bright stars because of the degrees of freedom (DOF) issue \citep[Appendix A]{DR2-DPACP-51}. 

We built the \gaia HRDs by simply estimating the absolute \gaia magnitude in the G band for individual stars using $M_G = G+5+5 \log_{10}(\varpi/1000.)$, with $\varpi$ the parallax in milliarcseconds (plus the extinction, see next section). This is valid only when the relative precision on the parallax is lower than about 20\% \citep{DR2-DPACP-38}. We aim here to examine the fine structures in the HRD revealed by \gaia and therefore adopt a 10\% relative precision criterion, which corresponds to an uncertainty on $M_G$ smaller than 0.22~mag: \gacs{parallax\_over\_error$>$10}. 

Similarly, we apply filters on the relative flux error on the \gmag, \gbp, and \grp\ photometry: \gacs{phot\_g\_mean\_flux\_over\_error$>$50} ($\sigma_G<0.022$~mag), \gacs{phot\_rp\_mean\_flux\_over\_error$>$20,} and \gacs{phot\_bp\_mean\_flux\_over\_error$>$20} ($\sigma_{G_{XP}}<0.054$~mag). These criteria may remove variable stars, which are specifically studied in \cite{DR2-DPACP-35}. 

The processing of the photometric data in DR2 has not treated blends in the windows of the blue and red photometers (BP and RP). 
As a consequence, the measured BP and RP fluxes may include the contribution of flux from nearby sources, the highest impact being in sky areas of high stellar density,
such as the inner regions of globular clusters, the Magellanic Clouds, or the Galactic Bulge. 
During the validation process, misdeterminations of the local background have also been identified.  
In some cases, this background is due to nearby bright sources with long wings of the point spread function that have not been properly subtracted. In other cases, the background has a solar type spectrum, which indicates that the modelling of the background flux is not good enough. The faint sources are most strongly affected. 
For details, see \cite{DR2-DPACP-40} and \cite{DR2-DPACP-39}.  
Here, we have limited our analysis to the sources within the empirically defined locus of the $(I_{BP}+I_{RP})/I_G$ fluxes ratio as a function of \bpmrp\ colour: \gacs{phot\_bp\_rp\_excess\_factor$>1.0+0.015\ (G_{\rm BP}-G_{\rm RP})^2$} and \gacs{phot\_bp\_rp\_excess\_factor$<1.3+0.06\ (G_{\rm BP}-G_{\rm RP})^2$}.
The Gaia archive query combining all the filters presented here is provided in Appendix~\ref{sec:query}.

\subsection{Extinction\label{sec:extinction}}

\begin{figure}
\centering
\includegraphics[width=9cm]{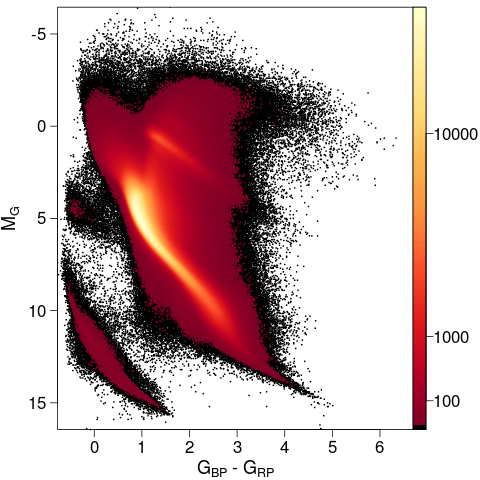}
\caption{Full {\gaia} colour-magnitude diagram of sources with the filters described in Sect.~\ref{sec:filters} applied (65,921,112 stars). The colour scale represents the square root of the relative density of stars. }\label{fig:fullHRDwithext}
\end{figure}

\begin{table*}[t]
\caption{Parameters used to derive the \gaia extinction coefficients as a function of colour and extinction (Eq. \ref{eq:extcoefs}).}
\centering
\begin{tabular}{lrrrrrrr}
\hline\hline
 & $c_1$ & $c_2$ &$c_3$ & $c_4$ & $c_5$ & $c_6$ & $c_7$ \\
 \hline
$k_G$ & 0.9761 & -0.1704 & 0.0086 & 0.0011 & -0.0438 & 0.0013 & 0.0099\\
$k_{BP}$ & 1.1517 & -0.0871 & -0.0333 & 0.0173 & -0.0230 & 0.0006 & 0.0043\\
$k_{RP}$ & 0.6104 & -0.0170 & -0.0026 & -0.0017 & -0.0078 & 0.00005 & 0.0006\\
\hline
\end{tabular}
\label{tab:extcoefs}
\end{table*}

The dust that is present along the line of sight towards the stars leads to a  dimming and reddening of their observed light. In the full colour - absolute magnitude diagram presented in Fig.~\ref{fig:fullHRDwithext}, the effect of the extinction is particularly striking for the red clump. The de-reddened HRD using the extinction provided together with DR2 is presented in \cite{DR2-DPACP-43}. To study the fine structures of the \gaia HRD for field stars, we selected here only low-extinction stars. High galactic latitude and close-by stars located within the local bubble (the reddening is almost negligible within $\sim$60~pc of the Sun \citep{2003A&A...411..447L}) are affected less from the extinction, and we did not apply further selection for them. To select low-extinction stars away from these simple cases, we followed \cite{2017arXiv171005803R} and used the 3D extinction map of \cite{2017A&A...606A..65C}\footnote{http://stilism.obspm.fr/}, which is particularly well adapted to finding holes in the interstellar medium and to select field stars with $E(B-V)<0.015$. 

For globular clusters we used literature extinction values (Sect.~\ref{sec:gcdata}), while for open clusters, they are derived together with the ages (Sect.~\ref{sec:ocdata}). Detailed comparisons of these global cluster extinctions with those that can be derived from the extinctions
provided by  \gaia DR2 can be found in \cite{DR2-DPACP-39}. To transform the global cluster extinction easily into the \gaia passbands while taking into account the extinction coefficients dependency on colour and extinction itself in these large passbands \cite[e.g.][]{2010A&A...523A..48J}, we used the same formulae as \cite{Danielski18} to compute the extinction coefficients $k_X = A_X / A_0$:
\begin{equation}
\begin{split}
        k_X = & c_1 + c_2 (G_{BP}-G_{RP})_0 + c_3 (G_{BP}-G_{RP})_0^2 + c_4  (G_{BP}-G_{RP})_0^3 \\
       & + c_5 A_0 + c_6 A_0^2 + c_7 (G_{BP}-G_{RP})_0 A_0 
 \end{split}
 \label{eq:extcoefs}
.\end{equation}
As in \cite{Danielski18}, this formula was fitted on a grid of extinctions convolving the latest \gaia passbands presented in \cite{DR2-DPACP-40} with Kurucz spectra \citep{2003IAUS..210P.A20C} and the \cite{2007ApJ...663..320F} extinction law for 3500~K$<T_{\rm eff}<$10000~K by steps of 250~K, 0.01$<A_0<5$~mag by steps of 0.01~mag and two surfaces gravities: log$g=2.5$ and 4. The resulting coefficients are provided in Table \ref{tab:extcoefs}. We assume in the following $A_0 = 3.1 E(B-V)$.

Some clusters show high differential extinction across their field, which broadens their colour-magnitude diagrams.
These clusters have been discarded from this analysis.

\section{Cluster data\label{sec:clustersdata}}

Star clusters can provide observational isochrones for a range of ages and chemical compositions. Most suitable are clusters with low and uniform reddening values and whose magnitude range is wide, which would limit our sample to the nearest clusters. 
Such a sample would, however, present a rather limited range in age and chemical composition. 

\subsection{Membership and astrometric solutions}

Two types of astrometric solutions were applied. The first type is applicable to nearby clusters. For the second \gaia data release, the nearby `limit' was set at 250 pc. Within this limit, the parallax and proper motion data for the individual cluster members are sufficiently accurate to reflect the effects of projection along the line of sight, thus enabling the 3D reconstruction of the cluster. This is further described in Appendix~\ref{sec:nearby}. 
        
For these nearby clusters, the size of the cluster relative to its distance will contribute a significant level of scatter to the HRD if parallaxes for individual cluster members are not taken into account. With a relative accuracy of about 1\%\ in the parallax measurement, an error contribution of around 0.02 in the absolute magnitude is possible. For a large portion of the \gaia photometry, the uncertainties are about 5 to 10 times lower, making the parallax measurement still the main contributor to the uncertainty in the absolute magnitude. The range of differences in parallax between the cluster centre and an individual cluster member depends on the ratio of the cluster radius over the cluster distance. At a radius of 15 pc, the 1\%\ level is found for a cluster at 1.5~kpc, or a parallax of 0.67~mas. In \gaia DR2, formal uncertainties on the parallaxes may reach levels of just lower than 10~$\mu$as, but the overall uncertainty from localised systematics is about 0.025~mas. If this value is considered the 1\%\ uncertainty level, then a resolution of a cluster along the line of sight, using \gaia DR2, becomes possible for clusters within 400~pc, and realistic for clusters within about 250~pc. 

For clusters at larger distances, the mean cluster proper motion and parallax are derived directly from the observed astrometric parameters for the individual cluster members. The details of this procedure are presented in Appendix~\ref{sec:distclust}.

\begin{figure*}
\centering
\includegraphics[width=17cm]{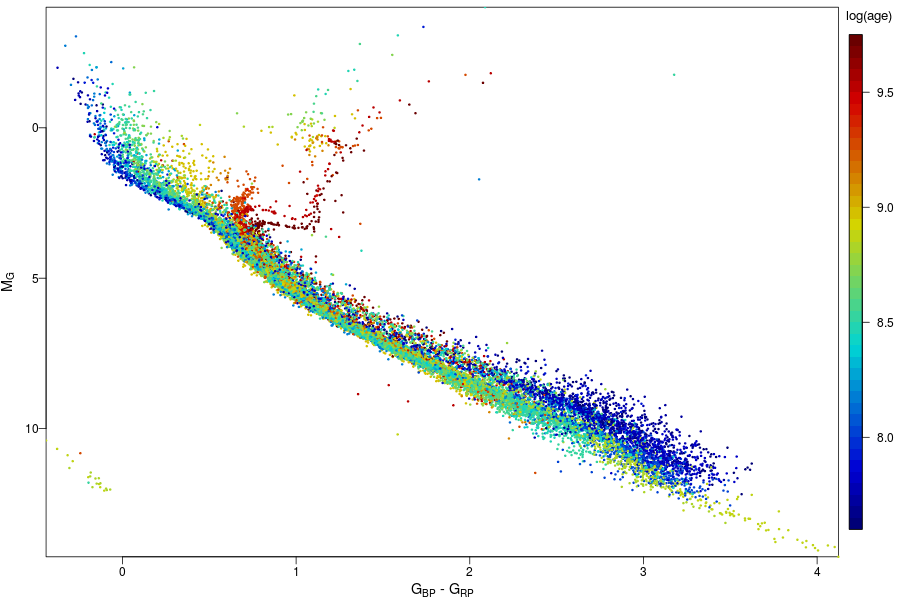}
\caption{Composite HRD for 32 open clusters, coloured according to log(age), using the extinction and distance moduli as determined from the \gaia data (Table~\ref{tab:openclref}). \label{fig:hr30open}}
\vspace{0.5cm}
\includegraphics[width=17cm]{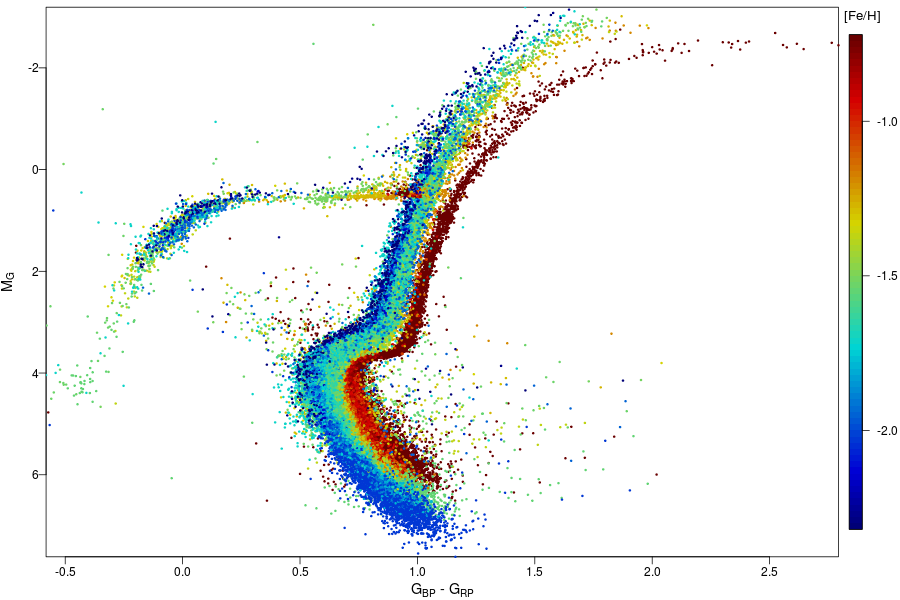}
\caption{Composite HRD for 14 globular clusters, coloured according to metallicity (Table~\ref{tab:globclust}). \label{fig:compgchrd}}
\end{figure*}

\subsection{Selection of open clusters\label{sec:ocdata}}

Our sample of open clusters consists of the mostly well-defined and fairly rich clusters within 250~pc, and a selection of mainly rich clusters at larger distances, covering a wider range of ages, mostly up to 1.5~kpc, with a few additional clusters at larger distances where these might supply additional information at more extreme ages. For very young clusters, the definition of the cluster is not always clear, as the youngest systems are mostly found embedded in OB associations, producing large samples of similar proper motions and parallaxes. Very few clusters appear to survive to an `old age', but those that do are generally rich, allowing good membership determination. The final selection consists of 9 clusters within 250~pc, and 37 clusters up to 5.3~kpc. Of the latter group, only 23 were finally used for construction of the colour-magnitude diagram; these clusters are listed together with the 9 nearby clusters in Table ~\ref{tab:openclref}. For the remaining 14 clusters, the colour-magnitude diagrams appeared to be too much affected by interstellar reddening variations. 
More details on the astrometric solutions are provided in Appendix~\ref{sec:ocsolu}; the solutions are presented for the nearby clusters in Table~\ref{tab:nearbyclusters} and for the more distant clusters in Table~\ref{tab:overview}.
Figure~\ref{fig:hr30open} shows the combined HRD of these clusters, coloured according to their ages as provided in Table~\ref{tab:openclref}. The main-sequence turn-off and red clump evolution with age is clearly visible. The age difference is also shown for lower mass stars, the youngest stars lie slightly above the main sequence
of the others. The white dwarf sequence is also visible.

\begin{table}[t]
\caption{Overview of reference values used in constructing the composite HRD for open clusters (Figure~\ref{fig:hr30open}).}
\centering
\begin{tabular}{lrrrlr}
\hline\hline
Cluster      & DM     & log(age) &[Fe/H] &E(B$-$V) & Memb\\
\hline
Hyades       &  3.389 & 8.90     &  0.13 & 0.001 &  480 \\ %
Coma Ber     &  4.669 & 8.81     &  0.00 & 0.000 &  127 \\ %
Pleiades     &  5.667 & 8.04     & -0.01 & 0.045 & 1059 \\ %
IC 2391      &  5.908 & 7.70     & -0.01 & 0.030 &  254 \\ %
IC 2602      &  5.914 & 7.60     & -0.02 & 0.031 &  391 \\ %
$\alpha$~Per &  6.214 & 7.85     &  0.14 & 0.090 &  598 \\ %
Praesepe     &  6.350 & 8.85     &  0.16 & 0.027 &  771 \\ %
NGC 2451A    &  6.433 & 7.78     & -0.08 & 0.000 &  311 \\ %
Blanco 1     &  6.876 & 8.06     &  0.03 & 0.010 &  361 \\ %
NGC 6475     &  7.234 & 8.54     &  0.02 & 0.049 &  874 \\ %
NGC 7092     &  7.390 & 8.54     &  0.00 & 0.010 &  248 \\ %
NGC 6774     &  7.455 & 9.30     &  0.16 & 0.080 &  165 \\ %
NGC 2232     &  7.575 & 7.70     &  0.11 & 0.031 &  241 \\ %
NGC 2547     &  7.980 & 7.60     & -0.14 & 0.040 &  404 \\ %
NGC 2516     &  8.091 & 8.48     &  0.05 & 0.071 & 1727 \\ %
Trumpler 10  &  8.223 & 7.78     & -0.12 & 0.056 &  407 \\ %
NGC 752      &  8.264 & 9.15     & -0.03 & 0.040 &  259 \\ %
NGC 6405     &  8.320 & 7.90     &  0.07 & 0.139 &  544 \\ %
IC 4756      &  8.401 & 8.98     &  0.02 & 0.128 &  515 \\ %
NGC 3532     &  8.430 & 8.60     &  0.00 & 0.022 & 1802 \\ %
NGC 2422     &  8.436 & 8.11     &  0.09 & 0.090 &  572 \\ %
NGC 1039     &  8.552 & 8.40     &  0.02 & 0.077 &  497 \\ %
NGC 6281     &  8.638 & 8.48     &  0.06 & 0.130 &  534 \\ %
NGC 6793     &  8.894 & 8.78     &       & 0.272 &  271 \\ %
NGC 2548     &  9.451 & 8.74     &  0.08 & 0.020 &  374 \\ %
NGC 6025     &  9.513 & 8.18     &       & 0.170 &  431 \\ %
NGC 2682     &  9.726 & 9.54     &  0.03 & 0.037 & 1194 \\ %
IC 4651      &  9.889 & 9.30     &  0.12 & 0.040 &  932 \\ %
NGC 2323     & 10.010 & 8.30     &       & 0.105 &  372 \\ %
NGC 2447     & 10.088 & 8.74     & -0.05 & 0.034 &  681 \\ %
NGC 2360     & 10.229 & 8.98     & -0.03 & 0.090 &  848 \\ %
NGC 188      & 11.490 & 9.74     &  0.11 & 0.085 &  956 \\ %
\hline 
\end{tabular}
\label{tab:openclref}
\tablefoot{Distance moduli (DM) as derived from the \gaia astrometry; ages and reddening values as derived from \gaia photometry (see Sect.~\ref{sec:ociso}), with distances fixed on astrometric determinations; metallicities from \cite{2016A&A...585A.150N}; Memb: the number of members with \gaia photometric data after application of the photometric filters.}
\end{table}

\subsection{Selection of globular clusters\label{sec:gcdata}}

The details of selecting globular clusters are presented in \cite{DR2-DPACP-34}. A major issue for the globular cluster data is the uncertainties on the parallaxes that result from the systematics, which is in most cases about one order of magnitude larger than the standard uncertainties on the mean parallax determinations for the globular clusters. The implication of this is that the parallaxes as determined with the \gaia data cannot be used to derive the  distance moduli needed to prepare the composite HRD for the globular clusters. Instead, we had to rely on distances as quoted in the literature, for which we used the tables (2010 edition) provided online by \cite{1996AJ....112.1487H}. The inevitable drawback is that these distances and reddening values have been obtained through isochrone fitting, and the application of these values to the \gaia data will provide only limited new information. The main advantage is the possibility of comparing the HRDs of all globular clusters within a single, accurate photometric system. The combined HRD for 14 globular clusters is shown in Fig.~\ref{fig:compgchrd}, the summary data for these clusters is presented in Table~\ref{tab:globclust}. The photometric data originate predominantly from the outskirts of the clusters, as in the cluster centres the crowding often affects the colour index determination.  
Figure~\ref{fig:compgchrd} shows the blue horizontal branch populated with the metal-poor clusters and the move of the giant branch towards the blue with decreasing metallicity. 

\begin{figure}[t]
\centering
\includegraphics[width=8.5cm]{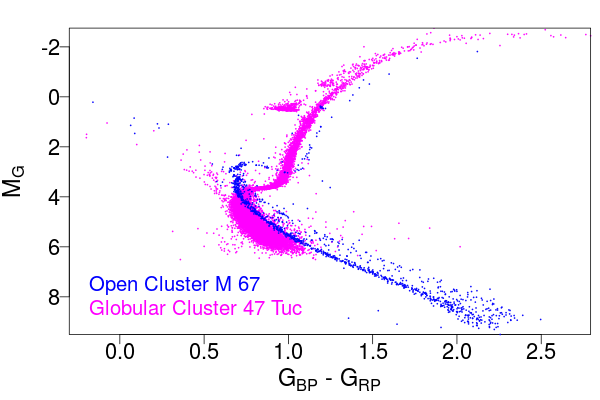}
\caption{Comparison between the HRDs of 47~Tuc (NGC~104, Age=12.75~Gyr, [Fe/H]=-0.72), one of the most metal-rich globular clusters (magenta dots), and M~67 (NGC~2682, Age=3.47~Gyr, [Fe/H]=0.03), one of the oldest open clusters (blue dots). }
\label{fig:m6747tuc}
\end{figure}

An interesting comparison can be made between the most metal-rich well-populated globular cluster of our sample, 47~Tuc (NGC~104), and one of the oldest open clusters, M67 (NGC~2682) (Fig.~\ref{fig:m6747tuc}). This provides the closest comparison between the HRDs of an open and a globular cluster. Most open clusters are much younger, while most globular clusters are much less metal rich.

\begin{table}[t]
\caption{Reference data for 14 globular clusters used in the construction of the combined HRD (Figure~\ref{fig:compgchrd}).}
\centering
\begin{tabular}{rrrlrr}
\hline\hline
NGC &DM  &Age&[Fe/H] &E(B$-$V) & Memb \\
&&(Gyr)&&&\\
\hline
 104 & 13.266 & 12.75$^1$ & -0.72 & 0.04 & 21580 \\ 
 288 & 14.747 & 12.50$^1$ &-1.31 & 0.03 & 1953 \\
 362 & 14.672 & 11.50$^1$ &-1.26 & 0.05 & 1737 \\
1851 & 15.414 & 13.30$^3$ & -1.18 & 0.02 & 744 \\
5272 & 15.043 & 12.60$^2$ & -1.50 & 0.01 & 9086 \\
5904 & 14.375 & 12.25$^1$ & -1.29 & 0.03 & 3476 \\
6205 & 14.256 & 13.00$^1$ &-1.53 & 0.02 & 10311 \\
6218 & 13.406 & 13.25$^1$ & -1.37 & 0.19 & 3127 \\
6341 & 14.595 & 13.25$^1$ & -2.31 & 0.02 & 1432 \\
6397 & 11.920 & 13.50$^1$ &-2.02 & 0.18 & 10055 \\
6656 & 12.526 & 12.86$^3$ &-1.70 & 0.35 & 9542 \\
6752 & 13.010 & 12.50$^1$ & -1.54 & 0.04 & 10779 \\
6809 & 13.662 & 13.50$^1$ & -1.94 & 0.08 & 8073 \\
7099 & 14.542 & 13.25$^1$ & -2.27 & 0.03 & 1016 \\
\hline
\end{tabular}
\label{tab:globclust}
\tablefoot{Data on distance moduli (DM), [Fe/H] and E(B$-$V) from \cite{1996AJ....112.1487H}, 2010 edition, (1): \cite{2010ApJ...708..698D}, (2) \cite{2017ApJ...849..159D}, (3) \cite{2017ApJ...844..104P} for age estimates. Memb: cluster members with photometry after application of photometric filters.}
\end{table}

\section{Details of the \gaia HRDs\label{sec:HRDdetails}}

\begin{figure}
\centering
\includegraphics[width=9cm]{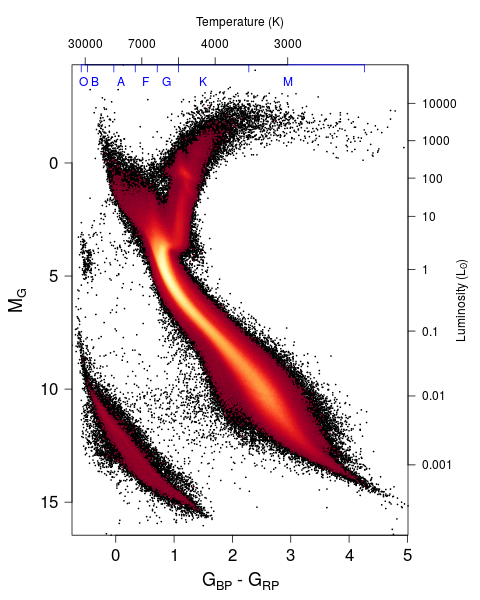}
\caption{\gaia HRD of sources with low extinction ($E(B-V)<0.015$~mag) satisfying the filters described in Sect.~\ref{sec:filters} (4,276,690 stars). 
The colour scale represents the square root of the density of stars.
Approximate temperature and luminosity equivalents for main-sequence stars are provided at the top and right axis, respectively, to guide the eye. }\label{fig:fullHRD}
\end{figure}

In the following, several field star HRDs are presented. Unless otherwise stated, the filters presented in Sect.~\ref{sec:filters}, including the $E(B-V)<0.015$~mag criteria, were applied. The HRDs use a red colour scale that represents the square root of the density of stars.
The \gaia DR2 HRD of the low-extinction stars is represented in Fig. \ref{fig:fullHRD}. The approximate equivalent temperature and luminosity to the \bpmrp\ colour and the absolute \gaia $M_G$ magnitude  provided in the figure were determined using the PARSEC isochrones \citep{2017ApJ...835...77M} for main-sequence stars. 

\begin{figure*}
    \centering
   \includegraphics[width=0.325\linewidth]{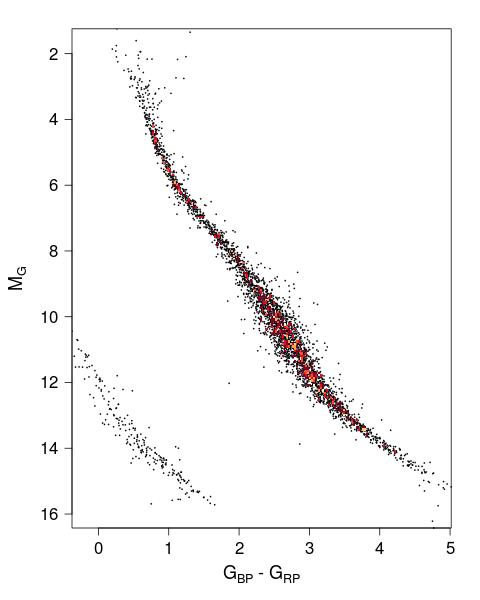}
   \includegraphics[width=0.325\linewidth]{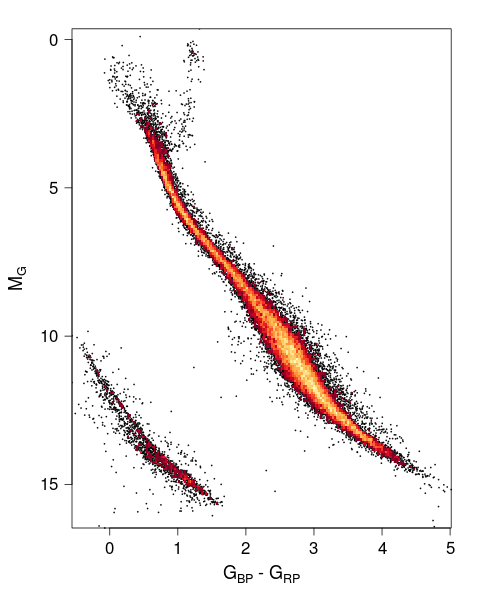}
   \includegraphics[width=0.325\linewidth]{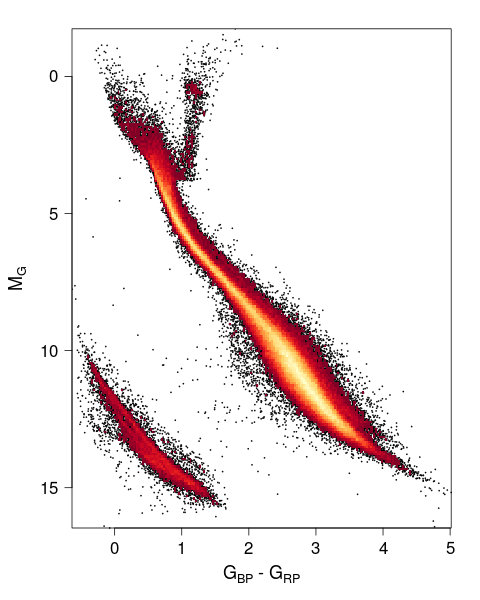}
   \caption{Solar neighbourhood {\gaia} HRDs for a) $\varpi>40$~mas (25 pc, 3,724 stars), b) $\varpi>20$~mas (50 pc, 29,683 stars),
and c) $\varpi>10$~mas (100 pc, 212,728 stars). }
   \label{fig:localHRD}
\end{figure*}

Figure \ref{fig:localHRD} shows the local \gaia HRDs using several cuts in parallax, still with the filters of Sect.~\ref{sec:filters}, but without the need to apply the $E(B-V)<0.015$~mag extinction criteria, as these sources mostly lie within the local bubble. 

\subsection{Main sequence}

\begin{figure}[t]
\centering
\includegraphics[width=9cm]{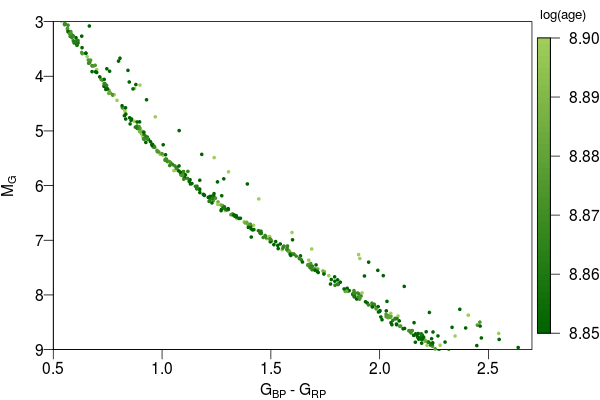}
\caption{Extract of the HRD for the Hyades and Praesepe clusters, showing the detailed agreement between the main sequences of the two clusters, the narrowness of the combined main sequence, and a scattering of double stars up to 0.75~mag\ above the main sequence.}
\label{fig:hyadpreashrd}
\end{figure}

The main sequence is very thin, both in fields and in clusters. This is very clearly visible in Fig.~\ref{fig:hyadpreashrd},
which shows the HRDs of the Hyades and Praesepe clusters (ages $\sim$700 Myr), which accurately overlap, as has previously been noticed in \cite{2009A&A...497..209V} and confirmed in \cite{2017A&A...601A..19G}. This figure shows the very narrow sequence described by the stars in both clusters, as well as the scattering of double stars up to 0.75 magnitudes above the main sequence. The remaining width of the main sequence is still largely explained as due to the uncertainties in the parallax of the individual stars, and the underlying main sequence is likely to be even narrower. 

\begin{figure}
     \centering
    \includegraphics[width=8cm]{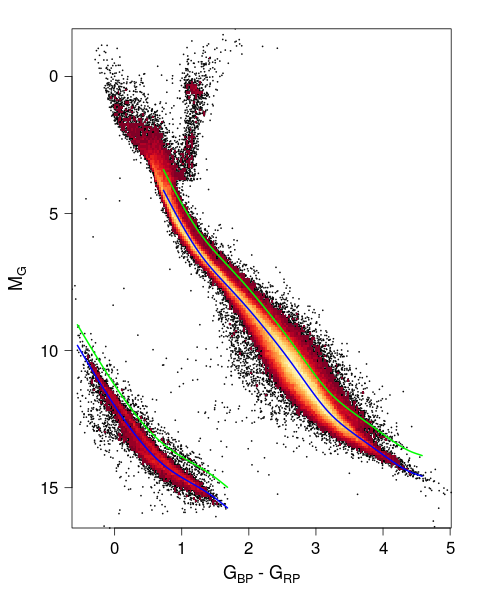}
    \caption{Same as Fig.~\ref{fig:localHRD}c, overlaid in blue with the median fiducial and in green with the same fiducial shifted by -0.753~mag, corresponding to an unresolved binary system of two identical stars.}
    \label{fig:fieldbinaries}
 \end{figure}

The binary sequence spread is visible throughout the main sequence (Figs.~\ref{fig:fullHRD} and~\ref{fig:localHRD}), and most clearly in open clusters (Fig.~\ref{fig:hyadpreashrd}, see also Sect.~\ref{sec:ociso}). It is most preeminent for field stars below $M_G=13$. Figure~\ref{fig:fieldbinaries} shows  the main-sequence fiducial of the local HRD shifted by 0.753~mag, which corresponds to two identical stars in an unresolved binary system observed with the same colour but twice the luminosity of the equivalent single star. See \cite{1998MNRAS.300..977H},
for instance, for a discussion of this strong sequence. Binaries with a main-sequence primary and a giant companion would lie much higher in the diagram, while binaries with a late-type main-sequence primary and a white dwarf companion lie between the white dwarf and the main sequence, as is shown in Fig.~\ref{fig:fullHRD},
for example. 

The main sequence is thicker between $10<M_G<13$ (Figs.~\ref{fig:hr30open}, \ref{fig:fullHRD}, \ref{fig:localHRD}). The youngest main-sequence stars lie on the upper part of the main sequence (in blue in Fig.~\ref{fig:hr30open}). The subdwarfs, which are metal-poor stars associated with the halo, are visible in the lower part of the local HRD (in red in Fig.~\ref{fig:hr30open}, see also Sect.~\ref{sec:kine}). 

The main-sequence turn-off variation with age is clearly illustrated in Fig.~\ref{fig:hr30open}, and the variation with metallicity is shown in Fig.~\ref{fig:compgchrd}. Blue stragglers are also visible over the main-sequence turn-off (Figs.~\ref{fig:m6747tuc}). 

Between the main sequence and the subgiants lies a tail of stars around $M_G=4$ and \bpmrp=1.5. These stars shows variability and may be associated with RS Canum Venaticorum variables, which are close binary stars \citep{DR2-DPACP-35}.

\subsection{Brown dwarfs}\label{sec:brown_dwarfs}

\begin{figure*}
    \centering 
   \includegraphics[height=7cm]{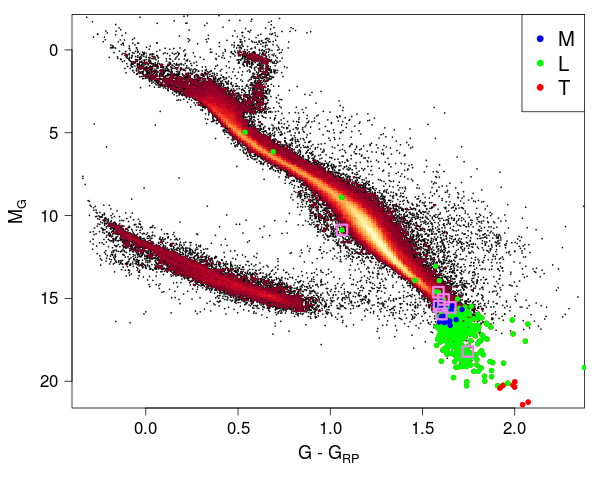}
   \includegraphics[height=7cm]{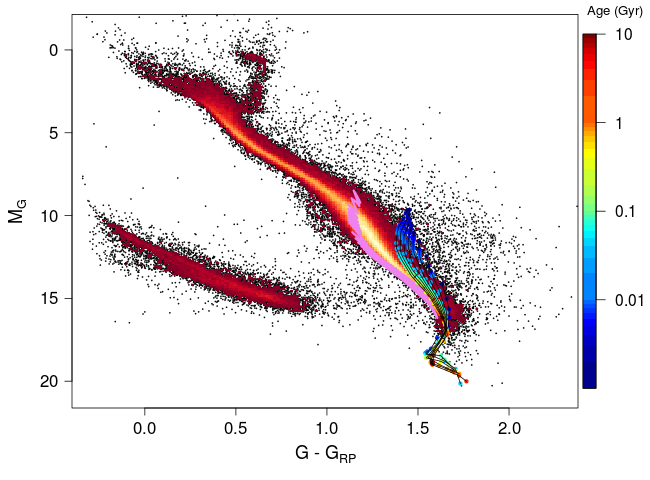} 
   \includegraphics[height=7cm]{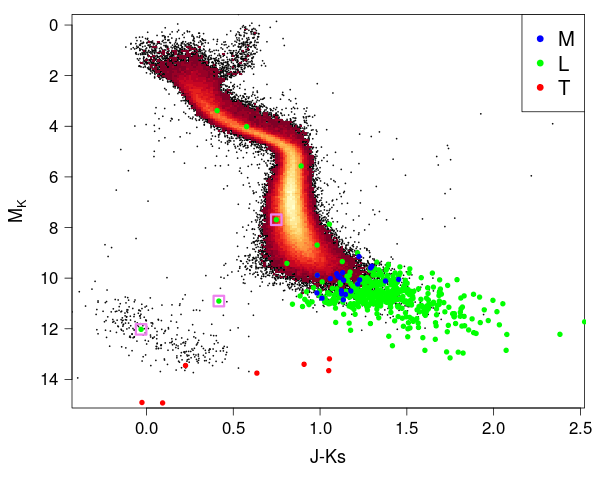}
   \includegraphics[height=7cm]{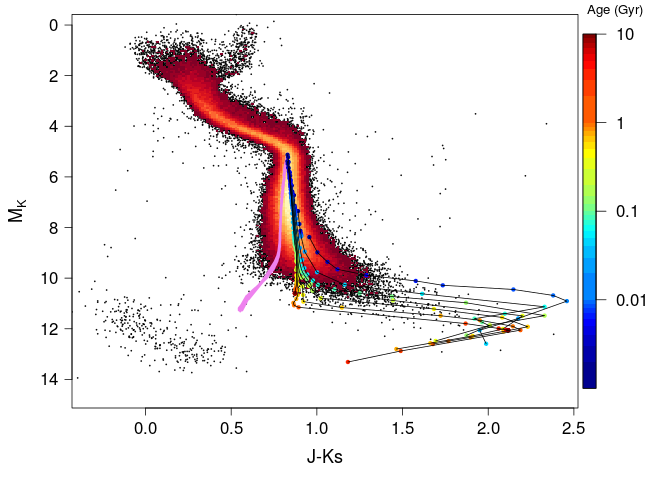}
   \caption{a): {\gaia} HRD of the stars with $\varpi>10$~ mas with adapted  photometric filters (see text, 240,703 stars) overlaid with all cross-matched GUCDS \citep{2017MNRAS.469..401S} stars with $\sigma_\varpi/\varpi<10\%$ in blue (M type), green (L type), and red (T type). Pink squares are added around stars with tangential velocity \VT$>$200\kms. 
   b) BT-Settl tracks \citep{2015A&A...577A..42B} of solar metallicity for masses from 0.01 \Msol\ to 0.08 \Msol\ in steps of 0.01 (the upper tracks correspond to lower masses) plus in pink the same tracks for [M/H]= -1.0.
   Panels c) and d): Same diagrams using the 2MASS colours.
   }
   \label{fig:BDs}
\end{figure*}

To study the location of the low-mass objects in the \gaia HRD, we used the \gaia ultracool dwarf sample (GUCDS) compiled by \cite{2017MNRAS.469..401S}. It includes 1886 brown dwarfs (BD) of L, T, and Y types, although a substantial fraction of them are too faint for \gaia. We note that the authors found 328 BDs in common with the \gaia DR1 catalogue \citep{DR1-DPACP-8}.

The crossmatch between the 2MASS catalogue \citep{2006AJ....131.1163S} and \gaia DR2 provided within the \gaia archive \citep{DR2-DPACP-41} has been used to identify GUCDS entries. 
The resulting sample includes 601 BDs. Of these, 527 have five-parameter solutions (coordinates, proper motions, and parallax) and full photometry ($G$, $G_{BP}$, and $G_{RP}$). Most of these BDs have parallaxes higher than 4 mas (equivalent to 250 pc in distance) and relative parallax errors smaller than 25\%. They also have astrometric excess noise larger than 1~mas and a high $(I_{BP}+I_{RP})/I_G$ flux ratio. They are faint red objects with very low flux in the BP wavelength range of their spectrum. Any background under-estimation causes the measured BP flux to increase to more than it should be, yielding high flux ratios, the highest ratios are derived for the faintest BDs.
The filters presented in Sect.~\ref{sec:filters} therefore did not allow us to retain them.

We accordingly adapted our filters for the background stars of Fig.~\ref{fig:BDs}. We plot the HRD using the $G-G_{\rm RP}$ colour instead of $G_{\rm BP}-G_{\rm RP}$ because of the poor quality of \gbp\ for these faint red sources. We applied the same astrometric filters as for Fig. \ref{fig:localHRD}c, but we did not filter the fluxes ratio or the \gbp\ photometric uncertainties. More dispersion is present in this diagram than in Fig. \ref{fig:localHRD}c because
of this missing filter, but the faint red sources we study here are represented better. 

The 470 BDs for which DR2 provides parallaxes better than 10\%,
and the \gmag\ and \grp\ magnitudes are overlaid in Fig.~\ref{fig:BDs} without any filtering.
The sequence of BDs follows the sequence of low-mass stars. The absolute magnitudes of four stars are too bright, most probably because of a cross-match issue. In Fig.~\ref{fig:BDs}a the M-, L- and T-type BDs are sorted according to the classification in GUCDS. There are 21, 443, and 7 of each type, respectively. 
We also present in Fig.~\ref{fig:BDs}c the corresponding HRD using 2MASS colours with the 2MASS photometric quality flag AAA (applied to background and GUCDS stars). Figures~\ref{fig:BDs}b,d includes BT-Settl tracks\footnote{https://phoenix.ens-lyon.fr/Grids/BT-Settl/CIFIST2011bc\label{note:BTSettl}} \citep{2015A&A...577A..42B} for masses $<0.08$~{\Msol} that were computed using the nominal \gaia passbands. With the \gaia $G-G_{\rm RP}$ colour, the sequence of M, L, T types is continuous and relatively thin. Conversely, the spread in the near-infrared is larger and the L/T transition feature is strongly seen with a shift of $J-K_s$ to the blue, which is due to a drastic change in the brown dwarf cloud properties \citep[e.g.][]{2008ApJ...689.1327S}. 
Some GUCDS L-type stars with very blue 2MASS colours seem at first sight intriguing, but their location might be consistent with metal-poor tracks (Fig.~\ref{fig:BDs}d). Following \cite{2009AJ....137....1F}, we studied their kinematics, which are indeed consistent with the halo kinematic cut of the tangential velocity \VT$>$200\kms (see Sect. \ref{sec:kine}). A kinematic selection of the global HRD as done in Sect. \ref{sec:kine} but using the 2MASS colours confirms the blue tail of the bottom of the main sequence in the near-infrared for the halo kinematic selection.

\subsection{Giant branch\label{sec:giants}} 

\begin{figure}
    \centering
   \includegraphics[width=9cm]{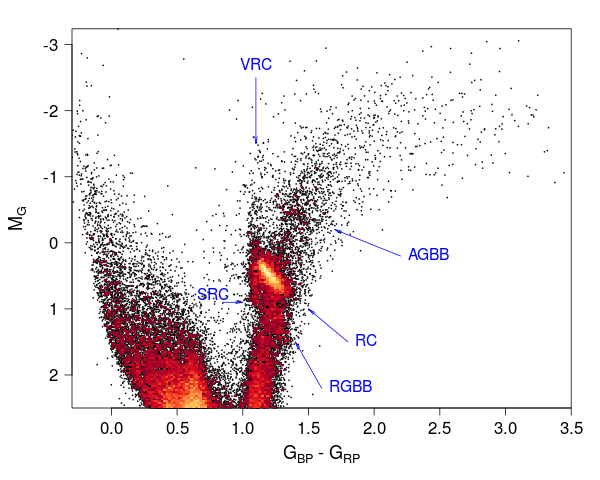}
   \caption{\gaia HRD of low-extinction nearby giants: $\varpi>2$~mas (500~pc), $E(B-V)<0.015$ and $M_G<2.5$ (29288 stars), with labels to the features discussed in the text.}
   \label{fig:giants}
\end{figure}

The clusters clearly illustrate the change in global shape of the giant branch with age and metallicity (Figs.~\ref{fig:hr30open} and \ref{fig:compgchrd}). For field stars, there are fewer giants
than dwarfs in the first 100~pc. To observe the field giant branch
in more detail, we therefore extended our selection to 500~pc with the low-extinction selection ($E(B-V)<0.015$, see Sect.~\ref{sec:extinction}) for Fig.~\ref{fig:giants}. 

\begin{figure*}
    \centering
   \includegraphics[width=4.5cm]{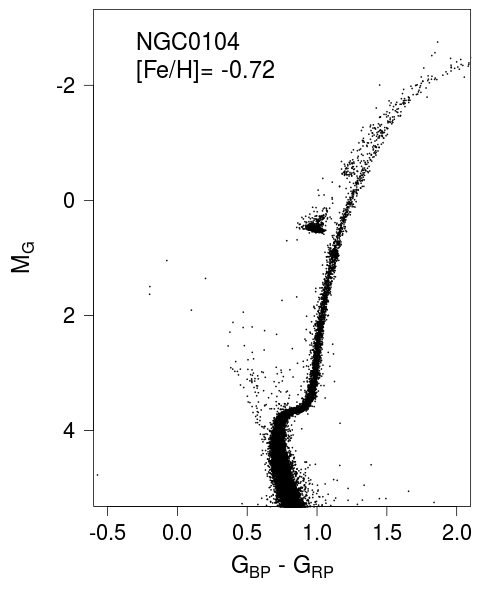}
   \includegraphics[width=4.5cm]{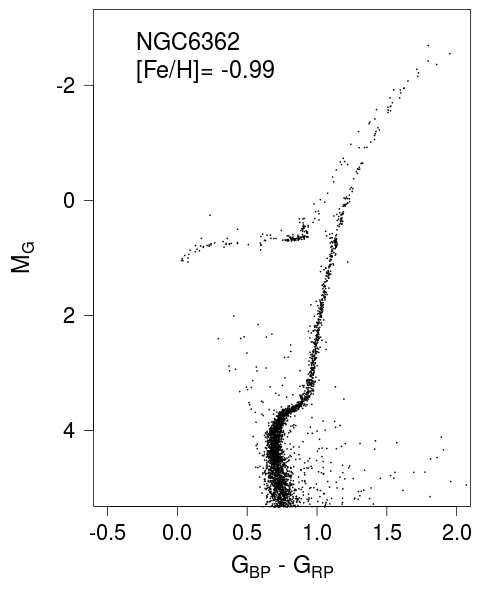}
   \includegraphics[width=4.5cm]{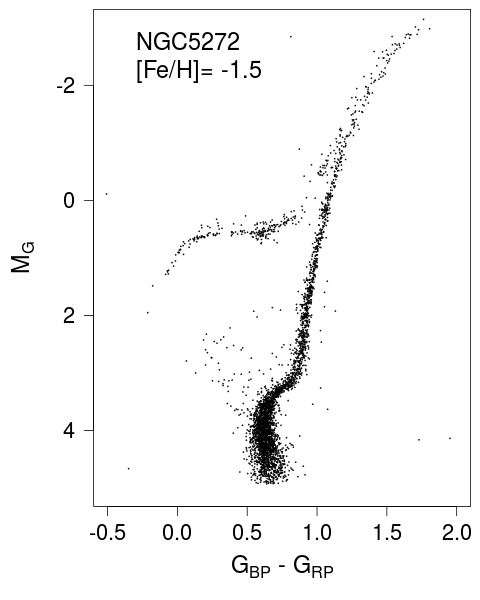}
   \includegraphics[width=4.5cm]{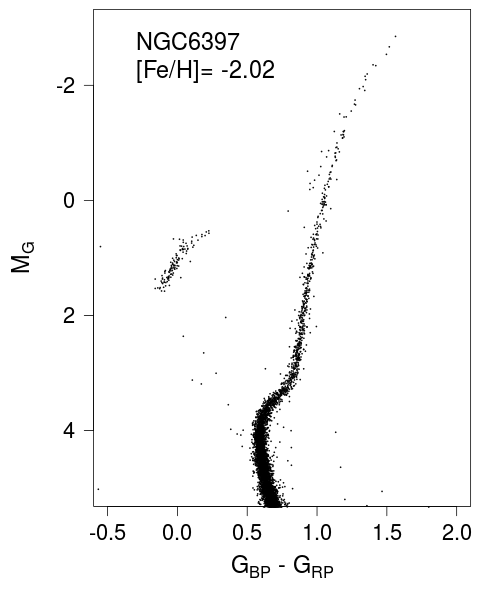}
   \caption{Several globular clusters selected to show a clearly defined and very different horizontal branch, sorted by decreasing metallicity. a) NGC 104 (47 Tuc), b) NGC 6362, c) NGC 5272, and d) NGC 6397.}
   \label{fig:GCs}
\end{figure*}

The most prominent feature of the giant branch is the Red Clump (RC in Fig.~\ref{fig:giants}, around  \bpmrp$=1.2$, $M_G=0.5$~mag). It corresponds to low-mass stars that burn helium in their core \citep[e.g.][]{2016ARA&A..54...95G}. 
The colour of core-helium burning stars is strongly dependent on metallicity and age. The more metal-rich, the redder, which
leads to this red clump feature in the local HRD. For more metal-poor populations, these stars are bluer and lead to the horizontal-branch (HB) feature that is clearly visible in globular clusters (Fig.~\ref{fig:GCs}). 

The secondary red clump (SRC in Fig.~\ref{fig:giants}, around \bpmrp$=1.1$, $M_G=0.6$) is more extended in its bluest part to fainter magnitudes than the red clump. It corresponds to younger more massive red clump stars \citep{1999MNRAS.308..818G} and is therefore mostly visible in the local HRD (Fig.~\ref{fig:localHRD}c).
Core-helium burning stars that are even more massive are more luminous than the red clump and lie still on the blue part of it, leading to a vertical structure that is sometimes called the Vertical Red Clump (VRC in Fig.~\ref{fig:giants}). 

On the red side and fainter than the clump lies the RGB bump (RGBB in Fig.~\ref{fig:giants}). 
This bump is caused by a brief interruption of the stellar luminosity increase as a star evolves on the red giant branch by burning its hydrogen shell, which creates an accumulation of stars at this HRD position \citep[e.g.][]{2015MNRAS.453..666C}. Its luminosity changes more with metallicity and age than the red clump. 
Brighter than the red clump, at $M_G\sim-0.5$, lies the AGB bump (AGBB in Fig.~\ref{fig:giants}), which corresponds to the start of the asymptotic giant branch (AGB) where stars are burning their helium shell \citep[e.g.][]{1998ApJ...495L..43G}. The AGB bump is much less densely populated than the RGB bump. It is also clearly visible in the HRD of 47 Tuc (Fig.~\ref{fig:GCs}a). 

 The globular clusters in Fig.~\ref{fig:GCs} clearly illustrate the diversity of the HB  morphology. Some have predominantly blue HB (NGC~6397), some just red HB (NGC~104), and some a mixed HB showing bimodal distribution (NGC~5272 and NGC~6362). The HB morphology is  explained in the framework of the multiple populations; it is regulated by age, metallicity, and first/second generation abundances  \citep{2009A&A...505..117C}.
NGC~6362 is the least massive globular that presents multiple populations. \cite{2016ApJ...824...73M} concluded that most of the stars that populate the red HB are Na poor and belong to the first generation, while the blue side of the HB is populated by the Na-rich stars belonging to the second generation. The same kind of correlation is shown in general by the globular clusters. We quote among others the studies of 47 Tuc \citep{2013A&A...549A..41G} and NGC~6397 \citep{2009A&A...505..117C}.
The role of the He abundances is still under discussion \citep{2016A&A...589A.126V, 2014MNRAS.437.1609M}. He-enhanced stars  are indeed  expected  to  populate  the  blue  side of the instability strip because
they are still   O  depleted  and  Na  enhanced,
as observed in the  second-generation stars. How significant the He enhancement is is still unclear.

Figure~\ref{fig:compgchrd} shows that the globular cluster HB can extend towards the extreme horizontal branch (EHB) region. They are in the same region of the HRD as the hot subdwarfs, which creates a clump at $M_G$=4 and \bpmrp=-0.5 that is well visible in Fig.~\ref{fig:fullHRDwithext} and Fig.~\ref{fig:fullHRD}. These stars are also nicely characterised in terms of variability, including binary-induced variability, in \cite{DR2-DPACP-35}. These hot subdwarfs are considered to be red giants that lost their outer hydrogen layers before the core began to fuse helium, which might be due to the interaction with a low-mass companion, although other processes might be at play \citep[e.g.][]{2009ARA&A..47..211H}. {\gaia} will allow detailed studies of the differences between cluster and field hot subdwarfs.

\subsection{Planetary nebulae}

\begin{figure}
    \centering
   \includegraphics[width=8cm]{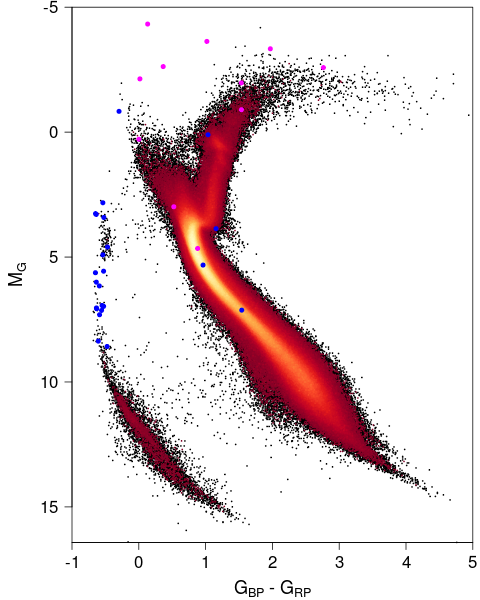}
   \caption{North Galactic Pole HRD ($b>50\degr$, 2,077,925 stars) with literature central planetary nebula stars (blue) and post-AGB stars (magenta).}
   \label{fig:PN}
\end{figure}

At the end of the AGB phase, the star has lost most of its hydrogen envelope. The gas expands while the central star first grows hotter at constant luminosity, contracting and fusing hydrogen in the shell around its core (post-AGB phase), then it slowly cools when the hydrogen shell is exhausted, to reach the white dwarf phase. This planetary nebulae phase is very short, about 10,000 years, and is therefore quite difficult to observe in the HRD. The \gaia DR2 contains many observations of nearby planetary nebulae as their expanding gas create excess flux over the mean sky background that triggers the on-board detection. We here wish to follow the route of the central star in the HRD. While some central planetary nebula stars are visible in the Galactic Pole HRD (Fig.~\ref{fig:PN}), post-AGB stars are too rare to appear in this diagram. We used catalogue compilations to highlight the position of the two types in the \gaia HRD. 

We used the \cite{2003A&A...408.1029K} catalogue of Galactic planetary nebulae, selecting only sources classified as central stars that are clearly separated from the nebula. 
With a cross-match radius of 1$\arcsec$ and using all our filter criteria of Sect.~\ref{sec:filters}, only four stars remain. We therefore relaxed the extinction criteria to $E(B-V)<0.05$  and the parallax relative uncertainty to $\sigma_\varpi/\varpi<20$\%, leading to 23 stars.  

For post-AGB stars, we used the catalogue of \cite{2007A&A...469..799S} and the 2MASS identifier provided for the cross-match. We selected only stars that are classified as very likely post-AGB objects. Here we also relaxed the extinction criteria to $E(B-V)<0.05$ and the parallax relative uncertainty to $\sigma_\varpi/\varpi<20$\%, leading to 11 stars. 

While some outliers are seen in Fig.~\ref{fig:PN}, either due to cross-match or misclassification issues, the global position of these stars in the HRD closely follows the expected track from the AGB to the white dwarf sequence. We note that this path crosses the hot subdwarf region we discussed in the previous section. 

\section{\label{sec:WDs}White dwarfs}

\begin{figure}
    \centering
   \includegraphics[width=8cm]{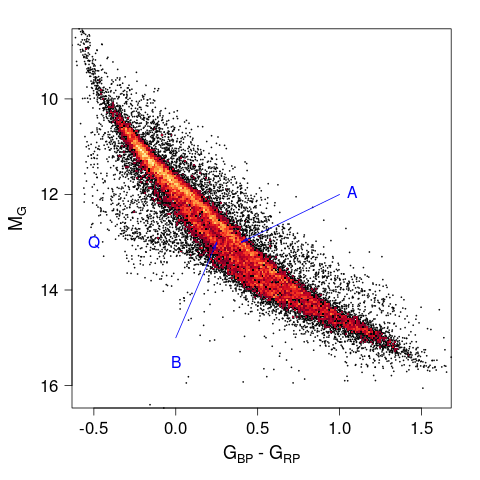}
   \caption{\gaia HRD of white dwarfs with $\sigma_\varpi/\varpi<5$\% (26,264 stars), with letter labels to the features discussed in the text.}
   \label{fig:gaiaWDsAnno}
\end{figure}

The Sloan Digital Sky Survey \citep[SDSS,][]{SDSS9} has produced the largest spectroscopic catalogue of white dwarfs so far \citep[e.g.][]{2013ApJS..204....5K}.
This data set has greatly aided our understanding of white dwarf classification and evolution. For example, it has allowed determining the white dwarf mass distribution for large statistical samples of different white dwarf spectral types. However, much of this work is model dependent and relies upon theoretical mass-radius relationships and stellar atmosphere models, whose precision has only been tested in a limited way. These tests have been limited by the relatively small number of white dwarfs for which accurate parallaxes are available \citep[e.g.][]{1998ApJ...494..759P} and by the precision of the parallaxes for these faint stars. This work was updated using the \gaia DR1 catalogue \citep{2017MNRAS.465.2849T}, which included more stars, but the uncertainties remain too large to constrain the theoretical mass-radius relations. Only in a few cases, where the white dwarf resides in a binary system, have mass radius measurements begun to approach the accuracy required to constrain the core composition and H layer mass of individual stars \citep[e.g.][]{2005MNRAS.362.1134B, 2017MNRAS.470.4473P, 2017ASPC..509..389J}. Even then, some of these white dwarfs may not be representative of the general population because common envelope evolution may have caused them to depart from the normal white dwarf evolutionary paths.

\begin{figure}
    \centering
   \includegraphics[width=8cm]{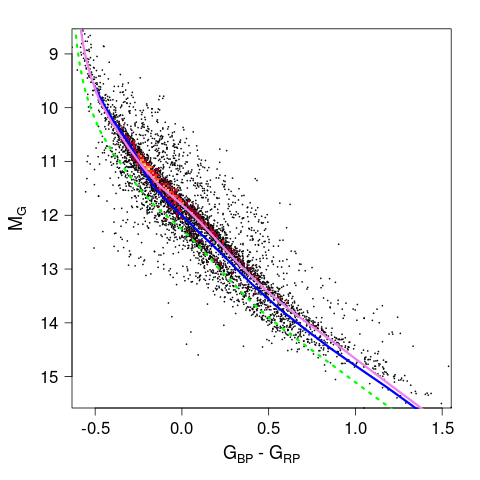}
   \includegraphics[width=8cm]{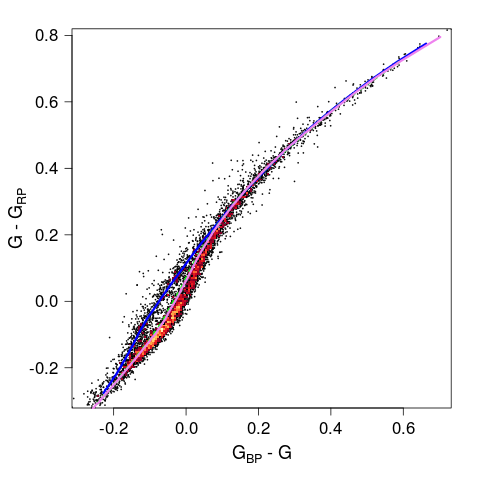}
   \caption{\gaia HRD of white dwarfs with $\sigma_\varpi/\varpi<5$\% and $\sigma_{G_{BP}}<0.01$ and $\sigma_{G_{RP}}<0.01$ (5,781 stars) overlaid with white dwarf evolutionary models.
   {\it  Magenta:}  0.6 \Msol\ pure H; 
   {\it green dashed:}  0.8 \Msol\ pure H;  and
   {\it blue:}  0.6 \Msol\ pure He.
   a): HRD. b): Colour-colour diagram. 
   }
   \label{fig:gaiaWDs}
\end{figure}

The publication of \gaia DR2 presents the opportunity to apply accurate parallaxes, with uncertainties of 1\%\ or smaller, to the study of white dwarf stars. The availability of these data, coupled with the accurate \gaia photometry, yields the absolute magnitude, with which the white dwarfs can be clearly located in the expected region of the HRD (Figs.~\ref{fig:fullHRD}, \ref{fig:localHRD}). Figure \ref{fig:gaiaWDsAnno} shows the white dwarf region of the HRD alone. This sample was selected with $G_{\rm BP}-G_{\rm RP}<2$ and $G-10+5\log_{10}\varpi>10+2.6\ (G_{\rm BP}-G_{\rm RP})$ and by applying the filters described in Sect.~\ref{sec:building}, including the low-extinction $E(B-V)<0.015$ criterion, but with a stronger constraint on the parallax relative uncertainty of 5\%. This yields a catalogue of 26,264 objects. We overplot in Fig. \ref{fig:gaiaWDs} white dwarf evolutionary models\footnote{http://www.astro.umontreal.ca/\textasciitilde bergeron/CoolingModels} for C/O cores \citep{2006AJ....132.1221H, 2006ApJ...651L.137K, 2011A&A...531L..19T,2011ApJ...737...28B} with colours computed using the revised \gaia DR2 passbands \citep{DR2-DPACP-40}. 

Several features are clearly visible in Fig.~\ref{fig:gaiaWDsAnno}. First there is a clear main concentration of stars that is distributed continuously from left to right in the diagram \texttt{(A)} and coincides with the 0.6 M$_\odot$ hydrogen evolutionary tracks (in magenta). This is expected because the white dwarf mass distribution peaks very strongly near 0.6 M$_\odot $ \citep{2013ApJS..204....5K}. Interestingly, the concentration of white dwarfs departs from the cooling tracks towards the red end of the sequence. 

Just below the main 0.6 M$_\odot $ concentration of white dwarfs is a second, separate concentration \texttt{(B)} that seems to be separate from the 0.6 M$_\odot $  peak at $G_{\rm BP}-G_{\rm RP}\sim-0.1$ before again merging by $G_{\rm BP}-G_{\rm RP}\sim0.8$. At the maximum separation, this concentration is roughly aligned with the 0.8 M$_\odot $ hydrogen white dwarf cooling track (in green), which is not expected. While the SDSS mass distribution \citep{2013ApJS..204....5K} shows a significant upper tail that extends through 0.8 M$_\odot $ and up to almost 1.2 M$_\odot $, there is no evidence for a minimum between 0.6 and 0.8 M$_\odot $ like that seen in Fig.~\ref{fig:gaiaWDs}a. 
A mass difference should therefore not lead to this feature. 
However, for a given mass, the evolutionary tracks for different compositions (DA: hydrogen and DB: helium) and envelope masses are virtually coincident at the resolution of Fig.~\ref{fig:gaiaWDs} in the theoretical tracks, leading to no direct interpretation from the tracks in the HRD alone, but we describe below a different view from the colour-colour relation and the SDSS comparison.

A third, weaker concentration of white dwarfs in Fig.~\ref{fig:gaiaWDsAnno} lies below the main groups \texttt{(Q)}. It does not follow an obvious evolutionary constant mass curve, which would be parallel to those shown in the plot. Beginning at approximately $M_G = 13$ and $G_{\rm BP}-G_{\rm RP} = - 0.3$, it follows a shallower curve that converges with the other concentrations near $G_{\rm BP} - G_{\rm RP} = 0.2$.

White dwarfs are also seen to lie above the main concentration \texttt{A}. This can be explained as a mix between natural white dwarf mass distributions and binarity (see Fig.~\ref{fig:fieldbinaries}).

Selecting only the most precise \gbp~ and \grp~ photometry ($\sigma_{G_{\rm BP}}<0.01$ and $\sigma_{G_{\rm RP}}<0.01$), we examined the colour-colour relation in Fig.~\ref{fig:gaiaWDs}b. The sequence is also split into two parts in this diagram. We verified that the two splits coincide, meaning that the stars in the lower part of Fig.~\ref{fig:gaiaWDs}a lie in the upper part of Fig.~\ref{fig:gaiaWDs}b. The mass is not expected to lead to significant differences
in this colour-colour diagram, and the theoretical tracks coincide with the observed splits, pointing towards a difference between helium and hydrogen white dwarfs. 
It also recalls the split in the SDSS colour-colour diagram \citep{2003AJ....126.1023H}.

\begin{figure}
    \centering
   \includegraphics[width=8cm]{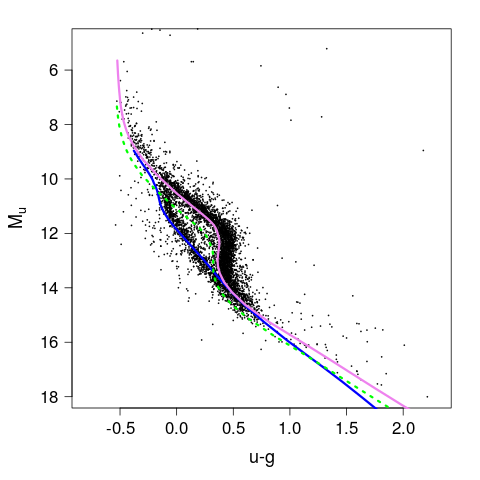}
   \caption{SDSS white dwarfs (5,237 stars) with evolutionary models. $M_u$ is computed using the SDSS $u$ magnitude and the \gaia parallax.   
   {\it  Magenta:}  0.6 \Msol\ pure H; 
   {\it green dashed:}  0.8 \Msol\ pure H; and
   {\it blue} : 0.6 \Msol\ pure He.}
   \label{fig:sdssWDs}
\end{figure}

While \gaia identifies white dwarfs based on their location on the HRD, SDSS white dwarfs were identified spectroscopically, providing further information on the spectral type, $T_{\rm eff}$, and \logg\ as well as a classification. Therefore we cross-matched the two data sets to better understand the features observed in Fig.~\ref{fig:gaiaWDs}. 

We obtained a catalogue of spectroscopically identified SDSS white dwarfs from the Montreal White Dwarf Database\footnote{http://www.montrealwhitedwarfdatabase.org/} \citep{2017ASPC..509....3D} by downloading the whole catalogue and then filtering for SDSS identifier, which yielded 28,797 objects. Using the SDSS cross-match provided in the \gaia archive \citep{DR2-DPACP-41}, we found that there are 22,802 objects in common and 5,237 satisfying all the filters described in Sect.~\ref{sec:filters} and with single-star spectral type information. Figure \ref{fig:sdssWDs} shows the SDSS $u-g$ colour magnitude for the sample with the absolute $u$ magnitude calculated using the \gaia parallax. The distribution is clearly bifurcated. Evolutionary tracks for H and He atmospheres (0.6~M$_\odot $) are overplotted in the figure, indicating that this is due to the different atmospheric compositions. 
The \gaia counterparts of these SDSS white dwarfs are quite faint, and therefore the features seen in Fig.~\ref{fig:gaiaWDs}a are less well visible in this sample because of the larger noise in the parallaxes and the colours. 
Still, it allowed us to verify that the split of the SDSS white dwarfs corresponds to the location of the \gaia splits in Fig.~\ref{fig:gaiaWDs}.
The narrower filter bands of SDSS are more sensitive to atmospheric compositions than the broad BP and RP \gaia bands. In particular, the $u$ -band fluxes of H-rich DA white dwarfs are suppressed by the Balmer jump at 364.6 nm, which reddens the colours of these stars. The Balmer jump is in the wavelength range where the \gaia filters calibrated for DR2 differ most from the nominal filters \citep{DR2-DPACP-40}, which explains the importance of using tracks that are updated to the DR2 filters for the white dwarf studies instead of the nominal tracks provided by \cite{2014A&A...565A..11C}. 

\begin{figure}
\centering
   \includegraphics[width=4.3cm]{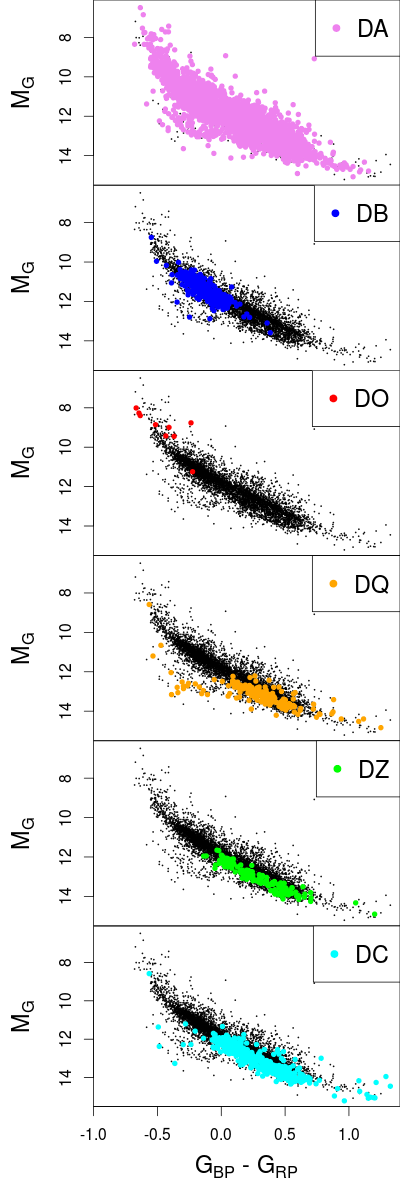}
   \includegraphics[width=4.3cm]{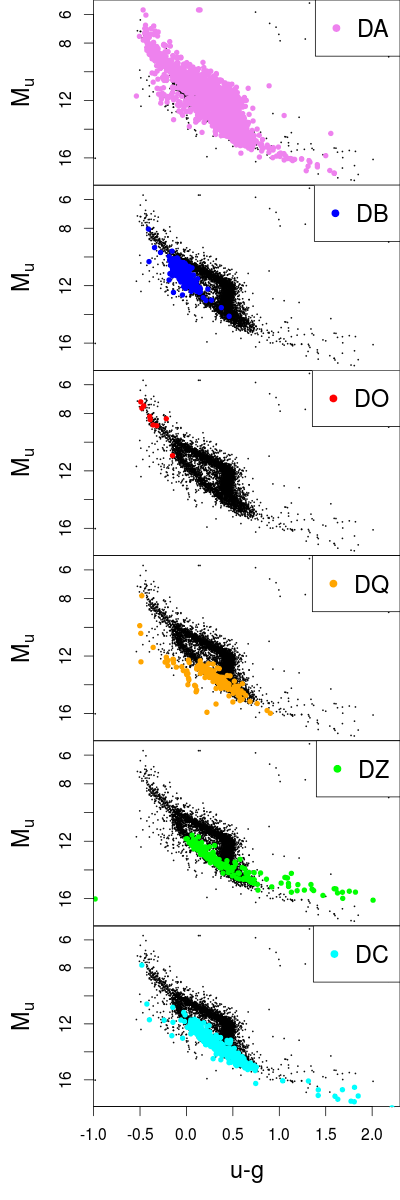}
   \caption{SDSS white dwarfs per spectral type (DA: hydrogen; DB: neutral helium; DO: ionised helium; DQ: carbon; DZ: metal
rich; and DC: no strong lines). panel a): \gaia photometry,  panel b): SDSS photometry.}
   \label{fig:wd_spectraltypes}
\end{figure}

Figure \ref{fig:wd_spectraltypes} shows the colour-magnitude diagrams in the \gaia and SDSS photometry bands, overlaid with the white dwarfs for specific spectral types. The locations of the various spectral types correspond well to the expected colours arising from their effective temperatures. For example, DQ (carbon), DZ (metal rich), and DC (no strong lines) stars are confined to the red end of the colour-magnitude diagram, while the DO stars (ionised helium) all lie at the blue end. DAs cover the whole diagram. Interestingly, in Fig.~\ref{fig:wd_spectraltypes}b, a significant number of classified DA white dwarfs appears to occupy the He-rich atmosphere branch that is indicated by the evolutionary track in Fig.~\ref{fig:sdssWDs}. The weaker  \texttt{Q} concentration seems to include stars of all types except for DO and DZ. However, the most numerous components are the DA and DQs.

\begin{figure*}
\centering
\includegraphics[width=0.46\linewidth]{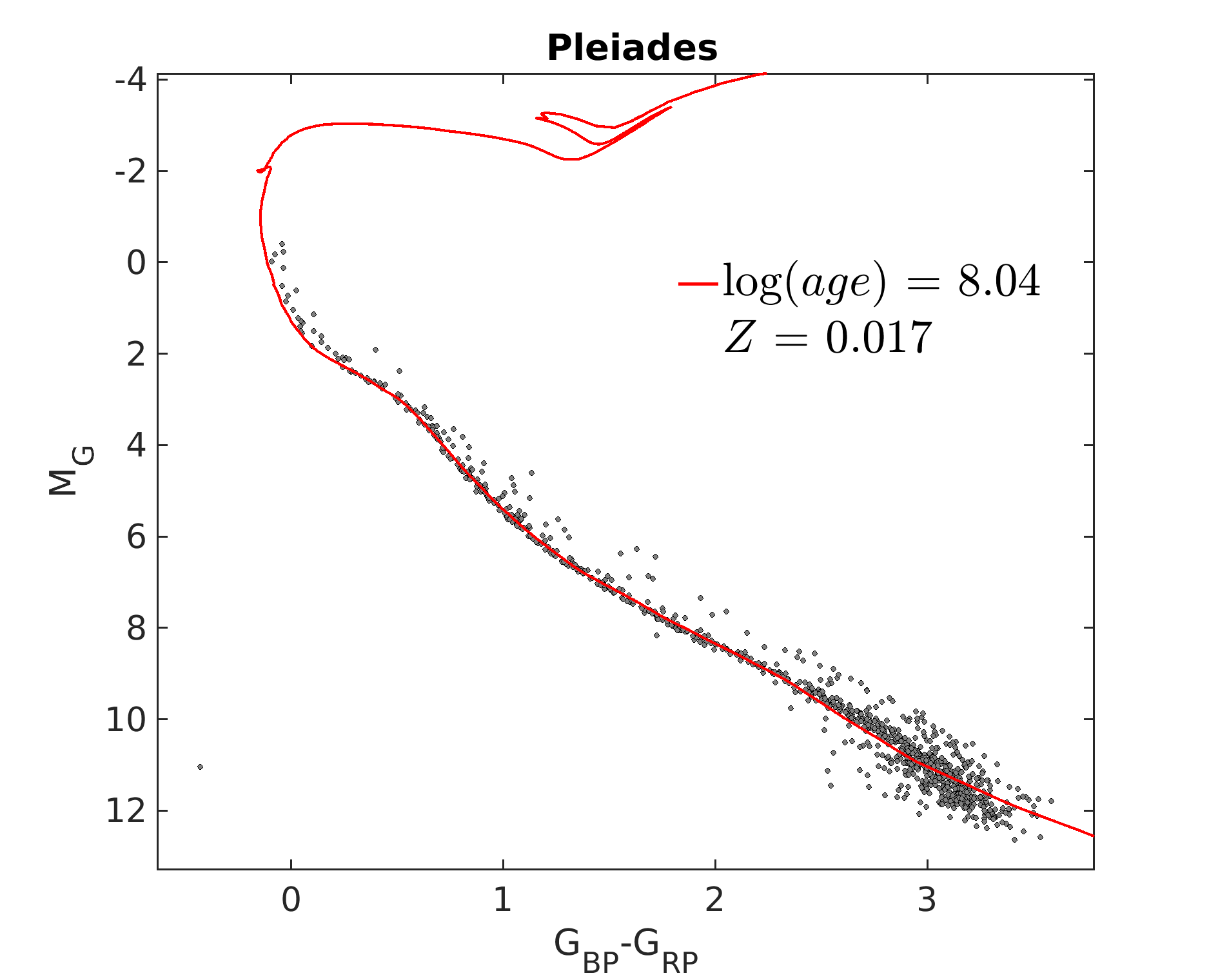}
\hspace{0.5cm}
\includegraphics[width=0.46\linewidth]{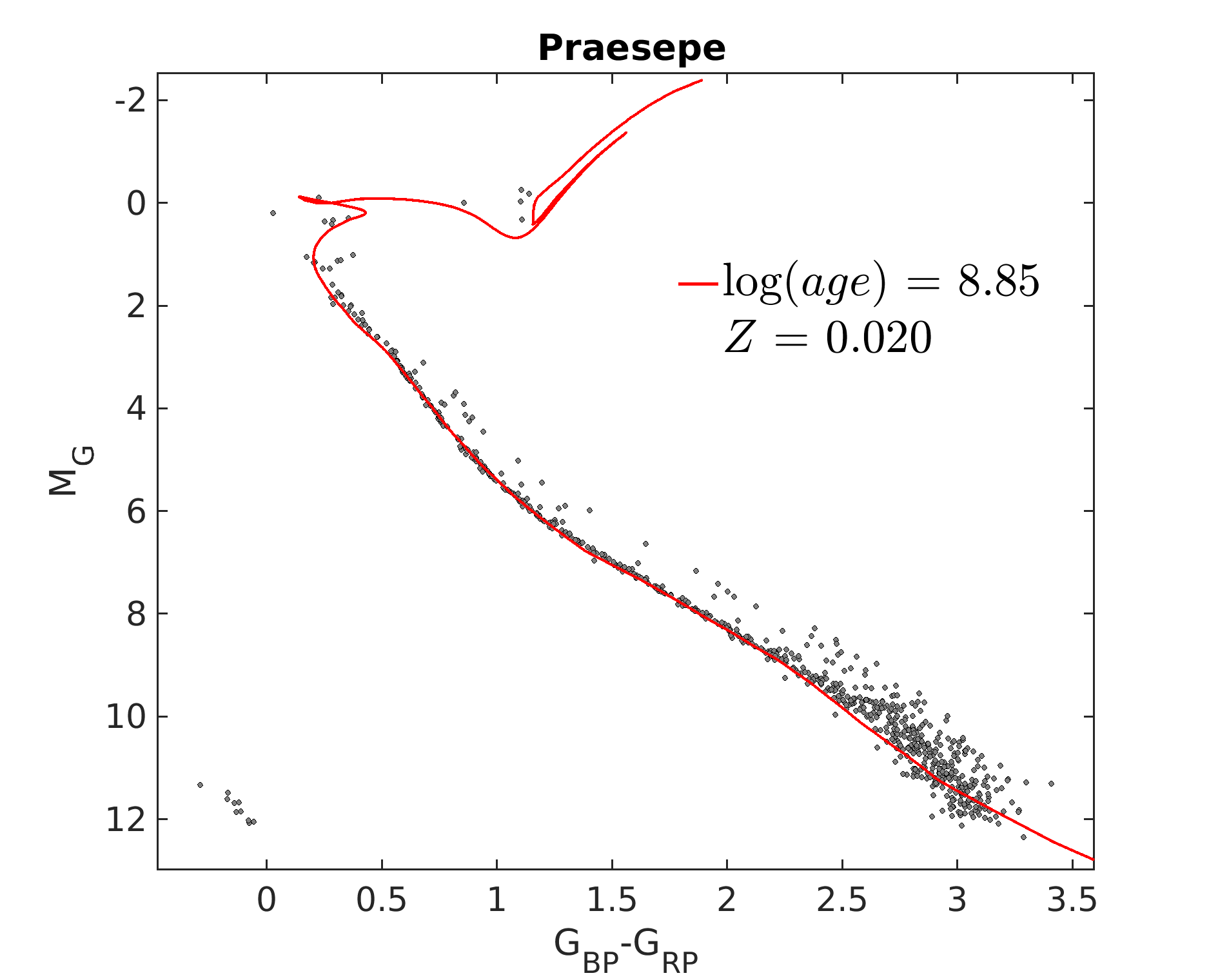}\\
\includegraphics[width=0.46\linewidth]{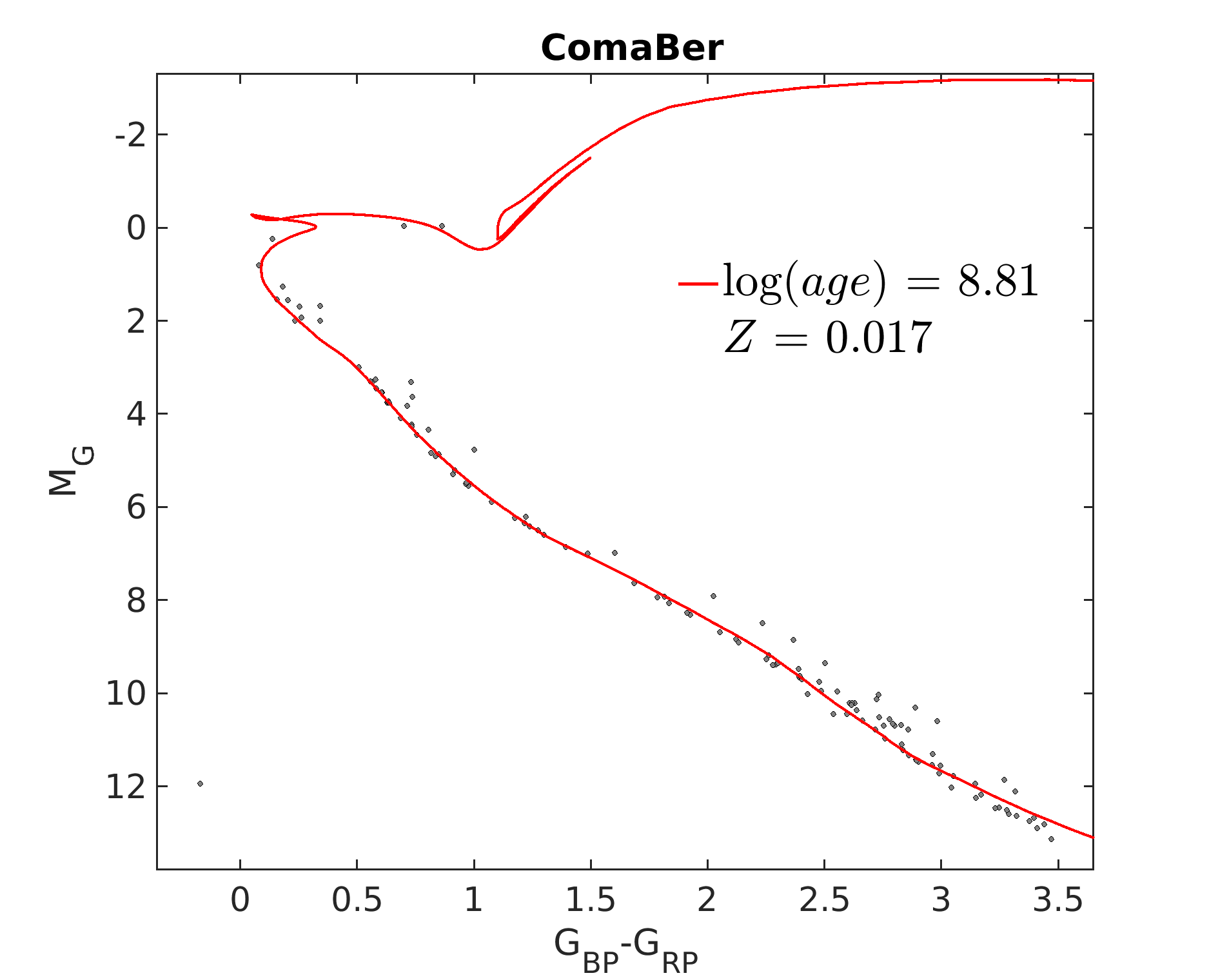}
\hspace{0.5cm}
\includegraphics[width=0.46\linewidth]{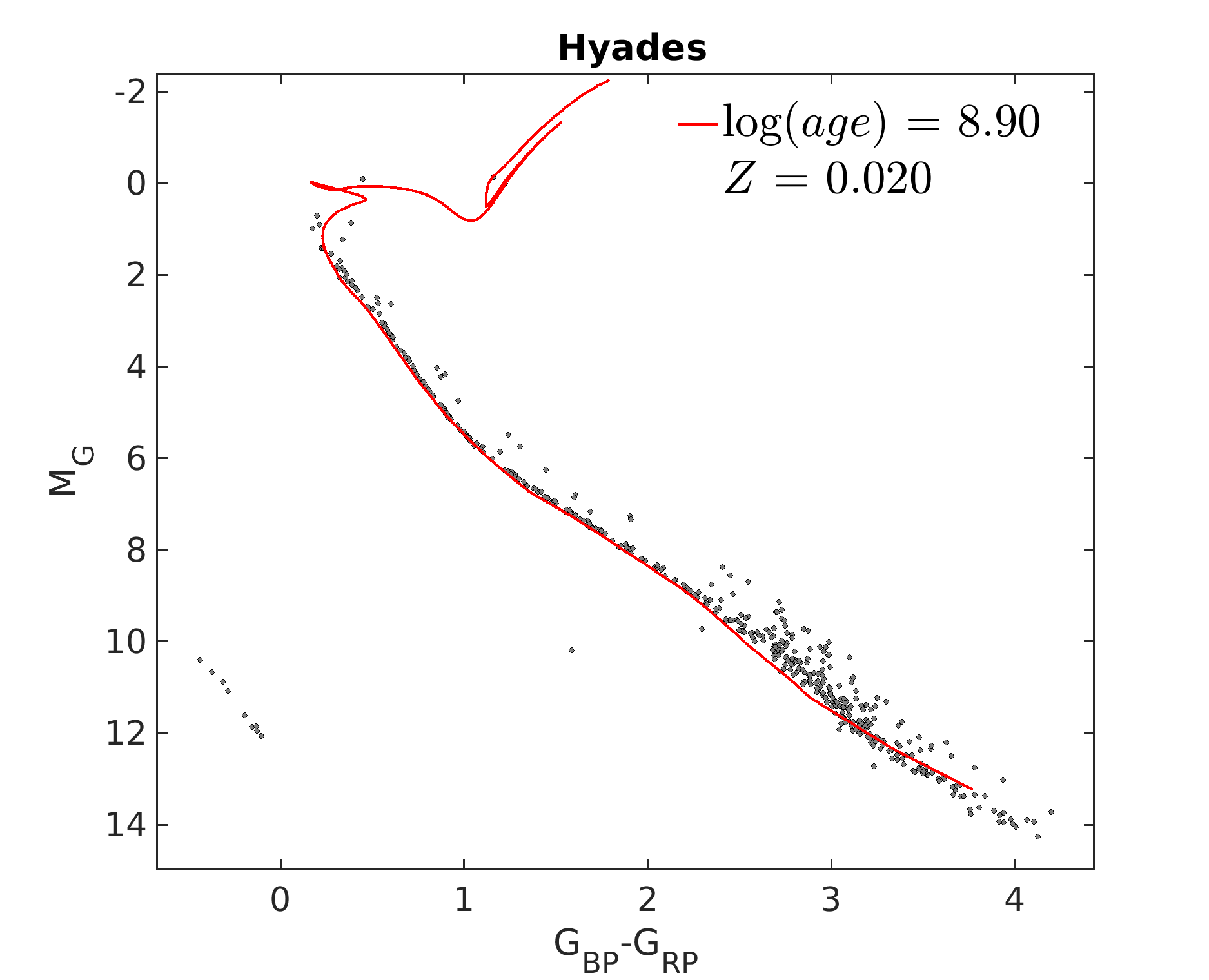}\\
\includegraphics[width=0.46\linewidth]{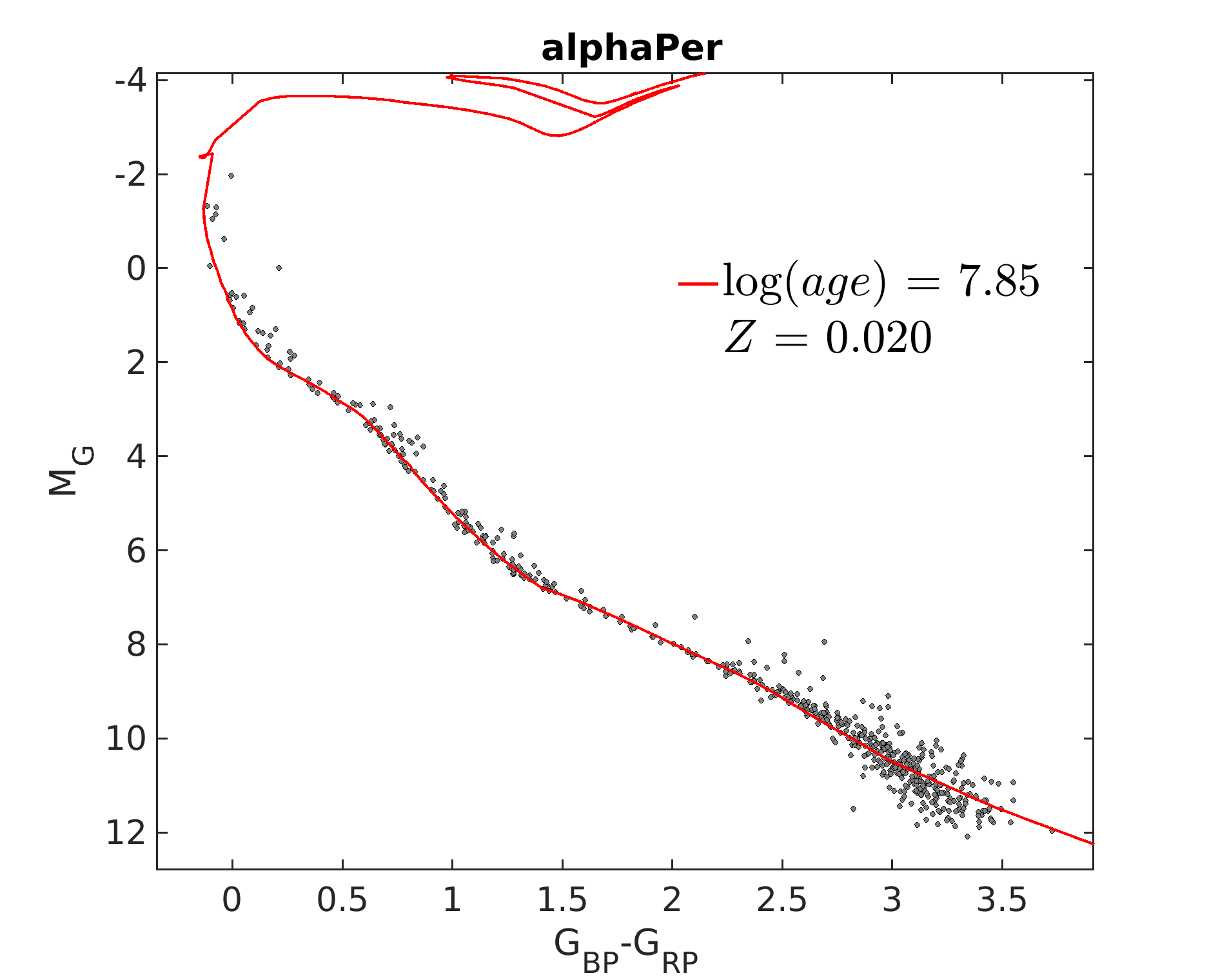}
\hspace{0.5cm}
\includegraphics[width=0.46\linewidth]{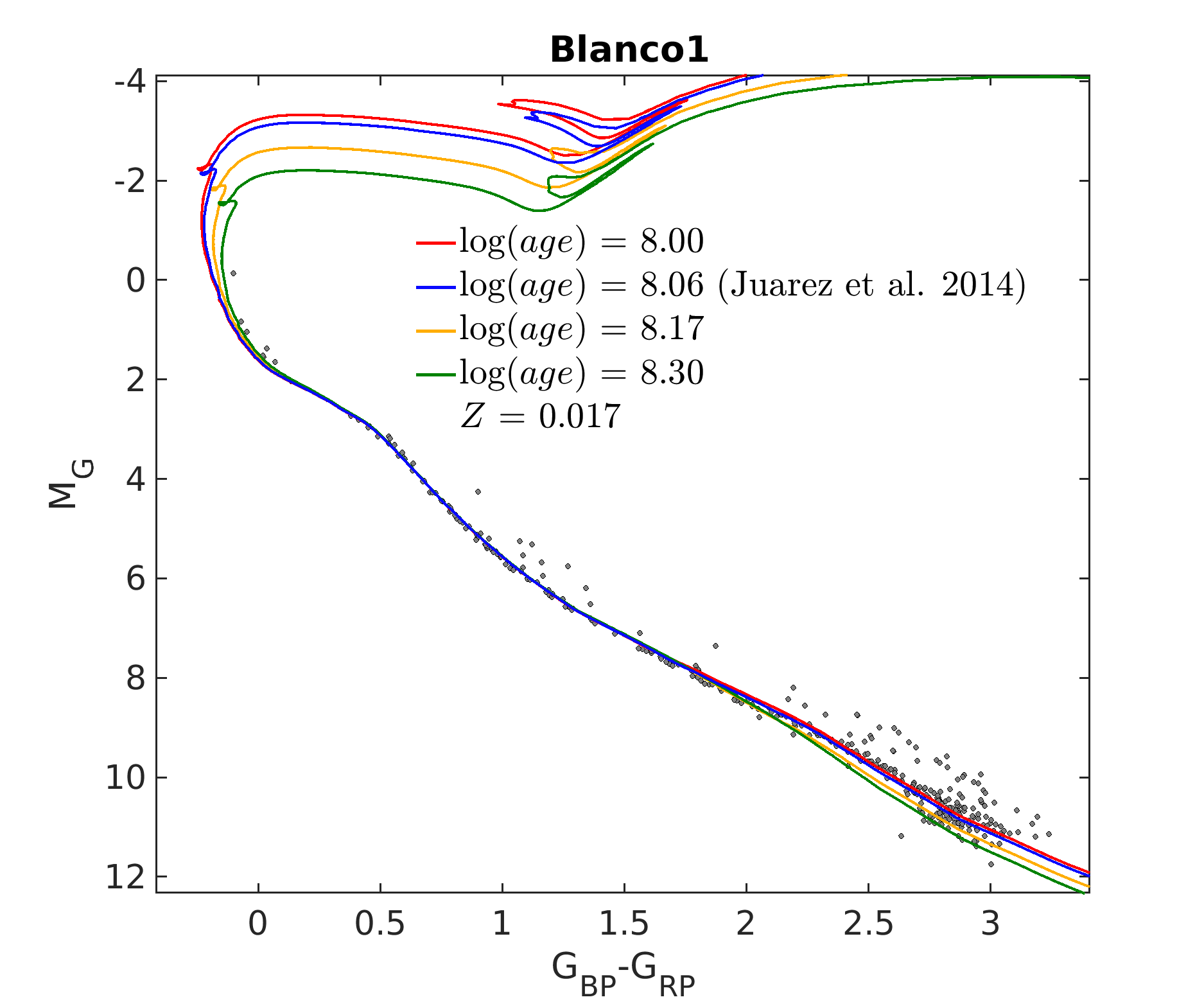}\\
\caption{\label{fig:oc_iso} HRDs of nearby clusters compared with PARSEC isochrones  (see text for details) of the Pleiades  (a), Praesepe (b), Coma~Ber (c), Hyades (d), Alpha Per (e), and Blanco~1 (f). Praesepe, Hyades, and Alpha Per are fitted with $Z=0.02$, while the others are reproduced using $Z=0.017$.}
\end{figure*}

\begin{figure*}
\centering
\includegraphics[width=0.46\linewidth]{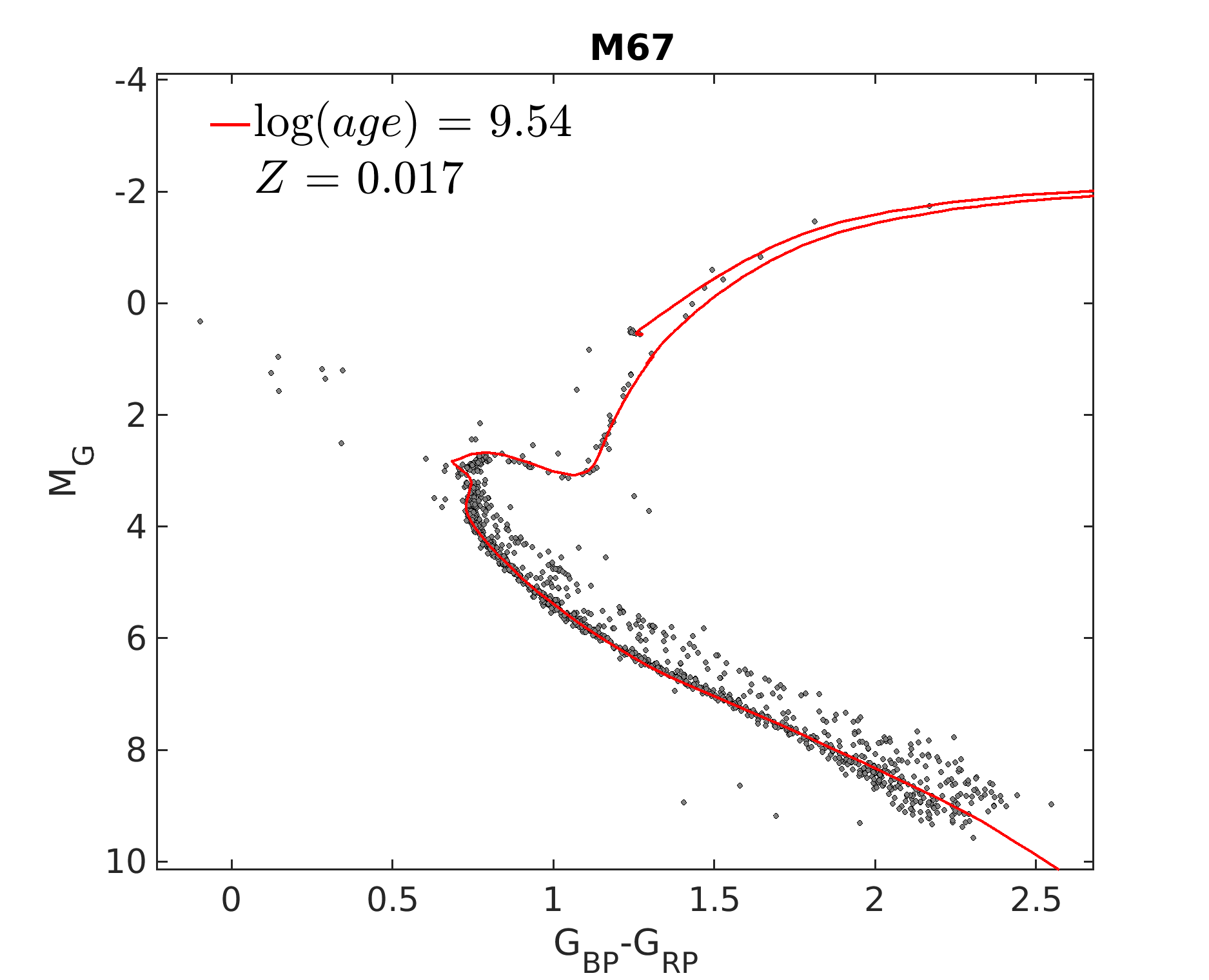}
\hspace{0.5cm}
\includegraphics[width=0.46\linewidth]{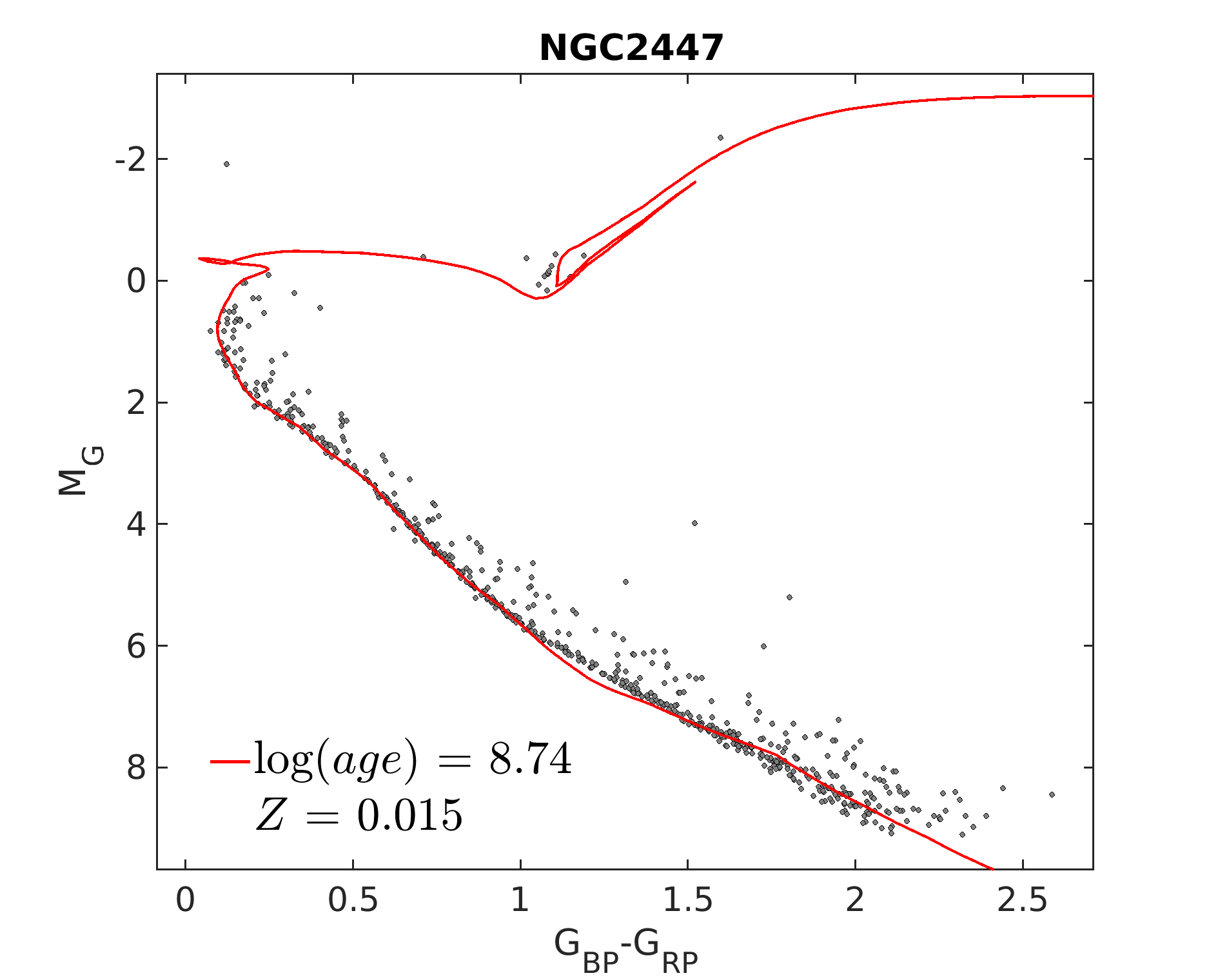}
\caption{\label{fig:oc_iso2} HRDs of  two distant clusters compared with PARSEC isochrones (see text for details): M67 (NGC~2682) (a), and NGC~ 2447(b). }
\end{figure*}

\begin{figure}
\centering
\includegraphics[width=8cm]{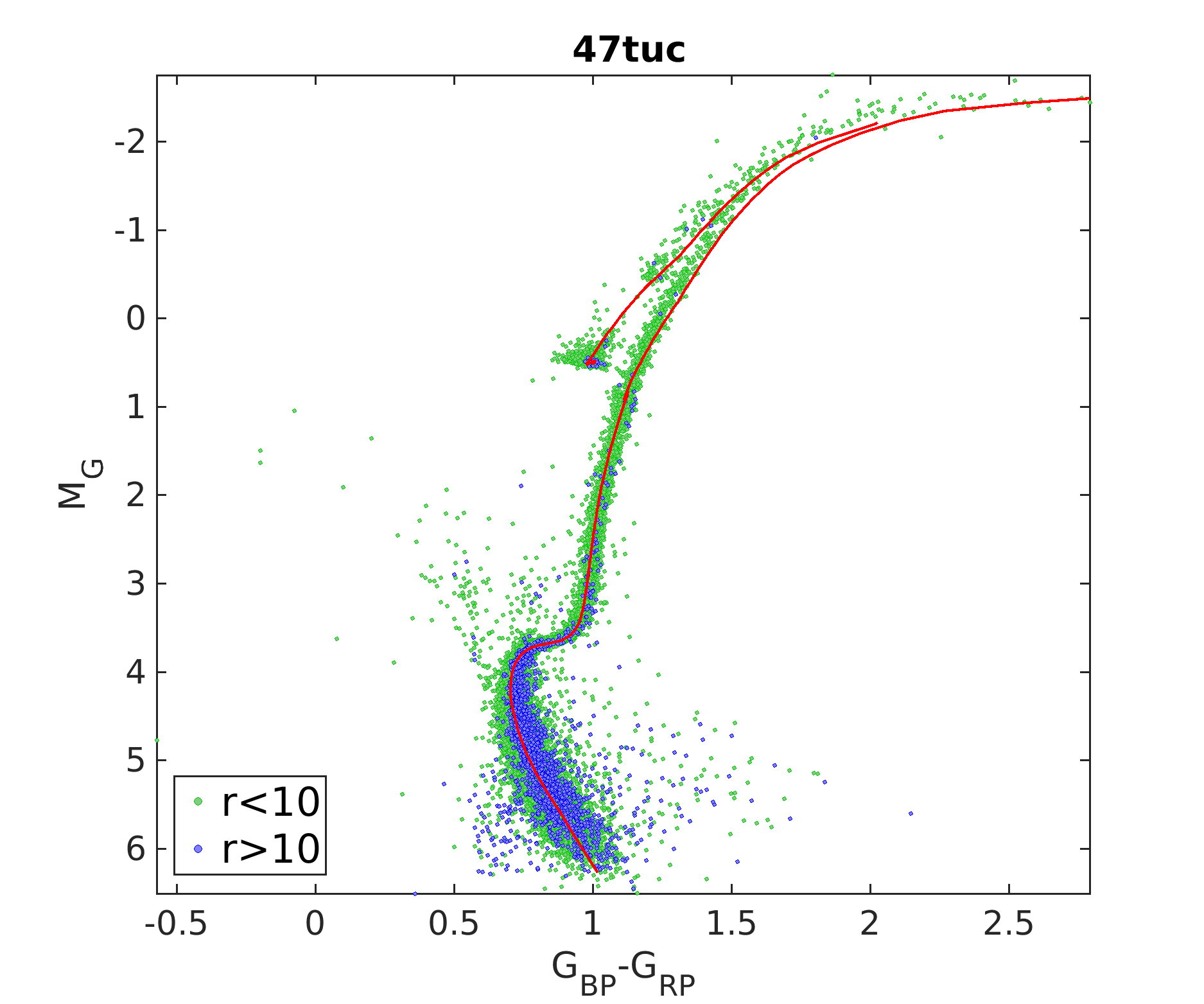}
\caption{\label{fig:GC_iso} HRD of  the globular cluster 47~Tuc compared with PARSEC isochrones (see text for details).  The
inner region
(radius < 10 arcmin, e.g. three times the half-light radius) is shown in green, while the external regions are plotted in blue. A maximum radius of 1.1 degrees was used. } 
\end{figure}

\section{Cluster as stellar parameter templates\label{sec:ociso}}
Clusters have long been considered as benchmarks with regard to the determination of the stellar properties.
Open cluster  stars share common properties, such as age and chemical abundances. The level of homogeneity of open clusters has been assessed in several papers 
\citep{2014A&A...569A..17C,2016ApJ...817...49B}.  By means of a high-precision differential abundance analysis, the Hyades have been proved to be  chemically in-homogeneous at the 0.02 dex level \citep{2016MNRAS.457.3934L} at maximum. Until now, the study of clusters was hampered by  the disk field contamination. This in turn results in difficult membership determination, and in highly uncertain parameters \citep{2015A&A...582A..19N}. Distance and age, together with chemical abundances, are the fundamental properties  for a meaningful description of the disk characteristics.  Their study complements  the field population studies that are based on Galactic surveys.
Globular clusters are fundamental tools for studying the properties of low-mass stars and the early chemical evolution of the Galaxy.
  Now \gaia DR2 data bring us into a completely new domain. High-accuracy parallaxes and exquisite photometry make the comparison with theoretical isochrones very fruitful, based on which, stellar properties can be defined. A detailed discussion of the uncertainties of stellar models is beyond the scope of this paper. Here we would like to recall that effects such as convection in the stellar core, mass loss, rotation, and magnetic fields are still poorly constrained and are often only parametrised in stellar models \citep{2016csss.confE.102B,2016MNRAS.459.3182P,2013EPJWC..4301002W}. Although very significant, seismic predictions depend on our poor knowledge of the relevant physics \citep{2015IAUGA..2251619M}. A calibration of these effects on star cluster photometry is mandatory and will complement  asteroseismology as a tool for testing stellar physics and will ultimately improve stellar models. In Table~\ref{tab:openclref} we present the ages and the extinction values derived by isochrone fitting for the sample of open clusters discussed in this paper. The uncertainties  are $\Delta({\rm log(age)})$ $_{-0.22}^{+0.14}$, $_{-0.13}^{+0.11}$, $_{-0.06}^{+0.08}$ for $6<{\rm log(age)} \leq 7$, $7<{\rm log(age)} \leq8$, ${\rm log(age)}> 8,$ respectively, and $\Delta E(B-R)=0.04$.
Here we made use of PARSEC isochrones \citep{2014MNRAS.444.2525C} for metallicities $Z=0.017$ and $Z=0.020$ {updated to the latest transmission curve calibrated on \gaia DR2 data \citep{DR2-DPACP-40}\footnote{PARSEC isochrones in \gaia DR2 passbands are available at \url{http://stev.oapd.inaf.it/cgi-bin/cmd}}. 
Praesepe, Hyades, Alpha Per, and NGC~6475 were fitted with $Z=0.02$ \citep{2017A&A...601A..19G, 2015yCat..35850101K}, while the others were reproduced using $Z=0.017$. This gave a relatively poor fit for clusters that are known to have sub-solar metallicity, such as  NGC~2158, which has [Fe/H]=-0.25 \citep{2015yCat..35850101K}. The PARSEC solar value is $Z=0.015$. This version of the PARSEC tracks makes use of  a modified 
relation between the effective temperature and Rosseland mean optical depth $\tau$ across the atmosphere 
that is derived  from PHOENIX \citep{2012RSPTA.370.2765A} and in particular from the set of BT-Settl models.  
With this modified relation, introduced to better reproduce the observed mass-radius relation in
nearby low mass stars \citep{2014MNRAS.444.2525C},
the models provide a good representation of the colour distribution
of very low mass stars in several passbands. 

Figure~\ref{fig:oc_iso} shows the HRD of a few nearby open clusters compared with PARSEC isochrones. The distance modulus (Table~\ref{tab:openclref}) was used, and the extinction was not corrected in the photometry,
but was applied on the isochrones. 

The fits are remarkably good in the upper and lower  main sequence. The high quality of {\gaia} photometry produces well-defined features, very clean main sequences, and a clear definition of the binary sequence. The agreement for the Pleiades is particularly
remarkable.
In spite of the impressively good agreement across a range of several magnitudes,
at about $M_G \sim 10,$  the model predictions and the observed main sequence still disagree. The slope of the theoretical main sequence initially seems   to be slightly steeper than the observed main sequence, while  at even lower magnitudes, the slope of the observed main sequence becomes steeper than the predicted main sequence. 
The latter effect might be due to the background subtraction,
which becomes challenging at these faint magnitudes \citep{DR2-DPACP-40,DR2-DPACP-39}.
Instead, the initial steepening of the isochrones, which is also
observed in other clusters in Fig.~\ref{fig:oc_iso},   might indicate that the adopted boundary conditions in the domain of very low mass stars in PARSEC need a further small revision. 
It is well known that current models and the colour transformations fail to reproduce the main sequence in the very low mass regime \citep{2016csss.confE.102B}, and the data  gathered by \gaia will certainly help to overcome this long-standing problem.

The age determination of Blanco~1  deserves further comments.
Blanco~1 has a slightly super-solar metallicity  ${\rm [Fe/H]}=+0.04\pm 0.04$ \citep{2005MNRAS.364..272F}.
Previous age determination placed Blanco~1 in the age range ${\rm log(age)}=8.0-8.17$  \citep{2007A&A...471..499M}. A determination of the lithium depletion boundary on very low mass stars gives  ${\rm log(age)} =8.06 \pm 0.13$ when a correction for magnetic activity is applied \citep{2014ApJ...795..143J}. From the main-sequence turnoff,
we obtain ${\rm log(age)}= 8.30$. However, fitting the main-sequence turnoff in such an
inconspicuous cluster might not lead to correct results, since the initial mass function disfavours higher mass stars. Using the lithium depletion boundary age of log(age) of $8.06 \pm 0.04$, we reproduce the lower main sequence,   with a marginal fit to the upper main sequence. Similar considerations apply to the Pleiades, whose log(age) is in the range $8.04 \pm 0.03$ - $8.10\pm 0.06$ and is derived from the lithium depletion boundary or from eclipsing binaries \citep[for a recent discussion, see][]{2016AJ....151..112D}. Using the lithium depletion boundary age of $8.04\pm 0.06$, we can reproduce the main sequence with PARSEC isochrones.

Figure~\ref{fig:oc_iso2} presents the comparison of two distant clusters, NGC~2682~(M67) and NGC~2447, with PARSEC isochrones.  M~67 is one of the best-studied  star clusters. It has a metallicity  near solar,  an accessible distance of about 1028 pc with low reddening \citep{2007AJ....133..370T}, and an age close to solar ($\sim 4$~Gyr). It is a very highly populated object that includes over 1000 members from  main-sequence dwarfs, a well-populated subgiant and red giant branch, white dwarfs, blue stragglers, sub-subgiants, X-ray sources, and cataclysmic variables.  \gaia identifies 1526 members. 
It was observed by almost all the most relevant spectroscopic surveys (\gaia-ESO, APOGEE, WIYN, etc.). Asteroseismologic data are available from the Kepler 2 mission \citep{2016ApJ...832..133S}. M67 is a cornerstone of stellar astrophysics, and it is a calibrator of age determination via gyrochronology \citep{2016ApJ...823...16B}. Its turn-off mass  is very close to the critical mass for the onset of
core convection. For this reason, the cluster is especially interesting for this specific regime of stellar models and their dependence on different parameters such as nuclear reaction rate and solar abundances.  The main-sequence termination presents a distinctive hook  and a  gap just above it. These features are used to distinguish between diffusive and non-diffusive evolutionary models. Atomic diffusion is very important for the morphology of isochrones in the vicinity
of the turn-off. The hook feature  traces the rapid contraction phase that
occurs at central H exhaustion in those stars that have convective
cores during their main-sequence phase. This hook is located
at somewhat higher luminosities and cooler temperatures
when diffusive processes are included  \citep{2004ApJ...606..452M}. \gaia photometry and parallax place the location of these features very precisely in the HRD. PARSEC isochrones, including overshoot and diffusion,  reproduce the main-sequence slope and termination point reasonably well, although additional overshoot calibration might be necessary.
A population of blue stragglers, a few  yellow giants, and two sub-subgiants are clearly visible among the members. The binary star sequence in M67 is clearly defined as well.

NGC~2447 is a younger object with an age of 0.55~Gyr and almost solar metallicity. 
Previous photometry is relatively poor \citep{2005BaltA..14..301C}. In \gaia DR2, photometry and membership of the cluster stand out very clearly. PARSEC isochrones 
reproduce the main sequence very well, while the red clump colour is slightly  redder.

Figure~\ref{fig:GC_iso} presents the HRD of the globular cluster 47~Tuc (see Table \ref{tab:globclust}), which is one prominent example of multiple populations in globular clusters. Hubble
Space Telescope (HST) photometry in the blue passbands has revealed a double
main sequence \citep{2012ApJ...744...58M} and distinct subgiant branches  \citep{2009ApJ...697L..58A}. These components are not visible in the high-accuracy \gaia photometry, since bluer colours would be necessary.
47~Tuc has a relatively high average metallicity of [Fe/H]=-0.72. We fit it with PARSEC isochrones with $Z=0.0056$, $Y=0.25$.  Since no alpha-enhanced tracks are available in the PARSEC data set, we use the \cite{1993ApJ...414..580S} relation  to account for the enhancement. 

\begin{figure*} 
    \centering
   \includegraphics[width=0.325\linewidth]{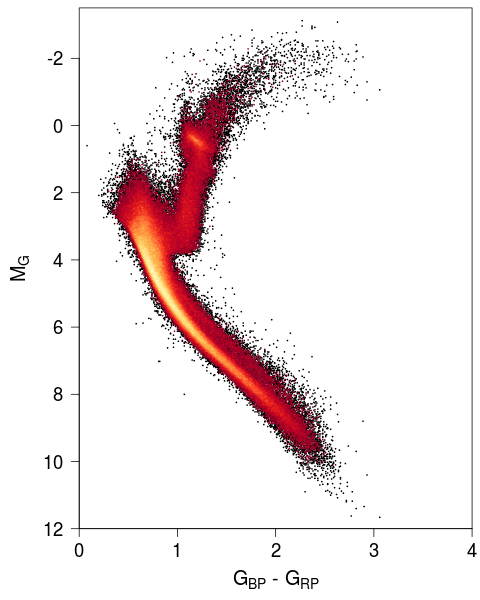}
   \includegraphics[width=0.325\linewidth]{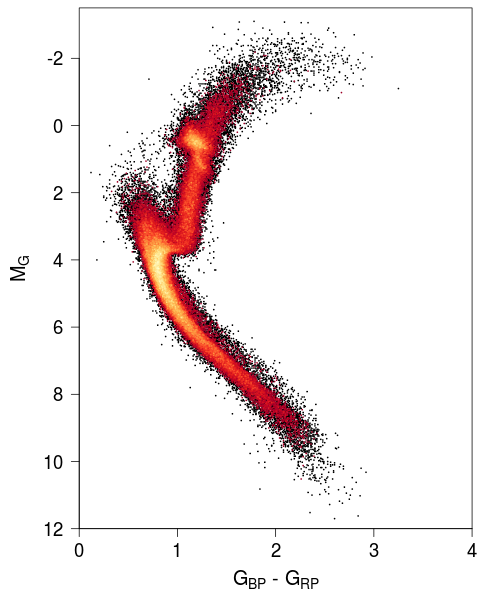}
   \includegraphics[width=0.325\linewidth]{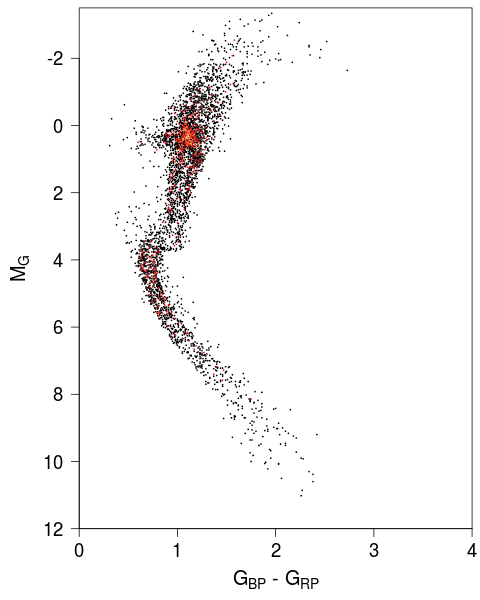}
   \caption{{\gaia} HRDs with kinematic selections based on the total velocity: a) \Vtot$<$50\kms\ (275,595 stars), b) 70$<$\Vtot$<$180\kms\ (116,198 stars), and c) \Vtot$>$200\kms\ (4,461 stars)\label{fig:kineVtot}. }
   
    \centering
   \includegraphics[width=0.325\linewidth]{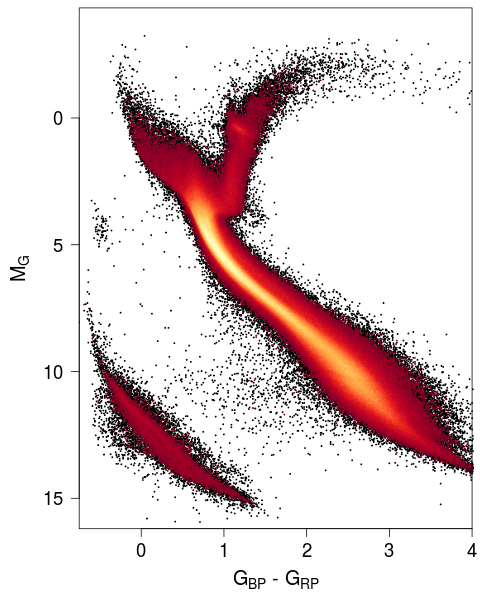}
   \includegraphics[width=0.325\linewidth]{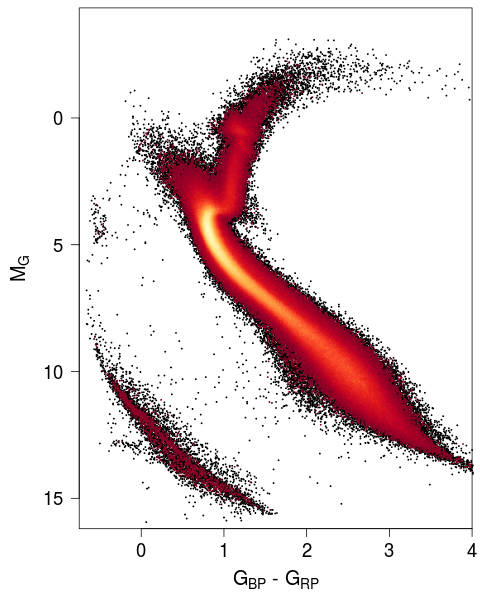}
   \includegraphics[width=0.325\linewidth]{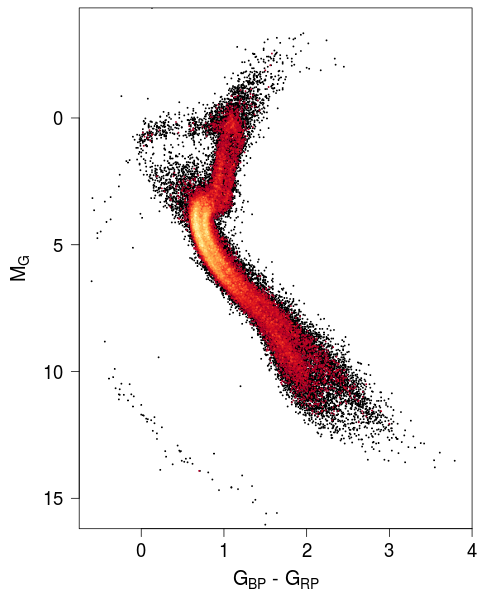}
   \caption{{\gaia} HRDs with kinematic selections based on the tangential velocity: a) \VT$<$40\kms\ (1,893,677 stars), b) 60$<$\VT$<$150\kms\ (1,303,558 stars), and c) \VT$>$200\kms\ (64,727 stars)\label{fig:kineVtan}. }
   
\end{figure*}

\begin{figure}
    \centering
   \includegraphics[width=8cm]{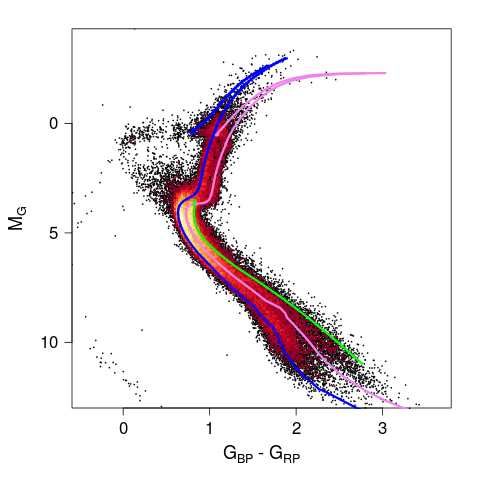} \\
   \includegraphics[width=4.3cm]{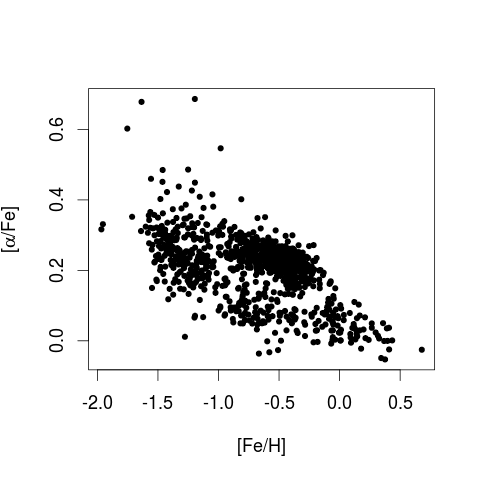}
   \includegraphics[width=4.3cm]{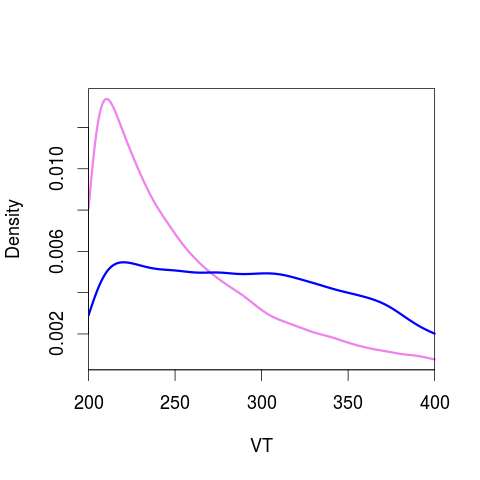}
   \caption{a) Same as Fig. \ref{fig:kineVtan}c (kinematic selection \VT$>$200\kms) overlaid with PARSEC isochrones for [M/H]$= -1.3$, age$= 13$~Gyr (blue), and [M/H]$=-0.5$, age$= 11$~Gyr (magenta) and [$\alpha$/Fe]$=0.23$;
   green line: median spline fit to the main sequence of the thick-disk kinematic selection (Fig. \ref{fig:kineVtan}b).
   b) [$\alpha$/Fe] vs. [Fe/H] of the corresponding APOGEE stars without extinction criterion applied. 
   c) Density distribution of the tangential velocity \VT~ on the blue and red sides of a median spline main-sequence fit. 
   }
   \label{fig:doublehalo}
\end{figure}

\section{\label{sec:kine}Variation of the HRD with kinematics} 

Thin disk, thick disk, and halo have different age and metallicity distributions as well as kinematics. The \gaia HRD is therefore
expected to vary with the kinematics properties. 
For stars with radial velocities, we apply classical cuts to broadly kinematically select thin-disk (\Vtot$<$50\kms), thick-disk (70$<$\Vtot$<$180\kms), and halo stars (\Vtot$>$200\kms) \citep[e.g.][]{2014A&A...562A..71B}, using $U$,$V$,$W$ computed within the framework of \cite{DR2-DPACP-33}, in which a global Toomre diagram is presented. This sample with radial velocities is limited to bright stars. To probe deeper into the HRD, we also made a selection using only tangential velocities, which we computed with \VT=$4.74 / \varpi \sqrt{\mu_{\alpha*}^2+\mu_\delta^2}$. We roughly adapted our kinematic cut to the fact that we now only have two components of the velocity instead of three:  we used \VT$<$40\kms for the thin disk and 60$<$\VT$<$150\kms for the thick disk, but still \VT$>$200\kms for the halo. To all our samples we also applied the $E(B-V)<0.015$ selection criterion. The results are presented in Fig. \ref{fig:kineVtot} and Fig. \ref{fig:kineVtan}. We note that hot star radial velocities are not included in \gaia DR2 \citep{DR2-DPACP-47}, which explains why they are missing in Fig. \ref{fig:kineVtot}.

The left figures associated with the thin disk show the same main features typical of a young population as the local HRD of Fig. \ref{fig:localHRD}: young hot main-sequence stars are present (Fig. \ref{fig:kineVtan}a), the secondary red clump as well as the AGB bump is visible (Fig. \ref{fig:kineVtot}a), and the turn-off region is diffusely populated. 
The middle figures associated with the thick disk show a more localised turn-off typical of an intermediate to old population. The median locus of the main sequence is similar to the thin-disk selection. 
The right figures associated with the halo show an extended horizontal branch, typical of old metal-poor populations, but also two very distinct main sequences and turn-offs. We note the presence of the halo white dwarfs.

We study the kinematic selection associated with the halo in Fig. \ref{fig:doublehalo} in more detail. The two main-sequence turn-offs are shifted by $\sim$0.1~mag in colour. The red main-sequence turn-off is shifted by $\sim$0.05~mag from the thick-disk kinematic selection main sequence (green line in Fig. \ref{fig:doublehalo}a). 
Comparison with isochrones clearly identifies the distinct main sequences as being driven by a metallicity difference of about 1~dex. 
To further confirm this, we cross-matched our selection with the APOGEE DR14 catalogue \citep{2015AJ....150..148H} using their 2MASS ID and the 2MASS cross-match provided in the \gaia archive \citep{DR2-DPACP-41}. There are 184 stars in common,  1168 if we relax the low-extinction criteria that mostly confine our HRD selection to the galactic poles.  The metallicity distribution is indeed double-peaked,  with peak metallicities of -1.3 and -0.5~dex. We superimpose in Fig.~\ref{fig:doublehalo} the corresponding PARSEC isochrones using the \cite{1993ApJ...414..580S} formula for the mean $\alpha$ enhancement of 0.23 for [M/H]=-1.3 and -0.5 and ages of 13 and 11~Gyr, respectively. While the extent of the horizontal branch does not correspond to the isochrones used here, it can be compared to the empirical horizontal branches of the globular clusters presented in Fig.~\ref{fig:GCs}.

This bimodal metallicity distribution in the kinematic selection of the halo may recall the globular cluster bimodal metallicity distribution with the same peaks at [Fe/H]$\sim$-0.5 and [Fe/H]$\sim$-1.5 \citep[e.g.][]{1985ApJ...293..424Z}, the more metal-rich part being associated with the thick disk and bulge. We verified with the globular cluster kinematics provided in \cite{DR2-DPACP-34} that 80\% of these globular clusters indeed fall into our halo kinematic selection, independently of their metallicity. 
The -0.5~dex peak also recalls the bulge metal-poor component \citep[e.g.][]{2011A&A...534A..80H}. 
However, it seems to be different from the double halo found at larger distances \citep{2007Natur.450.1020C,2010ApJ...714..663D}: while their inner-halo component at $\sim$-1.6 could correspond to our metal-poor component, their metal-poor component is at metallicity $\sim$-2.2 and is found in the outer Galaxy. 
This duality in the metallicity distribution of the kinematically selected halo stars has also been found using TGAS data with RAVE and APOGEE \citep{2017ApJ...845..101B}. Half of the stars are also found to have [M/H]$>$-1~dex with a dynamically selected halo sample in TGAS/RAVE by \cite{2017arXiv171104766P}. 

The $\alpha$ abundances of this APOGEE sample (Fig. \ref{fig:doublehalo}b)
let us recover the two sequences described by \cite{2010A&A...511L..10N} using an equivalent kinematic selection. 
We adjusted a median spline to the main sequence of the high-velocity HRD and present the velocity distribution of the stars on either side of this median spline in Fig.~\ref{fig:doublehalo}c. The magenta sequence looks like a velocity distribution tail towards high velocities, while the blue sequence has a flat velocity distribution. We do not see any difference in the sky distribution of these components, most probably because the sky distribution is fully dominated by our sample selection criteria. 
All these tests and comparisons with the literature seem to indicate a very different formation scenario for the two components of this kinematic selection of the halo.

\section{\label{sec:conclu}Summary}

The unprecedented all-sky precise and homogeneous astrometric and photometric content of \gaia DR2 allows us to see fine structures in both field star and cluster Hertzsprung-Russell diagrams to an extent that has never been reached before. 
We have described the main filtering of the data that is required for this purpose and provided membership for a selection of open clusters covering a wide range of ages. 

The variations with age and metallicity are clearly illustrated by the main sequence and the giant branches of a large set of open and globular clusters and kinematically selected stellar populations. 
The main sequence for nearby stars is extremely thin, for field and cluster stars both, with a clear scattering of double stars up to 0.75 magnitude visible above the main sequence.  
\gaia DR2 provides a very unique view of the bottom of the main sequence down to the brown dwarf regime, including L-type and halo BDs. We also see the post-AGB stars and the central stars of planetary nebulae, which follow the expected tracks down to the white dwarf sequence, as well as hot subdwarfs.

The split in the white dwarf sequence between hydrogen and helium white dwarfs, which was first detected in the SDSS colour-colour diagrams, is visible for the first time in an HRD, with very thin sequences that agree with the strong peak of their mass distribution around 0.6~\Msol. 

Kinematic selections clearly show the change in HRDs with stellar populations. It highlights the strong bimodality of the HRD of the classical halo kinematic selection, and gives evidence of two very different populations within this selection.  

All the features in the \gaia HRDs chiefly agree in general with the theoretical stellar evolution models. The differences that
are observed for the faintest brown dwarfs, the white dwarf hydrogen/helium split, or the very fine structures of the open cluster main sequences,
for example, are expected to bring new insight into stellar physics.

Numerous studies by the community are expected on the \gaia HRD. For example, rare stages of evolution will be extracted
from the archive, together with more clusters, and detailed comparisons with different stellar evolution models will be made. The
completeness of the data is a difficult question that we did not discuss here, but that will be studied by the community as it is a very important issue, in particular for determining
the local volume density and all the studies of the initial mass function and stellar evolution lifetimes.

The next \gaia release, DR3, will again be a new step for stellar studies. This will be achieved not only by the increase in completeness, precision, and accuracy of the data, but also by the additional spectrophotometry and spectroscopy, together with the binarity information that will be provided.

\begin{acknowledgements}
We thank Pierre Bergeron for providing the WD tracks in the \gaia DR2 passbands.
This work presents results from the European Space Agency (ESA) space mission \gaia. \gaia\ data are being processed by the \gaia\ Data Processing and Analysis Consortium (DPAC). Funding for the DPAC is provided by national institutions, in particular the institutions participating in the \gaia\ MultiLateral Agreement (MLA). The \gaia\ mission website is \url{https://www.cosmos.esa.int/gaia}. The \gaia\ archive website is \url{https://archives.esac.esa.int/gaia}.

The \gaia\ mission and data processing have financially been supported by, in alphabetical order by country:
 the Algerian Centre de Recherche en Astronomie, Astrophysique et G\'{e}ophysique of Bouzareah Observatory;
 the Austrian Fonds zur F\"{o}rderung der wissenschaftlichen Forschung (FWF) Hertha Firnberg Programme through grants T359, P20046, and P23737;
 the BELgian federal Science Policy Office (BELSPO) through various PROgramme de D\'eveloppement d'Exp\'eriences scientifiques (PRODEX) grants and the Polish Academy of Sciences - Fonds Wetenschappelijk Onderzoek through grant VS.091.16N;
 the Brazil-France exchange programmes Funda\c{c}\~{a}o de Amparo \`{a} Pesquisa do Estado de S\~{a}o Paulo (FAPESP) and Coordena\c{c}\~{a}o de Aperfeicoamento de Pessoal de N\'{\i}vel Superior (CAPES) - Comit\'{e} Fran\c{c}ais d'Evaluation de la Coop\'{e}ration Universitaire et Scientifique avec le Br\'{e}sil (COFECUB);
 the Chilean Direcci\'{o}n de Gesti\'{o}n de la Investigaci\'{o}n (DGI) at the University of Antofagasta and the Comit\'e Mixto ESO-Chile;
 the National Science Foundation of China (NSFC) through grants 11573054 and 11703065;  
 the Czech-Republic Ministry of Education, Youth, and Sports through grant LG 15010, the Czech Space Office through ESA PECS contract 98058, and Charles University Prague through grant PRIMUS/SCI/17;    
 the Danish Ministry of Science;
 the Estonian Ministry of Education and Research through grant IUT40-1;
 the European Commission’s Sixth Framework Programme through the European Leadership in Space Astrometry (\url{https://www.cosmos.esa.int/web/gaia/elsa-rtn-programme}{ELSA}) Marie Curie Research Training Network (MRTN-CT-2006-033481), through Marie Curie project PIOF-GA-2009-255267 (Space AsteroSeismology \& RR Lyrae stars, SAS-RRL), and through a Marie Curie Transfer-of-Knowledge (ToK) fellowship (MTKD-CT-2004-014188); the European Commission's Seventh Framework Programme through grant FP7-606740 (FP7-SPACE-2013-1) for the \gaia\ European Network for Improved data User Services (GENIUS) and through grant 264895 for the \gaia\ Research for European Astronomy Training (\url{https://www.cosmos.esa.int/web/gaia/great-programme}{GREAT-ITN}) network;
 the European Research Council (ERC) through grants 320360 and 647208 and through the European Union’s Horizon 2020 research and innovation programme through grants 670519 (Mixing and Angular Momentum tranSport of massIvE stars -- MAMSIE) and 687378 (Small Bodies: Near and Far);
 the European Science Foundation (ESF), in the framework of the \gaia\ Research for European Astronomy Training Research Network Programme (\url{https://www.cosmos.esa.int/web/gaia/great-programme}{GREAT-ESF});
 the European Space Agency (ESA) in the framework of the \gaia\ project, through the Plan for European Cooperating States (PECS) programme through grants for Slovenia, through contracts C98090 and 4000106398/12/NL/KML for Hungary, and through contract 4000115263/15/NL/IB for Germany;
 the European Union (EU) through a European Regional Development Fund (ERDF) for Galicia, Spain;    
 the Academy of Finland and the Magnus Ehrnrooth Foundation;
 the French Centre National de la Recherche Scientifique (CNRS) through action 'D\'efi MASTODONS', the Centre National d'Etudes Spatiales (CNES), the L'Agence Nationale de la Recherche (ANR) 'Investissements d'avenir' Initiatives D’EXcellence (IDEX) programme Paris Sciences et Lettres (PSL$\ast$) through grant ANR-10-IDEX-0001-02, the ANR 'D\'{e}fi de tous les savoirs' (DS10) programme through grant ANR-15-CE31-0007 for project 'Modelling the Milky Way in the Gaia era' (MOD4Gaia), the R\'egion Aquitaine, the Universit\'e de Bordeaux, and the Utinam Institute of the Universit\'e de Franche-Comt\'e, supported by the R\'egion de Franche-Comt\'e and the Institut des Sciences de l'Univers (INSU);
 the German Aerospace Agency (Deutsches Zentrum f\"{u}r Luft- und Raumfahrt e.V., DLR) through grants 50QG0501, 50QG0601, 50QG0602, 50QG0701, 50QG0901, 50QG1001, 50QG1101, 50QG1401, 50QG1402, 50QG1403, and 50QG1404 and the Centre for Information Services and High Performance Computing (ZIH) at the Technische Universit\"{a}t (TU) Dresden for generous allocations of computer time;
 the Hungarian Academy of Sciences through the Lend\"ulet Programme LP2014-17 and the J\'anos Bolyai Research Scholarship (L.~Moln\'ar and E.~Plachy) and the Hungarian National Research, Development, and Innovation Office through grants NKFIH K-115709, PD-116175, and PD-121203;
 the Science Foundation Ireland (SFI) through a Royal Society - SFI University Research Fellowship (M.~Fraser);
 the Israel Science Foundation (ISF) through grant 848/16;
 the Agenzia Spaziale Italiana (ASI) through contracts I/037/08/0, I/058/10/0, 2014-025-R.0, and 2014-025-R.1.2015 to the Italian Istituto Nazionale di Astrofisica (INAF), contract 2014-049-R.0/1/2 to INAF dedicated to the Space Science Data Centre (SSDC, formerly known as the ASI Sciece Data Centre, ASDC), and contracts I/008/10/0, 2013/030/I.0, 2013-030-I.0.1-2015, and 2016-17-I.0 to the Aerospace Logistics Technology Engineering Company (ALTEC S.p.A.), and INAF;
 the Netherlands Organisation for Scientific Research (NWO) through grant NWO-M-614.061.414 and through a VICI grant (A.~Helmi) and the Netherlands Research School for Astronomy (NOVA);
 the Polish National Science Centre through HARMONIA grant 2015/18/M/ST9/00544 and ETIUDA grants 2016/20/S/ST9/00162 and 2016/20/T/ST9/00170;
 the Portugese Funda\c{c}\~ao para a Ci\^{e}ncia e a Tecnologia (FCT) through grant SFRH/BPD/74697/2010; the Strategic Programmes UID/FIS/00099/2013 for CENTRA and UID/EEA/00066/2013 for UNINOVA;
 the Slovenian Research Agency through grant P1-0188;
 the Spanish Ministry of Economy (MINECO/FEDER, UE) through grants ESP2014-55996-C2-1-R, ESP2014-55996-C2-2-R, ESP2016-80079-C2-1-R, and ESP2016-80079-C2-2-R, the Spanish Ministerio de Econom\'{\i}a, Industria y Competitividad through grant AyA2014-55216, the Spanish Ministerio de Educaci\'{o}n, Cultura y Deporte (MECD) through grant FPU16/03827, the Institute of Cosmos Sciences University of Barcelona (ICCUB, Unidad de Excelencia 'Mar\'{\i}a de Maeztu') through grant MDM-2014-0369, the Xunta de Galicia and the Centros Singulares de Investigaci\'{o}n de Galicia for the period 2016-2019 through the Centro de Investigaci\'{o}n en Tecnolog\'{\i}as de la Informaci\'{o}n y las Comunicaciones (CITIC), the Red Espa\~{n}ola de Supercomputaci\'{o}n (RES) computer resources at MareNostrum, and the Barcelona Supercomputing Centre - Centro Nacional de Supercomputaci\'{o}n (BSC-CNS) through activities AECT-2016-1-0006, AECT-2016-2-0013, AECT-2016-3-0011, and AECT-2017-1-0020;
 the Swedish National Space Board (SNSB/Rymdstyrelsen);
 the Swiss State Secretariat for Education, Research, and Innovation through the ESA PRODEX programme, the Mesures d’Accompagnement, the Swiss Activit\'es Nationales Compl\'ementaires, and the Swiss National Science Foundation;
 the United Kingdom Rutherford Appleton Laboratory, the United Kingdom Science and Technology Facilities Council (STFC) through grant ST/L006553/1, the United Kingdom Space Agency (UKSA) through grant ST/N000641/1 and ST/N001117/1, as well as a Particle Physics and Astronomy Research Council Grant PP/C503703/1.

This publication has made use of SIMBAD and VizieR, both operated at the Centre de Donn\'ees astronomiques de Strasbourg ({CDS}, {http://cds.u-strasbg.fr/}).

This publication has made use of data products from the Two Micron All Sky Survey, which is a joint project of the University of Massachusetts and the Infrared Processing and Analysis Center/California Institute of Technology, funded by the National Aeronautics and Space Administration and the National Science Foundation.

This publication has made use of data products from SDSS-III. The Funding for SDSS-III has been provided by the Alfred P. Sloan Foundation, the Participating Institutions, the National Science Foundation, and the U.S. Department of Energy Office of Science. The SDSS-III web site is http://www.sdss3.org/. SDSS-III is managed by the Astrophysical Research Consortium for the Participating Institutions of the SDSS-III Collaboration.

\end{acknowledgements}

\bibliographystyle{aa} 
\bibliography{DR2HRD}

\appendix
 
\section{Open cluster membership and astrometric solutions\label{sec:ocsolu}}

\subsection{Nearby clusters\label{sec:nearby}}
\begin{table}[t]
\caption{Membership data for the open clusters. Only the first three lines with data for members of the Praesepe cluster are presented here. For the more distant clusters, the last two columns are not included. The astrometric and photometric extra filters presented in Sect.~\ref{sec:filters} and used in the figures of this paper are not applied in this table. The full table will be available in electronic form at the CDS. }
\scriptsize{
\centering
\begin{tabular}{rrrrrr}
\hline\hline
DR2 SourceId & cluster & $\alpha$ & $\delta$ & $\varpi_d$ & $\sigma\varpi_d$ \\ 
 & & (degr) & (degr) & mas& mas \\
\hline
    685747814353991296& Praesepe& 133.15933&  21.15502&  5.645&  0.033\\
        685805259540481664& Praesepe& 133.57003&  21.73443&  4.798&  0.100\\
        665141141087298688& Praesepe& 130.22501&  21.75663&  5.630&  0.020\\
    ... & ... & ... & ... & ... & ...\\
\hline
\end{tabular}
\label{tab:nearbyexample}
}
\end{table}
The nearby clusters were analysed with the method described and applied to the Hyades cluster in the first \gaia data release \citep{2017A&A...601A..19G}. By combining the information from the measured proper motions and parallaxes for individual cluster members, it is possible to derive a higher precision measurement for the relative parallax of these cluster members. The proper motion observed for an individual cluster member represents the local projection on the sky of the baricentric velocity of the cluster. It is therefore affected by the angular separation on the sky of the member star from the projection of the cluster centre and the baricentric distance of the star, again relative to that of the cluster centre. Similarly, the measured parallax for the star can be significantly different from the mean parallax of the cluster.

The primary aim of the present paper is to provide high-precision HRDs, for which these accurate relative parallaxes contribute important information by reducing the actual differential distance modulus variations of cluster members. The effectiveness of this procedure is limited by the amplitude of the proper motion of the cluster centre and the ratio of the diameter over the distance of the cluster. The standard uncertainties in the individual parallaxes and proper motions of the cluster members in the second \gaia data release allow for this procedure to be applied for clusters within 250~pc. Table~\ref{tab:nearbyexample} shows an example of an extract from the cluster member files produced for each of the nine clusters treated in this way.

\begin{figure}[t]
\centering
\includegraphics[width=7.5cm]{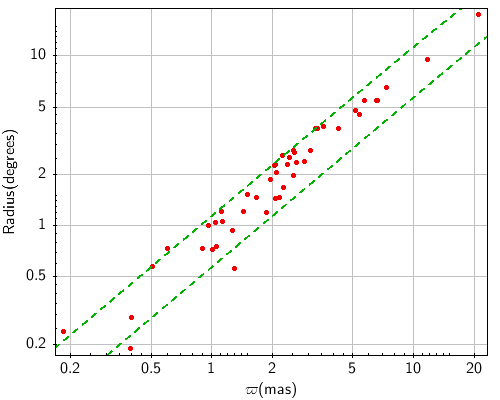}
\caption{Maximum radius in degrees in DR2 for the 46 open clusters as a function of parallax. The two diagonal lines represent maximum radii of 10 (bottom) and 20 (top) pc.}
\label{fig:rmaxopen}
\end{figure}

\begin{figure}[t]
\centering
\includegraphics[width=8cm]{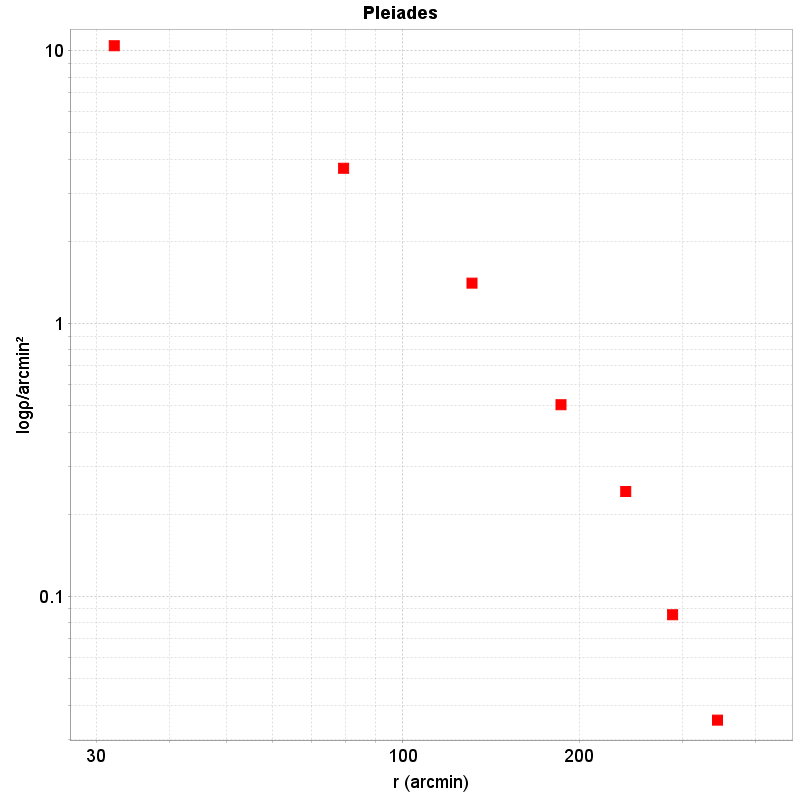}
\caption{Surface-density profile for the Pleiades cluster, based on 1332 identified cluster members. }
\label{fig:pleidens}
\end{figure}
Nine clusters within 250~pc from the Sun were analysed as nearby clusters. The analysis is iterative, and consists of two elements: 1.\ determinations of the space velocity vector at the cluster centre, and 2.\ determination of the cluster centre. A first selection is made of stars contained in a sphere with a radius of around 15~pc around the assumed centre of the cluster. A summary of the observed radii for the nearby and more distant clusters is shown in Fig.~\ref{fig:rmaxopen}. The radius can be adjusted based on the derived surface density distribution (Fig.~\ref{fig:pleidens}), where the outermost radius is set at the point beyond which the density of contaminating field stars starts to dominate. The selected stars are further filtered on their agreement between the observed proper motion and the predicted projection of the assumed space motion at the 3D position of the star, using the measured stellar parallax, and taking into account the uncertainties on the observed proper motion and parallax. The solution for the space motion follows Eq.~A13 in \cite{2017A&A...601A..19G}. Although it is in principle possible to solve also for the radial velocity using only the astrometric data, this effectively only works for the Hyades cluster. Instead, a single equation for the observed radial velocity of the cluster was added, where the observed radial velocity is based on the weighted mean of the \gaia radial velocities of cluster members for which these data are available. 

To stabilise the solution, it is important to align the coordinate system with the line of sight towards the cluster centre, minimising the mixing of the contributions from the proper motions and the additional information from the radial velocity. The solution for the space motion does provide an estimate of the radial velocity component, but except for the Hyades and Coma~Ber, this is largely dominated by the radial velocity value and its accuracy that
is used as input to the solution. Small differences are therefore seen between the radial velocities as presented in Table~\ref{tab:gaiarvexp} (as directly derived from the \gaia spectroscopic data) and in Table~\ref{tab:nearbyocuvw} (the summary data for the nine clusters in this selection), where the astrometric information on the radial velocity is also taken into account. Figure~\ref{fig:ic2602parcomp} shows an example of the level of agreement between the differential parallax and proper motion values in the cluster IC~2602. In Fig.~\ref{fig:ic2602xyz} we also show an example of the 3D distribution maps for this cluster; maps like this were prepared for all nearby clusters.

\begin{figure}[t]
\centering
\includegraphics[width=9cm]{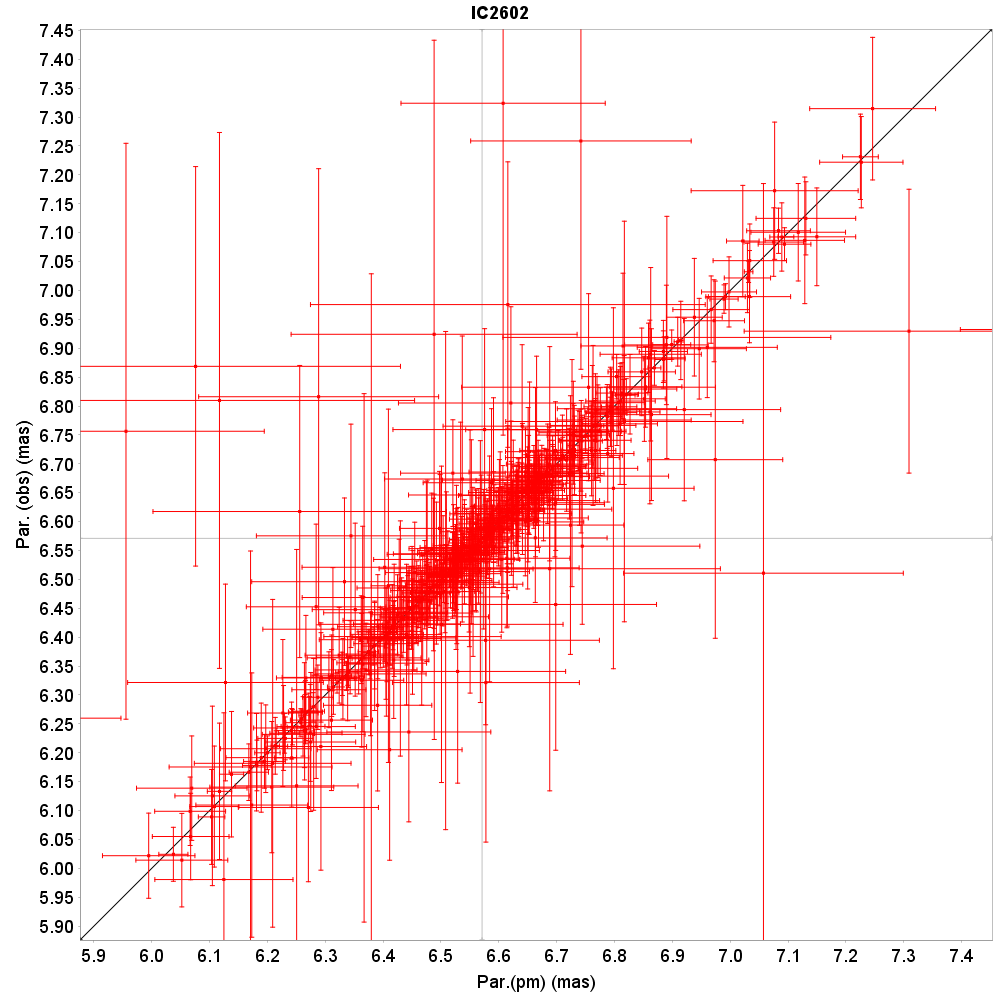}
\caption{Comparison between the directly measured parallaxes and the parallaxes obtained by including the relative proper motion data, for the cluster IC~2602. The clear linear relation shows the good agreement between proper motion and parallax offsets from the mean cluster values.}
\label{fig:ic2602parcomp}
\end{figure}

\begin{figure*}[t]
\centering
\includegraphics[width=14cm]{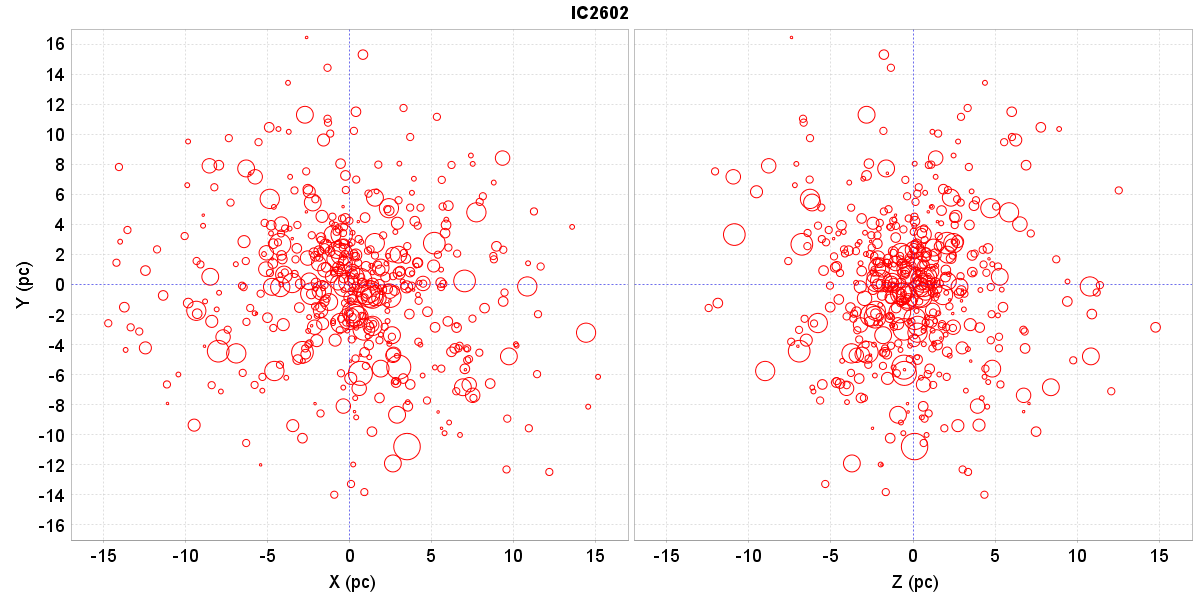}
\caption{Distribution of stars in IC~2602 in galactic rectangular coordinates, showing the flattening in the Z (galactic pole) direction.}
\label{fig:ic2602xyz}
\end{figure*}

Next to the astrometric data, the second \gaia data release also presents radial velocity measurements for a magnitude-limited sample. The radial velocities were compared with the projection of the cluster space velocity at the position of each star for which these data are available. This is particularly relevant for stars in the Hyades cluster, where the projection effects of the radial velocity can be of the order of several km~s$^{-1}$. Table~\ref{tab:gaiarvexp} presents the results for the nine nearby clusters. 

\begin{table}[t]
\centering
\caption{Mean radial velocity values as derived from the \gaia spectroscopic  data for nearby clusters.}
\label{tab:gaiarvexp}
\begin{tabular}{lrrrr}
\hline\hline
Name & V$_\mathrm{rad}$ & $\sigma$(V$_\mathrm{rad}$) & uwsd & Nobs \\
\hline
Hyades &  39.87 &  0.05 &  2.28 &  150 \\ 
ComaBer &   0.21 &  0.13 &  2.15 &   43 \\ 
Pleiades &   5.54 &  0.10 &  2.00 &  195 \\ 
IC2391 &  15.00 &  0.24 &  1.19 &   35 \\ 
IC2602 &  17.62 &  0.22 &  1.24 &   36 \\ 
alphaPer &  -0.32 &  0.17 &  1.38 &   71 \\ 
Praesepe &  34.84 &  0.07 &  1.45 &  176 \\ 
NGC2451A &  23.08 &  0.34 &  1.32 &   31 \\ 
Blanco1 &   6.01 &  0.15 &  1.03 &   51 \\ 
\hline
\end{tabular}
\tablefoot{Columns: 1. Cluster name; 2: weighted-mean radial velocity in km~s$^{-1}$; 3. standard uncertainty on radial velocity; 4. unit-weight standard deviation of mean velocity solution;
and 5. number of observations in mean velocity solution.}
\end{table} 

Figure~\ref{fig:hyadesrvcomp} shows the differences (observed $-$ predicted, where the predicted value is based on the local projection of the space velocity of the cluster) in the radial velocities for 191 stars in the Hyades cluster. Only stars for which the colour index \bpmrp\ is greater than 0.4~mag were used. The results for all 9 nearby clusters are shown in Table~\ref{tab:gaiarvexp}. 

\begin{figure}[t]
\centering
\includegraphics[width=9cm]{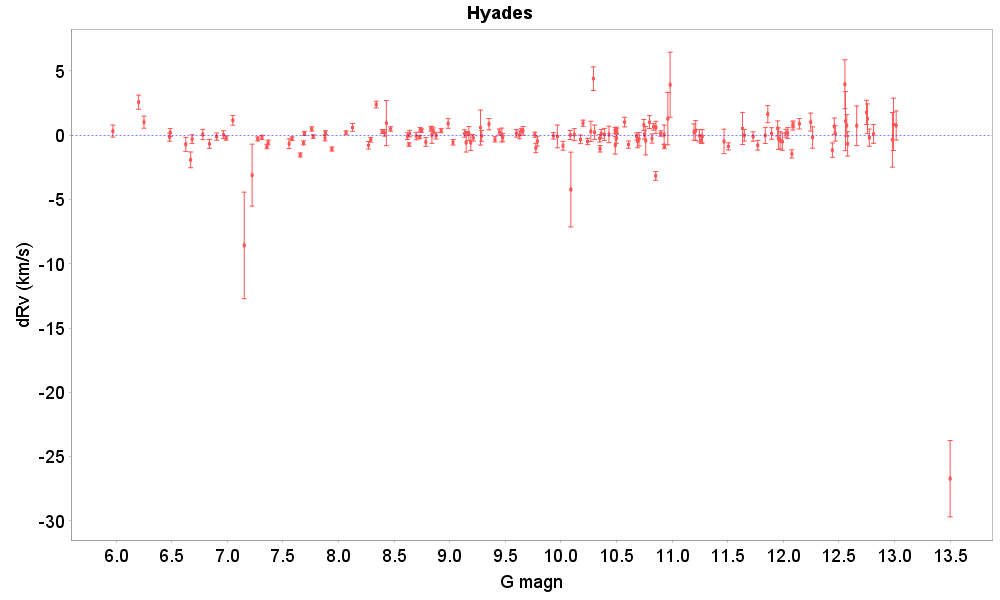}
\caption{Differences between the predicted and observed radial velocities in the Hyades cluster as a function of G magnitude.}
\label{fig:hyadesrvcomp}
\end{figure}

\begin{table*}[t]
\caption{Space velocity fitting results for nearby clusters \label{tab:nearbyocuvw}}
\centering
\begin{tabular}{lrrrrrrrrrrr}
\hline\hline
Name & U' & V' & W' & $c_{U'V'}$ & $c_{U'W'}$ & uwsd & $\alpha_c$  &$\varpi$ & V$_\textrm{rad}$ & $\mu_{\alpha *}$ & $\mu_\delta$\\
ClustId & $\sigma$U' & $\sigma$V' & $\sigma$W' & $c_{V'W'}$ & $\sigma$v & Observ. & $\delta_c$ &$\sigma\varpi$ & $\sigma$V$_\textrm{rad}$ & $\sigma\mu_{\alpha *}$ & $\sigma\mu_\delta$\\ 
& km/s & km/s & km/s & &&& degr. & mas & km/s & mas/yr & mas/yr \\
\hline
Hyades &  -6.059 &  45.691 &   5.544 &  0.33 &  0.35 &  0.67 &  97.5407 &  21.052  &  39.96 &  101.005 &  -28.490 \\
C0424+157 &   0.031 &   0.069 &   0.025 &  0.93 &  0.40 & 515 &   6.8148 &  0.065 &   0.06 &    0.171 &    0.137 \\
ComaBer &  -1.638 &   4.785 &  -3.528 &  0.35 & -0.86 &  0.48 & 110.1896 &  11.640  &  -0.52 &  -12.111 &   -8.996 \\
C1222+263 &   0.078 &   0.018 &   0.040 & -0.39 &  0.40 &  153 & -34.3206 &  0.034 &   0.07 &    0.048 &    0.121 \\
Pleiades &  -1.311 &  21.390 & -24.457 &  0.48 &  0.50 &  0.77 &  93.5183 &   7.364  &   5.65 &   19.997 &  -45.548 \\
C0344+239 &   0.070 &   0.105 &   0.057 &  0.90 &  0.40 & 1326 & -48.7831 &  0.005 &   0.09 &    0.127 &    0.101 \\
Praesepe &   0.339 &  49.097 &   1.200 & -0.50 & -0.60 &  0.76 &  89.5122 &   5.371  &  35.64 &  -36.047 &  -12.917 \\
C0937+201 &   0.090 &   0.106 &   0.050 &  0.92 &  0.40 & 938 &   1.3517 &  0.003 &   0.10 &    0.110 &    0.066 \\
alphaPer &  -5.110 &  24.183 & -14.122 &  0.25 &  0.40 &  0.68 & 101.9183 &   5.718  &  -0.29 &   22.929 &  -25.556 \\
C0318+484 &   0.053 &   0.067 &   0.097 &  0.59 &  0.40 & 740 & -29.7555 &  0.005 &   0.08 &    0.071 &    0.095 \\
IC2391 &  -0.751 &  28.459 &  -1.590 & -0.20 &  0.38 &  0.68 &  91.6471 &   6.597  &  14.59 &  -24.927 &   23.256 \\
C0838-528 &   0.054 &   0.062 &   0.105 & -0.52 &  0.40 &  325 &  -3.4126 &  0.007 &   0.09 &    0.080 &    0.110 \\
IC2602 &  -9.467 &  16.867 & -12.377 & -0.05 &  0.40 &  0.72 & 119.3285 &   6.571  &  17.43 &  -17.783 &   10.655 \\
C1041-641 &   0.056 &   0.024 &   0.120 & -0.16 &  0.40 & 492 & -32.7371 &  0.007 &   0.11 &    0.040 &    0.098 \\
Blanco1 &   6.176 &  21.150 &  -0.296 &  0.01 & -0.86 &  0.65 &  73.6042 &   4.216  &   5.78 &   18.724 &    2.650 \\
C0001-302 &   0.111 &   0.020 &   0.065 & -0.02 &  0.40 &  489 &  -0.8388 &  0.003 &   0.10 &    0.017 &    0.070 \\
NGC2451 &   5.806 &  32.440 &  -3.100 & -0.24 &  0.34 &  0.68 &  79.8905 &   5.163  &  22.85 &  -21.063 &   15.378 \\
C0743-378 &   0.048 &   0.095 &   0.084 & -0.76 &  0.40 &  400 &  -5.4202 &  0.005 &   0.09 &    0.065 &    0.093 \\
\\
\hline
\end{tabular}
\tablefoot{Columns: 1. Cluster identifiers; 2 to 4 U', V' and W' velocity components in the equatorial system; 5. U'V' error correlation (top) V'W' error correlation (bottom); 6. U'W' error correlation (top), applied internal velocity dispersion in km~s$^{-1}$ (bottom); 7. unit-weight standard deviation of solution (top), number of stars (bottom); 8. Coordinates of the convergent point; 9. parallax (mas); 10. radial velocity (km~s$^{-1}$); 11. proper motion in right ascension; and 12. proper motion in declination. 
}
\label{tab:nearbyclusters}
\end{table*}

\subsection{More distant open clusters\label{sec:distclust}}
\begin{figure*}[t]
\centering
\includegraphics[totalheight=5cm]{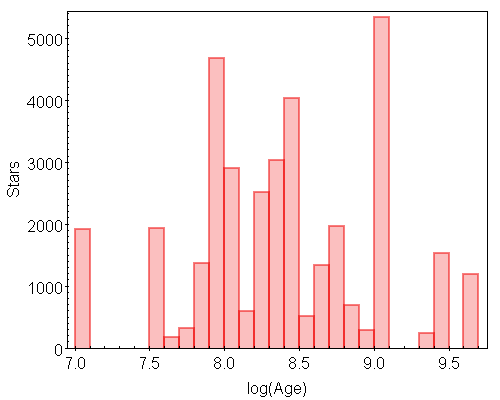}
\includegraphics[totalheight=5cm]{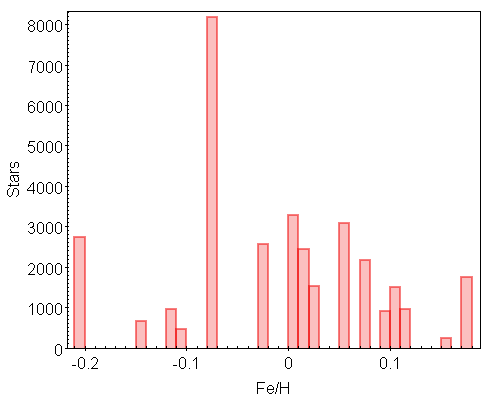}
\caption{Distributions over age and composition for stars in the 32 open clusters selected for the composite HRD, including the nearby clusters.}
\label{fig:ocdistr}
\end{figure*}

For the more distant clusters, a selection was made of 37 relatively rich clusters, generally only little reddened, and as far as
possible, covering a spread in ages and chemical composition (Fig.~\ref{fig:ocdistr}). These clusters were all analysed in a combined solution of the mean parallax and proper motion from the observed astrometric data of the member stars. This is an iterative procedure, where cluster membership determination is based on the solution for the astrometric parameters of the cluster. The combined solution for the astrometric parameters of a cluster takes into account noise contributions from three sources:
\begin{enumerate}
\item the covariance matrix of the astrometric solution for each star; 
\item the internal velocity dispersion of the cluster, affecting the dispersion of the proper motions;
\item the effect of the cluster size relative to its distance, which 
\begin{enumerate}
\item is reflected in a dispersion on the parallaxes of the cluster members; \item is reflected in a dispersion in proper motions in the direction of, and scaled by, the cluster proper motion.
\end{enumerate}
\end{enumerate}

\begin{figure}[t]
\centering
\includegraphics[width=8cm]{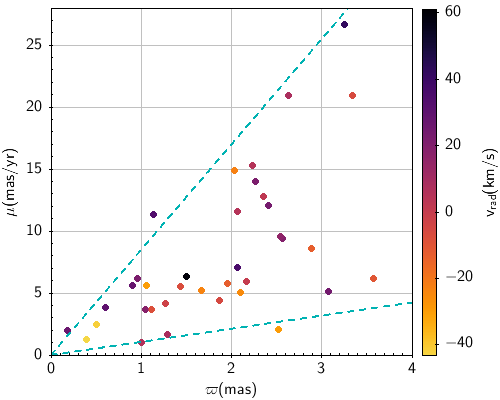}
\caption{Comparison between the parallaxes and proper motions for the 37 open clusters. The upper and lower diagonal lines represent tangential velocities of 40 and 5 km~s$^{-1}$ , respectively.}
\label{fig:pmpar}
\end{figure}
When we assume that the velocity distribution is isotropic within the measurement accuracy, then the second of these noise contributions will be diagonal. The first and third may also contain significant off-diagonal elements. Given a cluster parallax of $\varpi_c$ , a cluster proper motion of $(\mu_{\alpha,c},\mu_{\delta,c})$, and an average relative dispersion in the parallaxes of the cluster stars of $\sigma_\varpi/\varpi = \sigma_R/R$ (where $R$ is the distance to the cluster centre), the contribution to the dispersion in the proper motions of the cluster stars scales with the relative dispersion of the parallaxes and the proper motions of the cluster:
\begin{eqnarray}
\sigma\mu_{\alpha,s} &=& |\mu_{\alpha,c}| \times \sigma_\varpi/\varpi \\
\sigma\mu_{\delta,s} &=& |\mu_{\delta,c}| \times \sigma_\varpi/\varpi 
.\end{eqnarray}
For most of the clusters with distances beyond 250~pc, this contribution will be small to very small relative to other contributions. Figure~\ref{fig:pmpar} shows the overall relation between parallaxe and proper motion amplitudes for the selection of clusters we used.

The contributions are summed into a single noise matrix, of which an upper-triangular square root is used to normalise the observation equations  that describe the cluster proper motion and parallax as a function of the observed proper motions and parallaxes of the individual cluster members.

\begin{figure}
 \begin{center}
\includegraphics[width=8cm]{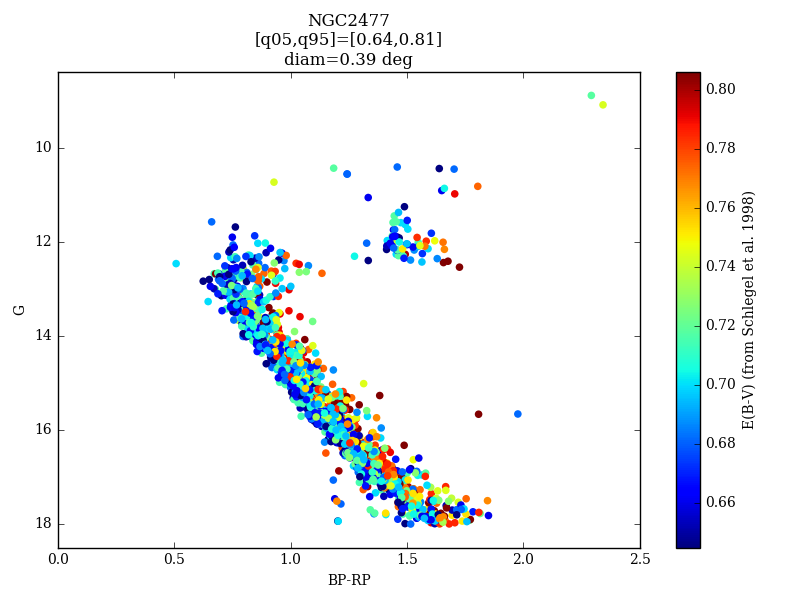}
\end{center}
\caption{Colour-magnitude diagram of NGC~2477, with each star colour-coded by the value of integrated extinction in the catalogue of \cite{Schlegel98}.}
\label{fig:NGC2477_cmdwithext}
\end{figure}

\begin{figure*}[t]
\centering
\centerline{
\includegraphics[width=6.5cm]{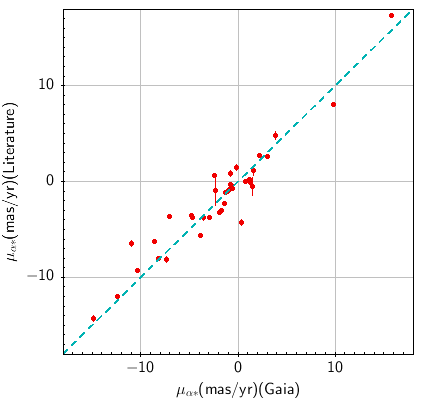}
\includegraphics[width=6.5cm]{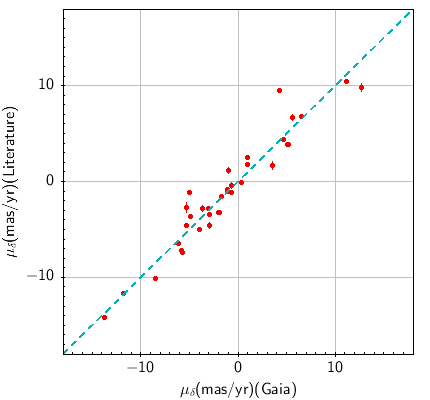}}
\centerline{
\includegraphics[width=7cm]{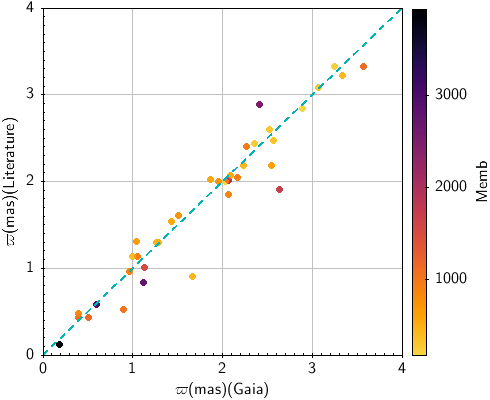}
\includegraphics[width=6.3cm]{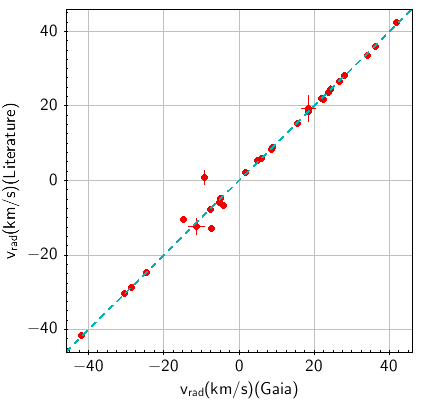}
}
\caption{Comparisons with values quoted in literature (see text) for (top) proper motions in right ascension, proper motions in declination, (bottom) parallaxes, and radial velocities for 37 open clusters with distances beyond 250 pc.}
\label{fig:astrcomp}
\end{figure*}

\begin{figure}[t]
\centering
\includegraphics[width=7.5cm]{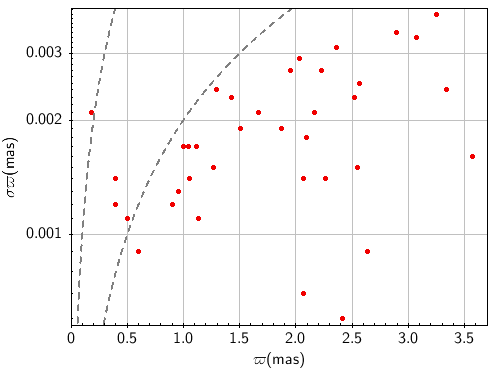}
\caption{Standard uncertainties on the mean cluster parallax determinations. The curves represent the 100 and 500 $\sigma$ significance levels when only the standard uncertainties are
considered.}
\label{fig:parerror}
\end{figure}

Table~\ref{tab:overview} presents an overview of the astrometric solutions for 37 open clusters, with mean radial velocities when available in the \gaia data. 
We note that some clusters are not included in Table~\ref{tab:openclref} because their colour-magnitude diagrams are too disturbed by interstellar extinction (see an illustration of the differential extinction effect in Fig. \ref{fig:NGC2477_cmdwithext}). 
The proper motions are compared with those presented by \cite{2003ARep...47....6L} in Fig.~\ref{fig:astrcomp}, and they agree well overall, but there is also an indication that errors on the data presented in \cite{2003ARep...47....6L} are underestimated. In the same figure the comparison between the parallaxes as derived from the DR2 data and parallax values derived from photometric distances as (mostly) presented in \cite{2005A&A...438.1163K} are shown, and again generally agree well \citep[see also the validation with more clusters in][]{DR2-DPACP-39}. The systematic difference of 0.029~mas, which can be observed for globular clusters \citep{DR2-DPACP-34}, is too small to be noticed here (Fig.~\ref{fig:parerror}), but the calibration noise on the DR2 parallaxes (0.025~mas), obtained in the same study, is significantly larger than the standard uncertainties on the mean cluster parallaxes and is therefore the main contributor to the uncertainties on the cluster parallaxes. In most cases, however, this amounts to less than 1\% in error on the parallax, or 0.02 mag\ in distance modulus.

The maximum radius for each cluster was determined from the contrast between the cluster and the field stars in the proper motion and parallax domain. In practice, this means that the density of field stars for which the combined information on the parallax and proper motion, combined with uncertainties and error correlations, leaves a significant possibility for a field star to be a cluster member. When the surface density of these field stars becomes similar to the surface density of the cluster stars, we have reached the maximum radius for the cluster in this particular data set and parameter space. It is well possible, however, that for a catalogue with higher accuracies on the astrometric parameters for the fainter stars in particular, this limit will be found still farther away from the cluster centre.  
Radial velocities for the clusters, mostly as given in \cite{2005A&A...438.1163K} or \cite{2014A&A...562A..54C}, were compared with the mean radial velocities as derived from the \gaia DR2 data. A limited spectral range was used, for which there is clear consistency of the radial velocity measurements. The summary of the results is shown in Fig.~\ref{fig:astrcomp} and generally agrees well \citep[see also the validation with more clusters in][]{DR2-DPACP-39}. The largest discrepancies are found for NGC~2516 \citep[RAVE measurements in][]{2014A&A...562A..54C} and Trumpler 2 \citep{2005A&A...438.1163K}.

\onecolumn
\begin{center}
\begin{longtable}{l|rrrrrrrrr}
\caption{Overview of the results for open clusters with distances beyond 250~pc\label{tab:overview}}
\\
\hline\hline 
Name & $\alpha$ &$\varpi$ & $\mu_{\alpha *}$ & $\mu_\delta$ & $c_{12}$ & $c_{23}$ & nMemb & Vrad & uwsd\\
ClustId & $\delta$ & $\sigma_\varpi$ & $\sigma_{\mu\alpha *}$ & 
$\sigma_{\mu\delta}$ & $c_{13}$ & r(max)\degr & st.dev. & $\sigma$ & Obs. \\ 
& degr. & mas & mas/yr & mas/yr & & & & km/s &\\
\hline 
\endfirsthead

\multicolumn{10}{c}%
{{\bfseries \tablename\ \thetable{} -- continued from previous page}} \\
\hline 
Name & $\alpha$ &$\varpi$ & $\mu_{\alpha *}$ & $\mu_\delta$ & $c_{12}$ & $c_{23}$ & nMemb & Vrad & uwsd \\
ClustId & $\delta$ & $\sigma_\varpi$ & $\sigma_{\mu\alpha *}$ & 
$\sigma_{\mu\delta}$ & $c_{13}$ & r(max)\degr & st.dev. & $\sigma$ & Obs. \\ 
& degr. & mas & mas/yr & mas/yr & & & & km/s &  \\
\hline 
\endhead

\hline \multicolumn{10}{r}{{Continued on next page}} \\ \hline
\endfoot

\hline \hline
\endlastfoot
NGC0188 &  11.7494 &    0.5053 &  -2.3087 &  -0.9565 & -0.04 & -0.02  & 1181 & -41.86 &  1.43  \\ 
C0039+850 &  85.2395 &   0.0011 &   0.0035 &   0.0030 &   0.16 &  0.58 &  0.84 &  0.13 &   20 \\ 
NGC0752 &  29.2054 &    2.2304 &   9.8092 & -11.7637 &  0.02 & -0.04  &  433 &   5.90 &  1.68  \\ 
C0154+374 &  37.7454 &   0.0027 &   0.0191 &   0.0180 &   0.04 &  2.58 &  0.86 &  0.11 &   76 \\ 
Stock2 &  33.8282 &    2.6367 &  15.8241 & -13.7669 &  0.01 &  0.01  & 1742 &   8.58 &  1.45  \\ 
C0211+590 &  59.5813 &   0.0009 &   0.0103 &   0.0104 &  -0.00 &  2.36 &  0.78 &  0.09 &  109 \\ 
NGC0869 &  34.7391 &    0.3942 &  -0.6943 &  -1.0831 &  0.14 &  0.10  &  829 &       &       \\ 
C0215+569 &  57.1339 &   0.0014 &   0.0038 &   0.0041 &   0.08 &  0.19 &  0.83 &      &    0 \\ 
NGC0884 &  35.5430 &    0.3976 &  -0.6021 &  -1.0616 &  0.16 &  0.11  & 1077 & -44.69 &  4.98  \\ 
C0218+568 &  57.1591 &   0.0012 &   0.0035 &   0.0036 &   0.10 &  0.29 &  0.86 &  0.73 &    2 \\ 
Trump02 &  39.1879 &    1.4316 &   1.5305 &  -5.3361 &  0.05 &  0.04  &  589 &  -4.06 &  0.75  \\ 
C0233+557 &  55.8846 &   0.0023 &   0.0116 &   0.0117 &   0.01 &  1.21 &  0.90 &  0.09 &    4 \\ 
NGC1039 &  40.5843 &    1.9536 &   0.7256 &  -5.7320 &  0.02 & -0.02  &  764 &  -7.27 &  1.44  \\ 
C0238+425 &  42.7027 &   0.0027 &   0.0109 &   0.0103 &   0.04 &  1.87 &  0.79 &  0.72 &   18 \\ 
NGC1901 &  79.6838 &    2.3582 &   1.5953 &  12.6920 &  0.03 &  0.10  &  290 &   1.62 &  1.60  \\ 
C0518-685 & -68.1627 &   0.0031 &   0.0276 &   0.0277 &  -0.03 &  2.30 &  1.04 &  0.56 &   16 \\ 
NGC2158 &  91.8751 &    0.1833 &  -0.1665 &  -1.9932 &  0.18 & -0.19  & 3942 &  26.64 &  2.30  \\ 
C0604+241 &  24.1163 &   0.0021 &   0.0035 &   0.0029 &   0.21 &  0.24 &  0.92 &  0.60 &   11 \\ 
NGC2168 &  92.2688 &    1.1301 &   2.2784 &  -2.9336 &  0.08 & -0.08  & 1794 &  -7.70 &  2.73  \\
C0605+243 &  24.3148 &   0.0013 &   0.0052 &   0.0050 &   0.05 &  1.20 &  0.87 &  0.27 &    6 \\ 
NGC2232 &  96.9973 &    3.0710 &  -4.7737 &  -1.9014 &  0.04 & -0.04  &  318 &  24.22 &  0.96  \\ 
C0624-047 &  -4.7929 &   0.0033 &   0.0185 &   0.0181 &   0.04 &  2.76 &  0.78 &  0.44 &    9 \\ 
Trump10 & 131.8982 &    2.2637 & -12.3536 &   6.5309 &  0.02 &  0.00  &  947 &  21.97 &  1.00  \\ 
C0646-423 & -42.5192 &   0.0014 &   0.0102 &   0.0104 &  -0.01 &  1.69 &  0.82 &  0.31 &   28 \\ 
NGC2323 & 105.7245 &    1.0012 &  -0.7977 &  -0.6540 &  0.06 & -0.03  &  382 &  11.55 &       \\ 
C0700-082 &  -8.3586 &   0.0017 &   0.0063 &   0.0063 &   0.00 &  0.73 &  0.87 &      &    1 \\ 
NGC2360 & 109.4452 &    0.9018 &   0.3853 &   5.5893 &  0.07 & -0.02  & 1037 &  28.02 &  1.74  \\ 
C0715-155 & -15.6317 &   0.0012 &   0.0048 &   0.0048 &  -0.05 &  0.74 &  0.79 &  0.19 &   15 \\ 
Coll140 & 111.0308 &    2.5685 &  -8.1285 &   4.7105 &  0.02 &  0.02  &  332 &  18.53 &  1.75  \\ 
C0722-321 & -32.1113 &   0.0025 &   0.0215 &   0.0220 &  -0.01 &  2.69 &  0.81 &  1.85 &    5 \\ 
NGC2423 & 114.2904 &    1.0438 &  -0.7343 &  -3.6333 &  0.09 & -0.00  &  694 &  18.50 &  2.04  \\ 
C0734-137 & -13.8348 &   0.0017 &   0.0070 &   0.0069 &  -0.04 &  1.04 &  0.81 &  0.17 &   19 \\ 
NGC2422 & 114.1463 &    2.0690 &  -7.0200 &   0.9592 &  0.05 &  0.01  &  907 &  36.21 &  1.42  \\ 
C0734-143 & -14.4844 &   0.0014 &   0.0098 &   0.0099 &  -0.02 &  1.45 &  0.74 &  0.57 &   30 \\ 
NGC2437 & 115.4358 &    0.6005 &  -3.8232 &   0.3729 &  0.11 &  0.01  & 3032 &  37.34 &       \\ 
C0739-147 & -14.8506 &   0.0009 &   0.0031 &   0.0031 &  -0.06 &  0.74 &  0.83 &      &    1 \\ 
NGC2447 & 116.1262 &    0.9603 &  -3.5680 &   5.0434 &  0.03 &  0.01  &  926 &  22.37 &  3.01  \\ 
C0742-237 & -23.8567 &   0.0013 &   0.0056 &   0.0057 &  -0.01 &  1.00 &  0.80 &  0.26 &   11 \\ 
NGC2516 & 119.5469 &    2.4118 &  -4.6579 &  11.1517 &  0.02 & -0.00  & 2518 &  23.78 &  1.39  \\ 
C0757-607 & -60.7749 &   0.0006 &   0.0075 &   0.0075 &  -0.01 &  2.54 &  0.83 &  0.11 &  156 \\ 
NGC2547 & 122.5654 &    2.5438 &  -8.5999 &   4.2542 &  0.02 &  0.00  &  644 &  15.46 &  2.47  \\ 
C0809-491 & -49.0498 &   0.0015 &   0.0148 &   0.0148 &  -0.00 &  2.79 &  0.78 &  0.83 &   22 \\ 
NGC2548 & 123.3834 &    1.2897 &  -1.3302 &   1.0164 &  0.13 &  0.00  &  509 &   8.83 &  1.77  \\ 
C0811-056 &  -5.7363 &   0.0024 &   0.0095 &   0.0093 &  -0.03 &  0.56 &  0.80 &  0.27 &    8 \\ 
NGC2682 & 132.8476 &    1.1325 & -10.9737 &  -2.9396 &  0.08 & -0.00  & 1520 &  34.05 &  1.94  \\ 
C0847+120 &  11.8369 &   0.0011 &   0.0064 &   0.0063 &  -0.01 &  1.06 &  0.76 &  0.10 &   66 \\ 
NGC3228 & 155.3791 &    2.0323 & -14.8800 &  -0.6498 &  0.03 &  0.03  &  222 &       &       \\ 
C1019-514 & -51.7693 &   0.0029 &   0.0220 &   0.0220 &  -0.01 &  2.27 &  0.81 &      &    0 \\ 
NGC3532 & 166.3975 &    2.0659 & -10.3790 &   5.1958 &  0.03 & -0.02  & 1879 &   4.85 &  2.24  \\ 
C1104-584 & -58.7335 &   0.0007 &   0.0079 &   0.0079 &   0.01 &  2.31 &  0.79 &  0.13 &  143 \\ 
NGC6025 & 240.7714 &    1.2646 &  -2.8846 &  -3.0222 & -0.02 &  0.01  &  452 &  -7.66 &       \\ 
C1559-603 & -60.4562 &   0.0015 &   0.0100 &   0.0099 &   0.03 &  0.94 &  0.75 &      &    1 \\ 
NGC6281 & 256.1638 &    1.8716 &  -1.8764 &  -3.9506 & -0.03 &  0.05  &  573 &  -5.02 &  2.17  \\ 
C1701-378 & -37.9180 &   0.0019 &   0.0144 &   0.0136 &   0.05 &  1.19 &  0.80 &  0.20 &   21 \\ 
IC4651 & 261.2035 &    1.0542 &  -2.4051 &  -5.0280 & -0.07 &  0.10  &  960 & -30.32 &  3.41  \\ 
C1720-499 & -49.9185 &   0.0014 &   0.0061 &   0.0060 &   0.06 &  0.76 &  0.80 &  0.19 &   56 \\ 
NGC6405 & 265.1220 &    2.1626 &  -1.3662 &  -5.8063 & -0.04 &  0.11  &  967 &  -9.20 &  5.39  \\ 
C1736-321 & -32.4135 &   0.0021 &   0.0140 &   0.0132 &   0.04 &  1.46 &  0.82 &  0.77 &   17 \\ 
IC4665 & 266.4978 &    2.8918 &  -0.8993 &  -8.5114 & -0.02 &  0.04  &  174 & -11.26 &  1.86  \\ 
C1743+057 &   5.5653 &   0.0034 &   0.0347 &   0.0345 &   0.02 &  2.39 &  0.75 &  2.12 &    6 \\ 
NGC6475 & 268.2736 &    3.5704 &   3.0722 &  -5.3157 & -0.02 &  0.04  & 1140 & -14.84 &  2.63  \\ 
C1750-348 & -34.6639 &   0.0016 &   0.0185 &   0.0184 &   0.02 &  3.86 &  0.82 &  0.17 &  113 \\ 
NGC6633 & 276.8737 &    2.5232 &   1.1584 &  -1.7371 & -0.03 &  0.09  &  321 & -28.59 &  1.83  \\ 
C1825+065 &   6.6081 &   0.0023 &   0.0199 &   0.0200 &   0.01 &  1.99 &  0.84 &  0.14 &   28 \\ 
IC4725 & 277.9462 &    1.5043 &  -1.7201 &  -6.1010 & -0.07 &  0.09  &  755 &       &       \\ 
C1828-192 & -19.1058 &   0.0019 &   0.0091 &   0.0091 &   0.04 &  1.53 &  0.89 &      &    0 \\ 
IC4756 & 279.6698 &    2.0943 &   1.2574 &  -4.9145 & -0.04 &  0.06  &  543 & -24.72 &  2.76  \\ 
C1836+054 &   5.3836 &   0.0018 &   0.0134 &   0.0134 &   0.02 &  2.05 &  0.84 &  0.17 &   38 \\ 
NGC6774 & 289.1055 &    3.2516 &  -0.9733 & -26.6464 & -0.03 &  0.11  &  234 &  41.79 &  3.36  \\ 
C1913-163 & -16.3901 &   0.0038 &   0.0367 &   0.0383 &   0.00 &  3.74 &  1.00 &  0.15 &   62 \\ 
NGC6793 & 290.7795 &    1.6672 &   3.8120 &   3.5622 & -0.03 &  0.06  &  465 & -10.85 &       \\ 
C1921+220 &  22.1400 &   0.0021 &   0.0131 &   0.0136 &  -0.02 &  1.47 &  0.81 &      &    1 \\ 
NGC7092 & 322.4220 &    3.3373 &  -7.3569 & -19.5993 & -0.02 & -0.00  &  433 &  -5.07 &  0.95  \\ 
C2130+482 &  48.1315 &   0.0024 &   0.0256 &   0.0260 &  -0.00 &  3.72 &  0.86 &  0.21 &   21 \\ 
\end{longtable}
\end{center}

\section{Gaia archive query\label{sec:query}}

The \gaia archive\footnote{https://gea.esac.esa.int/archive/} query corresponding to the filters described in Sect.~\ref{sec:filters} is the following (selecting here the first five stars):

{\small
\begin{verbatim}
SELECT TOP 5 phot_g_mean_mag+5*log10(parallax)-10 AS mg, bp_rp FROM gaiadr2.gaia_source 
   WHERE parallax_over_error > 10
   AND phot_g_mean_flux_over_error>50
   AND phot_rp_mean_flux_over_error>20
   AND phot_bp_mean_flux_over_error>20
   AND phot_bp_rp_excess_factor < 1.3+0.06*power(phot_bp_mean_mag-phot_rp_mean_mag,2)
   AND phot_bp_rp_excess_factor > 1.0+0.015*power(phot_bp_mean_mag-phot_rp_mean_mag,2)
   AND visibility_periods_used>8
   AND astrometric_chi2_al/(astrometric_n_good_obs_al-5)<1.44*greatest(1,exp(-0.4*(phot_g_mean_mag-19.5)))
\end{verbatim}
}

\end{document}

%% file: authors.tex
\author{
{\it Gaia} Collaboration
\and C.        ~Babusiaux                     \inst{\ref{inst:0002},\ref{inst:0001}}
\and F.        ~van Leeuwen                   \inst{\ref{inst:0003}}
\and M.A.      ~Barstow                       \inst{\ref{inst:0004}}
\and C.        ~Jordi                         \inst{\ref{inst:0005}}
\and A.        ~Vallenari                     \inst{\ref{inst:0006}}
\and D.        ~Bossini                       \inst{\ref{inst:0006}}
\and A.        ~Bressan                       \inst{\ref{inst:0008}}
\and T.        ~Cantat-Gaudin                 \inst{\ref{inst:0006},\ref{inst:0005}}
\and M.        ~van Leeuwen                   \inst{\ref{inst:0003}}
\and A.G.A.    ~Brown                         \inst{\ref{inst:0012}}
\and T.        ~Prusti                        \inst{\ref{inst:0013}}
\and J.H.J.    ~de Bruijne                    \inst{\ref{inst:0013}}
\and C.A.L.    ~Bailer-Jones                  \inst{\ref{inst:0015}}
\and M.        ~Biermann                      \inst{\ref{inst:0016}}
\and D.W.      ~Evans                         \inst{\ref{inst:0003}}
\and L.        ~Eyer                          \inst{\ref{inst:0018}}
\and F.        ~Jansen                        \inst{\ref{inst:0019}}
\and S.A.      ~Klioner                       \inst{\ref{inst:0020}}
\and U.        ~Lammers                       \inst{\ref{inst:0021}}
\and L.        ~Lindegren                     \inst{\ref{inst:0022}}
\and X.        ~Luri                          \inst{\ref{inst:0005}}
\and F.        ~Mignard                       \inst{\ref{inst:0024}}
\and C.        ~Panem                         \inst{\ref{inst:0025}}
\and D.        ~Pourbaix                      \inst{\ref{inst:0026},\ref{inst:0027}}
\and S.        ~Randich                       \inst{\ref{inst:0028}}
\and P.        ~Sartoretti                    \inst{\ref{inst:0001}}
\and H.I.      ~Siddiqui                      \inst{\ref{inst:0030}}
\and C.        ~Soubiran                      \inst{\ref{inst:0031}}
\and N.A.      ~Walton                        \inst{\ref{inst:0003}}
\and F.        ~Arenou                        \inst{\ref{inst:0001}}
\and U.        ~Bastian                       \inst{\ref{inst:0016}}
\and M.        ~Cropper                       \inst{\ref{inst:0035}}
\and R.        ~Drimmel                       \inst{\ref{inst:0036}}
\and D.        ~Katz                          \inst{\ref{inst:0001}}
\and M.G.      ~Lattanzi                      \inst{\ref{inst:0036}}
\and J.        ~Bakker                        \inst{\ref{inst:0021}}
\and C.        ~Cacciari                      \inst{\ref{inst:0040}}
\and J.        ~Casta\~{n}eda                 \inst{\ref{inst:0005}}
\and L.        ~Chaoul                        \inst{\ref{inst:0025}}
\and N.        ~Cheek                         \inst{\ref{inst:0043}}
\and F.        ~De Angeli                     \inst{\ref{inst:0003}}
\and C.        ~Fabricius                     \inst{\ref{inst:0005}}
\and R.        ~Guerra                        \inst{\ref{inst:0021}}
\and B.        ~Holl                          \inst{\ref{inst:0018}}
\and E.        ~Masana                        \inst{\ref{inst:0005}}
\and R.        ~Messineo                      \inst{\ref{inst:0049}}
\and N.        ~Mowlavi                       \inst{\ref{inst:0018}}
\and K.        ~Nienartowicz                  \inst{\ref{inst:0051}}
\and P.        ~Panuzzo                       \inst{\ref{inst:0001}}
\and J.        ~Portell                       \inst{\ref{inst:0005}}
\and M.        ~Riello                        \inst{\ref{inst:0003}}
\and G.M.      ~Seabroke                      \inst{\ref{inst:0035}}
\and P.        ~Tanga                         \inst{\ref{inst:0024}}
\and F.        ~Th\'{e}venin                  \inst{\ref{inst:0024}}
\and G.        ~Gracia-Abril                  \inst{\ref{inst:0058},\ref{inst:0016}}
\and G.        ~Comoretto                     \inst{\ref{inst:0030}}
\and M.        ~Garcia-Reinaldos              \inst{\ref{inst:0021}}
\and D.        ~Teyssier                      \inst{\ref{inst:0030}}
\and M.        ~Altmann                       \inst{\ref{inst:0016},\ref{inst:0064}}
\and R.        ~Andrae                        \inst{\ref{inst:0015}}
\and M.        ~Audard                        \inst{\ref{inst:0018}}
\and I.        ~Bellas-Velidis                \inst{\ref{inst:0067}}
\and K.        ~Benson                        \inst{\ref{inst:0035}}
\and J.        ~Berthier                      \inst{\ref{inst:0069}}
\and R.        ~Blomme                        \inst{\ref{inst:0070}}
\and P.        ~Burgess                       \inst{\ref{inst:0003}}
\and G.        ~Busso                         \inst{\ref{inst:0003}}
\and B.        ~Carry                         \inst{\ref{inst:0024},\ref{inst:0069}}
\and A.        ~Cellino                       \inst{\ref{inst:0036}}
\and G.        ~Clementini                    \inst{\ref{inst:0040}}
\and M.        ~Clotet                        \inst{\ref{inst:0005}}
\and O.        ~Creevey                       \inst{\ref{inst:0024}}
\and M.        ~Davidson                      \inst{\ref{inst:0079}}
\and J.        ~De Ridder                     \inst{\ref{inst:0080}}
\and L.        ~Delchambre                    \inst{\ref{inst:0081}}
\and A.        ~Dell'Oro                      \inst{\ref{inst:0028}}
\and C.        ~Ducourant                     \inst{\ref{inst:0031}}
\and J.        ~Fern\'{a}ndez-Hern\'{a}ndez   \inst{\ref{inst:0084}}
\and M.        ~Fouesneau                     \inst{\ref{inst:0015}}
\and Y.        ~Fr\'{e}mat                    \inst{\ref{inst:0070}}
\and L.        ~Galluccio                     \inst{\ref{inst:0024}}
\and M.        ~Garc\'{i}a-Torres             \inst{\ref{inst:0088}}
\and J.        ~Gonz\'{a}lez-N\'{u}\~{n}ez    \inst{\ref{inst:0043},\ref{inst:0090}}
\and J.J.      ~Gonz\'{a}lez-Vidal            \inst{\ref{inst:0005}}
\and E.        ~Gosset                        \inst{\ref{inst:0081},\ref{inst:0027}}
\and L.P.      ~Guy                           \inst{\ref{inst:0051},\ref{inst:0095}}
\and J.-L.     ~Halbwachs                     \inst{\ref{inst:0096}}
\and N.C.      ~Hambly                        \inst{\ref{inst:0079}}
\and D.L.      ~Harrison                      \inst{\ref{inst:0003},\ref{inst:0099}}
\and J.        ~Hern\'{a}ndez                 \inst{\ref{inst:0021}}
\and D.        ~Hestroffer                    \inst{\ref{inst:0069}}
\and S.T.      ~Hodgkin                       \inst{\ref{inst:0003}}
\and A.        ~Hutton                        \inst{\ref{inst:0103}}
\and G.        ~Jasniewicz                    \inst{\ref{inst:0104}}
\and A.        ~Jean-Antoine-Piccolo          \inst{\ref{inst:0025}}
\and S.        ~Jordan                        \inst{\ref{inst:0016}}
\and A.J.      ~Korn                          \inst{\ref{inst:0107}}
\and A.        ~Krone-Martins                 \inst{\ref{inst:0108}}
\and A.C.      ~Lanzafame                     \inst{\ref{inst:0109},\ref{inst:0110}}
\and T.        ~Lebzelter                     \inst{\ref{inst:0111}}
\and W.        ~L\"{ o}ffler                  \inst{\ref{inst:0016}}
\and M.        ~Manteiga                      \inst{\ref{inst:0113},\ref{inst:0114}}
\and P.M.      ~Marrese                       \inst{\ref{inst:0115},\ref{inst:0116}}
\and J.M.      ~Mart\'{i}n-Fleitas            \inst{\ref{inst:0103}}
\and A.        ~Moitinho                      \inst{\ref{inst:0108}}
\and A.        ~Mora                          \inst{\ref{inst:0103}}
\and K.        ~Muinonen                      \inst{\ref{inst:0120},\ref{inst:0121}}
\and J.        ~Osinde                        \inst{\ref{inst:0122}}
\and E.        ~Pancino                       \inst{\ref{inst:0028},\ref{inst:0116}}
\and T.        ~Pauwels                       \inst{\ref{inst:0070}}
\and J.-M.     ~Petit                         \inst{\ref{inst:0126}}
\and A.        ~Recio-Blanco                  \inst{\ref{inst:0024}}
\and P.J.      ~Richards                      \inst{\ref{inst:0128}}
\and L.        ~Rimoldini                     \inst{\ref{inst:0051}}
\and A.C.      ~Robin                         \inst{\ref{inst:0126}}
\and L.M.      ~Sarro                         \inst{\ref{inst:0131}}
\and C.        ~Siopis                        \inst{\ref{inst:0026}}
\and M.        ~Smith                         \inst{\ref{inst:0035}}
\and A.        ~Sozzetti                      \inst{\ref{inst:0036}}
\and M.        ~S\"{ u}veges                  \inst{\ref{inst:0015}}
\and J.        ~Torra                         \inst{\ref{inst:0005}}
\and W.        ~van Reeven                    \inst{\ref{inst:0103}}
\and U.        ~Abbas                         \inst{\ref{inst:0036}}
\and A.        ~Abreu Aramburu                \inst{\ref{inst:0139}}
\and S.        ~Accart                        \inst{\ref{inst:0140}}
\and C.        ~Aerts                         \inst{\ref{inst:0080},\ref{inst:0142}}
\and G.        ~Altavilla                     \inst{\ref{inst:0115},\ref{inst:0116},\ref{inst:0040}}
\and M.A.      ~\'{A}lvarez                   \inst{\ref{inst:0113}}
\and R.        ~Alvarez                       \inst{\ref{inst:0021}}
\and J.        ~Alves                         \inst{\ref{inst:0111}}
\and R.I.      ~Anderson                      \inst{\ref{inst:0149},\ref{inst:0018}}
\and A.H.      ~Andrei                        \inst{\ref{inst:0151},\ref{inst:0152},\ref{inst:0064}}
\and E.        ~Anglada Varela                \inst{\ref{inst:0084}}
\and E.        ~Antiche                       \inst{\ref{inst:0005}}
\and T.        ~Antoja                        \inst{\ref{inst:0013},\ref{inst:0005}}
\and B.        ~Arcay                         \inst{\ref{inst:0113}}
\and T.L.      ~Astraatmadja                  \inst{\ref{inst:0015},\ref{inst:0160}}
\and N.        ~Bach                          \inst{\ref{inst:0103}}
\and S.G.      ~Baker                         \inst{\ref{inst:0035}}
\and L.        ~Balaguer-N\'{u}\~{n}ez        \inst{\ref{inst:0005}}
\and P.        ~Balm                          \inst{\ref{inst:0030}}
\and C.        ~Barache                       \inst{\ref{inst:0064}}
\and C.        ~Barata                        \inst{\ref{inst:0108}}
\and D.        ~Barbato                       \inst{\ref{inst:0167},\ref{inst:0036}}
\and F.        ~Barblan                       \inst{\ref{inst:0018}}
\and P.S.      ~Barklem                       \inst{\ref{inst:0107}}
\and D.        ~Barrado                       \inst{\ref{inst:0171}}
\and M.        ~Barros                        \inst{\ref{inst:0108}}
\and S.        ~Bartholom\'{e} Mu\~{n}oz      \inst{\ref{inst:0005}}
\and J.-L.     ~Bassilana                     \inst{\ref{inst:0140}}
\and U.        ~Becciani                      \inst{\ref{inst:0110}}
\and M.        ~Bellazzini                    \inst{\ref{inst:0040}}
\and A.        ~Berihuete                     \inst{\ref{inst:0177}}
\and S.        ~Bertone                       \inst{\ref{inst:0036},\ref{inst:0064},\ref{inst:0180}}
\and L.        ~Bianchi                       \inst{\ref{inst:0181}}
\and O.        ~Bienaym\'{e}                  \inst{\ref{inst:0096}}
\and S.        ~Blanco-Cuaresma               \inst{\ref{inst:0018},\ref{inst:0031},\ref{inst:0185}}
\and T.        ~Boch                          \inst{\ref{inst:0096}}
\and C.        ~Boeche                        \inst{\ref{inst:0006}}
\and A.        ~Bombrun                       \inst{\ref{inst:0188}}
\and R.        ~Borrachero                    \inst{\ref{inst:0005}}
\and S.        ~Bouquillon                    \inst{\ref{inst:0064}}
\and G.        ~Bourda                        \inst{\ref{inst:0031}}
\and A.        ~Bragaglia                     \inst{\ref{inst:0040}}
\and L.        ~Bramante                      \inst{\ref{inst:0049}}
\and M.A.      ~Breddels                      \inst{\ref{inst:0194}}
\and N.        ~Brouillet                     \inst{\ref{inst:0031}}
\and T.        ~Br\"{ u}semeister             \inst{\ref{inst:0016}}
\and E.        ~Brugaletta                    \inst{\ref{inst:0110}}
\and B.        ~Bucciarelli                   \inst{\ref{inst:0036}}
\and A.        ~Burlacu                       \inst{\ref{inst:0025}}
\and D.        ~Busonero                      \inst{\ref{inst:0036}}
\and A.G.      ~Butkevich                     \inst{\ref{inst:0020}}
\and R.        ~Buzzi                         \inst{\ref{inst:0036}}
\and E.        ~Caffau                        \inst{\ref{inst:0001}}
\and R.        ~Cancelliere                   \inst{\ref{inst:0204}}
\and G.        ~Cannizzaro                    \inst{\ref{inst:0205},\ref{inst:0142}}
\and R.        ~Carballo                      \inst{\ref{inst:0207}}
\and T.        ~Carlucci                      \inst{\ref{inst:0064}}
\and J.M.      ~Carrasco                      \inst{\ref{inst:0005}}
\and L.        ~Casamiquela                   \inst{\ref{inst:0005}}
\and M.        ~Castellani                    \inst{\ref{inst:0115}}
\and A.        ~Castro-Ginard                 \inst{\ref{inst:0005}}
\and P.        ~Charlot                       \inst{\ref{inst:0031}}
\and L.        ~Chemin                        \inst{\ref{inst:0214}}
\and A.        ~Chiavassa                     \inst{\ref{inst:0024}}
\and G.        ~Cocozza                       \inst{\ref{inst:0040}}
\and G.        ~Costigan                      \inst{\ref{inst:0012}}
\and S.        ~Cowell                        \inst{\ref{inst:0003}}
\and F.        ~Crifo                         \inst{\ref{inst:0001}}
\and M.        ~Crosta                        \inst{\ref{inst:0036}}
\and C.        ~Crowley                       \inst{\ref{inst:0188}}
\and J.        ~Cuypers$^\dagger$             \inst{\ref{inst:0070}}
\and C.        ~Dafonte                       \inst{\ref{inst:0113}}
\and Y.        ~Damerdji                      \inst{\ref{inst:0081},\ref{inst:0225}}
\and A.        ~Dapergolas                    \inst{\ref{inst:0067}}
\and P.        ~David                         \inst{\ref{inst:0069}}
\and M.        ~David                         \inst{\ref{inst:0228}}
\and P.        ~de Laverny                    \inst{\ref{inst:0024}}
\and F.        ~De Luise                      \inst{\ref{inst:0230}}
\and R.        ~De March                      \inst{\ref{inst:0049}}
\and D.        ~de Martino                    \inst{\ref{inst:0232}}
\and R.        ~de Souza                      \inst{\ref{inst:0233}}
\and A.        ~de Torres                     \inst{\ref{inst:0188}}
\and J.        ~Debosscher                    \inst{\ref{inst:0080}}
\and E.        ~del Pozo                      \inst{\ref{inst:0103}}
\and M.        ~Delbo                         \inst{\ref{inst:0024}}
\and A.        ~Delgado                       \inst{\ref{inst:0003}}
\and H.E.      ~Delgado                       \inst{\ref{inst:0131}}
\and S.        ~Diakite                       \inst{\ref{inst:0126}}
\and C.        ~Diener                        \inst{\ref{inst:0003}}
\and E.        ~Distefano                     \inst{\ref{inst:0110}}
\and C.        ~Dolding                       \inst{\ref{inst:0035}}
\and P.        ~Drazinos                      \inst{\ref{inst:0244}}
\and J.        ~Dur\'{a}n                     \inst{\ref{inst:0122}}
\and B.        ~Edvardsson                    \inst{\ref{inst:0107}}
\and H.        ~Enke                          \inst{\ref{inst:0247}}
\and K.        ~Eriksson                      \inst{\ref{inst:0107}}
\and P.        ~Esquej                        \inst{\ref{inst:0249}}
\and G.        ~Eynard Bontemps               \inst{\ref{inst:0025}}
\and C.        ~Fabre                         \inst{\ref{inst:0251}}
\and M.        ~Fabrizio                      \inst{\ref{inst:0115},\ref{inst:0116}}
\and S.        ~Faigler                       \inst{\ref{inst:0254}}
\and A.J.      ~Falc\~{a}o                    \inst{\ref{inst:0255}}
\and M.        ~Farr\`{a}s Casas              \inst{\ref{inst:0005}}
\and L.        ~Federici                      \inst{\ref{inst:0040}}
\and G.        ~Fedorets                      \inst{\ref{inst:0120}}
\and P.        ~Fernique                      \inst{\ref{inst:0096}}
\and F.        ~Figueras                      \inst{\ref{inst:0005}}
\and F.        ~Filippi                       \inst{\ref{inst:0049}}
\and K.        ~Findeisen                     \inst{\ref{inst:0001}}
\and A.        ~Fonti                         \inst{\ref{inst:0049}}
\and E.        ~Fraile                        \inst{\ref{inst:0249}}
\and M.        ~Fraser                        \inst{\ref{inst:0003},\ref{inst:0266}}
\and B.        ~Fr\'{e}zouls                  \inst{\ref{inst:0025}}
\and M.        ~Gai                           \inst{\ref{inst:0036}}
\and S.        ~Galleti                       \inst{\ref{inst:0040}}
\and D.        ~Garabato                      \inst{\ref{inst:0113}}
\and F.        ~Garc\'{i}a-Sedano             \inst{\ref{inst:0131}}
\and A.        ~Garofalo                      \inst{\ref{inst:0272},\ref{inst:0040}}
\and N.        ~Garralda                      \inst{\ref{inst:0005}}
\and A.        ~Gavel                         \inst{\ref{inst:0107}}
\and P.        ~Gavras                        \inst{\ref{inst:0001},\ref{inst:0067},\ref{inst:0244}}
\and J.        ~Gerssen                       \inst{\ref{inst:0247}}
\and R.        ~Geyer                         \inst{\ref{inst:0020}}
\and P.        ~Giacobbe                      \inst{\ref{inst:0036}}
\and G.        ~Gilmore                       \inst{\ref{inst:0003}}
\and S.        ~Girona                        \inst{\ref{inst:0283}}
\and G.        ~Giuffrida                     \inst{\ref{inst:0116},\ref{inst:0115}}
\and F.        ~Glass                         \inst{\ref{inst:0018}}
\and M.        ~Gomes                         \inst{\ref{inst:0108}}
\and M.        ~Granvik                       \inst{\ref{inst:0120},\ref{inst:0289}}
\and A.        ~Gueguen                       \inst{\ref{inst:0001},\ref{inst:0291}}
\and A.        ~Guerrier                      \inst{\ref{inst:0140}}
\and J.        ~Guiraud                       \inst{\ref{inst:0025}}
\and R.        ~Guti\'{e}rrez-S\'{a}nchez     \inst{\ref{inst:0030}}
\and R.        ~Haigron                       \inst{\ref{inst:0001}}
\and D.        ~Hatzidimitriou                \inst{\ref{inst:0244},\ref{inst:0067}}
\and M.        ~Hauser                        \inst{\ref{inst:0016},\ref{inst:0015}}
\and M.        ~Haywood                       \inst{\ref{inst:0001}}
\and U.        ~Heiter                        \inst{\ref{inst:0107}}
\and A.        ~Helmi                         \inst{\ref{inst:0194}}
\and J.        ~Heu                           \inst{\ref{inst:0001}}
\and T.        ~Hilger                        \inst{\ref{inst:0020}}
\and D.        ~Hobbs                         \inst{\ref{inst:0022}}
\and W.        ~Hofmann                       \inst{\ref{inst:0016}}
\and G.        ~Holland                       \inst{\ref{inst:0003}}
\and H.E.      ~Huckle                        \inst{\ref{inst:0035}}
\and A.        ~Hypki                         \inst{\ref{inst:0012},\ref{inst:0310}}
\and V.        ~Icardi                        \inst{\ref{inst:0049}}
\and K.        ~Jan{\ss}en                    \inst{\ref{inst:0247}}
\and G.        ~Jevardat de Fombelle          \inst{\ref{inst:0051}}
\and P.G.      ~Jonker                        \inst{\ref{inst:0205},\ref{inst:0142}}
\and \'{A}.L.  ~Juh\'{a}sz                    \inst{\ref{inst:0316},\ref{inst:0317}}
\and F.        ~Julbe                         \inst{\ref{inst:0005}}
\and A.        ~Karampelas                    \inst{\ref{inst:0244},\ref{inst:0320}}
\and A.        ~Kewley                        \inst{\ref{inst:0003}}
\and J.        ~Klar                          \inst{\ref{inst:0247}}
\and A.        ~Kochoska                      \inst{\ref{inst:0323},\ref{inst:0324}}
\and R.        ~Kohley                        \inst{\ref{inst:0021}}
\and K.        ~Kolenberg                     \inst{\ref{inst:0326},\ref{inst:0080},\ref{inst:0185}}
\and M.        ~Kontizas                      \inst{\ref{inst:0244}}
\and E.        ~Kontizas                      \inst{\ref{inst:0067}}
\and S.E.      ~Koposov                       \inst{\ref{inst:0003},\ref{inst:0332}}
\and G.        ~Kordopatis                    \inst{\ref{inst:0024}}
\and Z.        ~Kostrzewa-Rutkowska           \inst{\ref{inst:0205},\ref{inst:0142}}
\and P.        ~Koubsky                       \inst{\ref{inst:0336}}
\and S.        ~Lambert                       \inst{\ref{inst:0064}}
\and A.F.      ~Lanza                         \inst{\ref{inst:0110}}
\and Y.        ~Lasne                         \inst{\ref{inst:0140}}
\and J.-B.     ~Lavigne                       \inst{\ref{inst:0140}}
\and Y.        ~Le Fustec                     \inst{\ref{inst:0341}}
\and C.        ~Le Poncin-Lafitte             \inst{\ref{inst:0064}}
\and Y.        ~Lebreton                      \inst{\ref{inst:0001},\ref{inst:0344}}
\and S.        ~Leccia                        \inst{\ref{inst:0232}}
\and N.        ~Leclerc                       \inst{\ref{inst:0001}}
\and I.        ~Lecoeur-Taibi                 \inst{\ref{inst:0051}}
\and H.        ~Lenhardt                      \inst{\ref{inst:0016}}
\and F.        ~Leroux                        \inst{\ref{inst:0140}}
\and S.        ~Liao                          \inst{\ref{inst:0036},\ref{inst:0351},\ref{inst:0352}}
\and E.        ~Licata                        \inst{\ref{inst:0181}}
\and H.E.P.    ~Lindstr{\o}m                  \inst{\ref{inst:0354},\ref{inst:0355}}
\and T.A.      ~Lister                        \inst{\ref{inst:0356}}
\and E.        ~Livanou                       \inst{\ref{inst:0244}}
\and A.        ~Lobel                         \inst{\ref{inst:0070}}
\and M.        ~L\'{o}pez                     \inst{\ref{inst:0171}}
\and S.        ~Managau                       \inst{\ref{inst:0140}}
\and R.G.      ~Mann                          \inst{\ref{inst:0079}}
\and G.        ~Mantelet                      \inst{\ref{inst:0016}}
\and O.        ~Marchal                       \inst{\ref{inst:0001}}
\and J.M.      ~Marchant                      \inst{\ref{inst:0364}}
\and M.        ~Marconi                       \inst{\ref{inst:0232}}
\and S.        ~Marinoni                      \inst{\ref{inst:0115},\ref{inst:0116}}
\and G.        ~Marschalk\'{o}                \inst{\ref{inst:0316},\ref{inst:0369}}
\and D.J.      ~Marshall                      \inst{\ref{inst:0370}}
\and M.        ~Martino                       \inst{\ref{inst:0049}}
\and G.        ~Marton                        \inst{\ref{inst:0316}}
\and N.        ~Mary                          \inst{\ref{inst:0140}}
\and D.        ~Massari                       \inst{\ref{inst:0194}}
\and G.        ~Matijevi\v{c}                 \inst{\ref{inst:0247}}
\and T.        ~Mazeh                         \inst{\ref{inst:0254}}
\and P.J.      ~McMillan                      \inst{\ref{inst:0022}}
\and S.        ~Messina                       \inst{\ref{inst:0110}}
\and D.        ~Michalik                      \inst{\ref{inst:0022}}
\and N.R.      ~Millar                        \inst{\ref{inst:0003}}
\and D.        ~Molina                        \inst{\ref{inst:0005}}
\and R.        ~Molinaro                      \inst{\ref{inst:0232}}
\and L.        ~Moln\'{a}r                    \inst{\ref{inst:0316}}
\and P.        ~Montegriffo                   \inst{\ref{inst:0040}}
\and R.        ~Mor                           \inst{\ref{inst:0005}}
\and R.        ~Morbidelli                    \inst{\ref{inst:0036}}
\and T.        ~Morel                         \inst{\ref{inst:0081}}
\and D.        ~Morris                        \inst{\ref{inst:0079}}
\and A.F.      ~Mulone                        \inst{\ref{inst:0049}}
\and T.        ~Muraveva                      \inst{\ref{inst:0040}}
\and I.        ~Musella                       \inst{\ref{inst:0232}}
\and G.        ~Nelemans                      \inst{\ref{inst:0142},\ref{inst:0080}}
\and L.        ~Nicastro                      \inst{\ref{inst:0040}}
\and L.        ~Noval                         \inst{\ref{inst:0140}}
\and W.        ~O'Mullane                     \inst{\ref{inst:0021},\ref{inst:0095}}
\and C.        ~Ord\'{e}novic                 \inst{\ref{inst:0024}}
\and D.        ~Ord\'{o}\~{n}ez-Blanco        \inst{\ref{inst:0051}}
\and P.        ~Osborne                       \inst{\ref{inst:0003}}
\and C.        ~Pagani                        \inst{\ref{inst:0004}}
\and I.        ~Pagano                        \inst{\ref{inst:0110}}
\and F.        ~Pailler                       \inst{\ref{inst:0025}}
\and H.        ~Palacin                       \inst{\ref{inst:0140}}
\and L.        ~Palaversa                     \inst{\ref{inst:0003},\ref{inst:0018}}
\and A.        ~Panahi                        \inst{\ref{inst:0254}}
\and M.        ~Pawlak                        \inst{\ref{inst:0408},\ref{inst:0409}}
\and A.M.      ~Piersimoni                    \inst{\ref{inst:0230}}
\and F.-X.     ~Pineau                        \inst{\ref{inst:0096}}
\and E.        ~Plachy                        \inst{\ref{inst:0316}}
\and G.        ~Plum                          \inst{\ref{inst:0001}}
\and E.        ~Poggio                        \inst{\ref{inst:0167},\ref{inst:0036}}
\and E.        ~Poujoulet                     \inst{\ref{inst:0416}}
\and A.        ~Pr\v{s}a                      \inst{\ref{inst:0324}}
\and L.        ~Pulone                        \inst{\ref{inst:0115}}
\and E.        ~Racero                        \inst{\ref{inst:0043}}
\and S.        ~Ragaini                       \inst{\ref{inst:0040}}
\and N.        ~Rambaux                       \inst{\ref{inst:0069}}
\and M.        ~Ramos-Lerate                  \inst{\ref{inst:0422}}
\and S.        ~Regibo                        \inst{\ref{inst:0080}}
\and C.        ~Reyl\'{e}                     \inst{\ref{inst:0126}}
\and F.        ~Riclet                        \inst{\ref{inst:0025}}
\and V.        ~Ripepi                        \inst{\ref{inst:0232}}
\and A.        ~Riva                          \inst{\ref{inst:0036}}
\and A.        ~Rivard                        \inst{\ref{inst:0140}}
\and G.        ~Rixon                         \inst{\ref{inst:0003}}
\and T.        ~Roegiers                      \inst{\ref{inst:0430}}
\and M.        ~Roelens                       \inst{\ref{inst:0018}}
\and M.        ~Romero-G\'{o}mez              \inst{\ref{inst:0005}}
\and N.        ~Rowell                        \inst{\ref{inst:0079}}
\and F.        ~Royer                         \inst{\ref{inst:0001}}
\and L.        ~Ruiz-Dern                     \inst{\ref{inst:0001}}
\and G.        ~Sadowski                      \inst{\ref{inst:0026}}
\and T.        ~Sagrist\`{a} Sell\'{e}s       \inst{\ref{inst:0016}}
\and J.        ~Sahlmann                      \inst{\ref{inst:0021},\ref{inst:0439}}
\and J.        ~Salgado                       \inst{\ref{inst:0440}}
\and E.        ~Salguero                      \inst{\ref{inst:0084}}
\and N.        ~Sanna                         \inst{\ref{inst:0028}}
\and T.        ~Santana-Ros                   \inst{\ref{inst:0310}}
\and M.        ~Sarasso                       \inst{\ref{inst:0036}}
\and H.        ~Savietto                      \inst{\ref{inst:0445}}
\and M.        ~Schultheis                    \inst{\ref{inst:0024}}
\and E.        ~Sciacca                       \inst{\ref{inst:0110}}
\and M.        ~Segol                         \inst{\ref{inst:0448}}
\and J.C.      ~Segovia                       \inst{\ref{inst:0043}}
\and D.        ~S\'{e}gransan                 \inst{\ref{inst:0018}}
\and I-C.      ~Shih                          \inst{\ref{inst:0001}}
\and L.        ~Siltala                       \inst{\ref{inst:0120},\ref{inst:0453}}
\and A.F.      ~Silva                         \inst{\ref{inst:0108}}
\and R.L.      ~Smart                         \inst{\ref{inst:0036}}
\and K.W.      ~Smith                         \inst{\ref{inst:0015}}
\and E.        ~Solano                        \inst{\ref{inst:0171},\ref{inst:0458}}
\and F.        ~Solitro                       \inst{\ref{inst:0049}}
\and R.        ~Sordo                         \inst{\ref{inst:0006}}
\and S.        ~Soria Nieto                   \inst{\ref{inst:0005}}
\and J.        ~Souchay                       \inst{\ref{inst:0064}}
\and A.        ~Spagna                        \inst{\ref{inst:0036}}
\and F.        ~Spoto                         \inst{\ref{inst:0024},\ref{inst:0069}}
\and U.        ~Stampa                        \inst{\ref{inst:0016}}
\and I.A.      ~Steele                        \inst{\ref{inst:0364}}
\and H.        ~Steidelm\"{ u}ller            \inst{\ref{inst:0020}}
\and C.A.      ~Stephenson                    \inst{\ref{inst:0030}}
\and H.        ~Stoev                         \inst{\ref{inst:0470}}
\and F.F.      ~Suess                         \inst{\ref{inst:0003}}
\and J.        ~Surdej                        \inst{\ref{inst:0081}}
\and L.        ~Szabados                      \inst{\ref{inst:0316}}
\and E.        ~Szegedi-Elek                  \inst{\ref{inst:0316}}
\and D.        ~Tapiador                      \inst{\ref{inst:0475},\ref{inst:0476}}
\and F.        ~Taris                         \inst{\ref{inst:0064}}
\and G.        ~Tauran                        \inst{\ref{inst:0140}}
\and M.B.      ~Taylor                        \inst{\ref{inst:0479}}
\and R.        ~Teixeira                      \inst{\ref{inst:0233}}
\and D.        ~Terrett                       \inst{\ref{inst:0128}}
\and P.        ~Teyssandier                   \inst{\ref{inst:0064}}
\and W.        ~Thuillot                      \inst{\ref{inst:0069}}
\and A.        ~Titarenko                     \inst{\ref{inst:0024}}
\and F.        ~Torra Clotet                  \inst{\ref{inst:0485}}
\and C.        ~Turon                         \inst{\ref{inst:0001}}
\and A.        ~Ulla                          \inst{\ref{inst:0487}}
\and E.        ~Utrilla                       \inst{\ref{inst:0103}}
\and S.        ~Uzzi                          \inst{\ref{inst:0049}}
\and M.        ~Vaillant                      \inst{\ref{inst:0140}}
\and G.        ~Valentini                     \inst{\ref{inst:0230}}
\and V.        ~Valette                       \inst{\ref{inst:0025}}
\and A.        ~van Elteren                   \inst{\ref{inst:0012}}
\and E.        ~Van Hemelryck                 \inst{\ref{inst:0070}}
\and M.        ~Vaschetto                     \inst{\ref{inst:0049}}
\and A.        ~Vecchiato                     \inst{\ref{inst:0036}}
\and J.        ~Veljanoski                    \inst{\ref{inst:0194}}
\and Y.        ~Viala                         \inst{\ref{inst:0001}}
\and D.        ~Vicente                       \inst{\ref{inst:0283}}
\and S.        ~Vogt                          \inst{\ref{inst:0430}}
\and C.        ~von Essen                     \inst{\ref{inst:0501}}
\and H.        ~Voss                          \inst{\ref{inst:0005}}
\and V.        ~Votruba                       \inst{\ref{inst:0336}}
\and S.        ~Voutsinas                     \inst{\ref{inst:0079}}
\and G.        ~Walmsley                      \inst{\ref{inst:0025}}
\and M.        ~Weiler                        \inst{\ref{inst:0005}}
\and O.        ~Wertz                         \inst{\ref{inst:0507}}
\and T.        ~Wevers                        \inst{\ref{inst:0003},\ref{inst:0142}}
\and \L{}.     ~Wyrzykowski                   \inst{\ref{inst:0003},\ref{inst:0408}}
\and A.        ~Yoldas                        \inst{\ref{inst:0003}}
\and M.        ~\v{Z}erjal                    \inst{\ref{inst:0323},\ref{inst:0514}}
\and H.        ~Ziaeepour                     \inst{\ref{inst:0126}}
\and J.        ~Zorec                         \inst{\ref{inst:0516}}
\and S.        ~Zschocke                      \inst{\ref{inst:0020}}
\and S.        ~Zucker                        \inst{\ref{inst:0518}}
\and C.        ~Zurbach                       \inst{\ref{inst:0104}}
\and T.        ~Zwitter                       \inst{\ref{inst:0323}}
}
\institute{
     Univ. Grenoble Alpes, CNRS, IPAG, 38000 Grenoble, France\relax                                                                                                                                          \label{inst:0002}
\and GEPI, Observatoire de Paris, Universit\'{e} PSL, CNRS, 5 Place Jules Janssen, 92190 Meudon, France\relax                                                                                                \label{inst:0001}
\and Institute of Astronomy, University of Cambridge, Madingley Road, Cambridge CB3 0HA, United Kingdom\relax                                                                                                \label{inst:0003}
\and Leicester Institute of Space and Earth Observation and Department of Physics and Astronomy, University of Leicester, University Road, Leicester LE1 7RH, United Kingdom\relax                           \label{inst:0004}
\and Institut de Ci\`{e}ncies del Cosmos, Universitat  de  Barcelona  (IEEC-UB), Mart\'{i} i  Franqu\`{e}s  1, 08028 Barcelona, Spain\relax                                                                  \label{inst:0005}
\and INAF - Osservatorio astronomico di Padova, Vicolo Osservatorio 5, 35122 Padova, Italy\relax                                                                                                             \label{inst:0006}
\and SISSA - Scuola Internazionale Superiore di Studi Avanzati, via Bonomea 265, 34136 Trieste, Italy\relax                                                                                                  \label{inst:0008}
\and Leiden Observatory, Leiden University, Niels Bohrweg 2, 2333 CA Leiden, The Netherlands\relax                                                                                                           \label{inst:0012}
\and Science Support Office, Directorate of Science, European Space Research and Technology Centre (ESA/ESTEC), Keplerlaan 1, 2201AZ, Noordwijk, The Netherlands\relax                                       \label{inst:0013}
\and Max Planck Institute for Astronomy, K\"{ o}nigstuhl 17, 69117 Heidelberg, Germany\relax                                                                                                                 \label{inst:0015}
\and Astronomisches Rechen-Institut, Zentrum f\"{ u}r Astronomie der Universit\"{ a}t Heidelberg, M\"{ o}nchhofstr. 12-14, 69120 Heidelberg, Germany\relax                                                   \label{inst:0016}
\and Department of Astronomy, University of Geneva, Chemin des Maillettes 51, 1290 Versoix, Switzerland\relax                                                                                                \label{inst:0018}
\and Mission Operations Division, Operations Department, Directorate of Science, European Space Research and Technology Centre (ESA/ESTEC), Keplerlaan 1, 2201 AZ, Noordwijk, The Netherlands\relax          \label{inst:0019}
\and Lohrmann Observatory, Technische Universit\"{ a}t Dresden, Mommsenstra{\ss}e 13, 01062 Dresden, Germany\relax                                                                                           \label{inst:0020}
\and European Space Astronomy Centre (ESA/ESAC), Camino bajo del Castillo, s/n, Urbanizacion Villafranca del Castillo, Villanueva de la Ca\~{n}ada, 28692 Madrid, Spain\relax                                \label{inst:0021}
\and Lund Observatory, Department of Astronomy and Theoretical Physics, Lund University, Box 43, 22100 Lund, Sweden\relax                                                                                    \label{inst:0022}
\and Universit\'{e} C\^{o}te d'Azur, Observatoire de la C\^{o}te d'Azur, CNRS, Laboratoire Lagrange, Bd de l'Observatoire, CS 34229, 06304 Nice Cedex 4, France\relax                                        \label{inst:0024}
\and CNES Centre Spatial de Toulouse, 18 avenue Edouard Belin, 31401 Toulouse Cedex 9, France\relax                                                                                                          \label{inst:0025}
\and Institut d'Astronomie et d'Astrophysique, Universit\'{e} Libre de Bruxelles CP 226, Boulevard du Triomphe, 1050 Brussels, Belgium\relax                                                                 \label{inst:0026}
\and F.R.S.-FNRS, Rue d'Egmont 5, 1000 Brussels, Belgium\relax                                                                                                                                               \label{inst:0027}
\and INAF - Osservatorio Astrofisico di Arcetri, Largo Enrico Fermi 5, 50125 Firenze, Italy\relax                                                                                                            \label{inst:0028}
\and Telespazio Vega UK Ltd for ESA/ESAC, Camino bajo del Castillo, s/n, Urbanizacion Villafranca del Castillo, Villanueva de la Ca\~{n}ada, 28692 Madrid, Spain\relax                                       \label{inst:0030}
\and Laboratoire d'astrophysique de Bordeaux, Univ. Bordeaux, CNRS, B18N, all{\'e}e Geoffroy Saint-Hilaire, 33615 Pessac, France\relax                                                                       \label{inst:0031}
\and Mullard Space Science Laboratory, University College London, Holmbury St Mary, Dorking, Surrey RH5 6NT, United Kingdom\relax                                                                            \label{inst:0035}
\and INAF - Osservatorio Astrofisico di Torino, via Osservatorio 20, 10025 Pino Torinese (TO), Italy\relax                                                                                                   \label{inst:0036}
\and INAF - Osservatorio di Astrofisica e Scienza dello Spazio di Bologna, via Piero Gobetti 93/3, 40129 Bologna, Italy\relax                                                                                \label{inst:0040}
\and Serco Gesti\'{o}n de Negocios for ESA/ESAC, Camino bajo del Castillo, s/n, Urbanizacion Villafranca del Castillo, Villanueva de la Ca\~{n}ada, 28692 Madrid, Spain\relax                                \label{inst:0043}
\and ALTEC S.p.a, Corso Marche, 79,10146 Torino, Italy\relax                                                                                                                                                 \label{inst:0049}
\and Department of Astronomy, University of Geneva, Chemin d'Ecogia 16, 1290 Versoix, Switzerland\relax                                                                                                      \label{inst:0051}
\and Gaia DPAC Project Office, ESAC, Camino bajo del Castillo, s/n, Urbanizacion Villafranca del Castillo, Villanueva de la Ca\~{n}ada, 28692 Madrid, Spain\relax                                            \label{inst:0058}
\and SYRTE, Observatoire de Paris, Universit\'{e} PSL, CNRS,  Sorbonne Universit\'{e}, LNE, 61 avenue de l’Observatoire 75014 Paris, France\relax                                                          \label{inst:0064}
\and National Observatory of Athens, I. Metaxa and Vas. Pavlou, Palaia Penteli, 15236 Athens, Greece\relax                                                                                                   \label{inst:0067}
\and IMCCE, Observatoire de Paris, Universit\'{e} PSL, CNRS,  Sorbonne Universit\'{e}, Univ. Lille, 77 av. Denfert-Rochereau, 75014 Paris, France\relax                                                      \label{inst:0069}
\and Royal Observatory of Belgium, Ringlaan 3, 1180 Brussels, Belgium\relax                                                                                                                                  \label{inst:0070}
\and Institute for Astronomy, University of Edinburgh, Royal Observatory, Blackford Hill, Edinburgh EH9 3HJ, United Kingdom\relax                                                                            \label{inst:0079}
\and Instituut voor Sterrenkunde, KU Leuven, Celestijnenlaan 200D, 3001 Leuven, Belgium\relax                                                                                                                \label{inst:0080}
\and Institut d'Astrophysique et de G\'{e}ophysique, Universit\'{e} de Li\`{e}ge, 19c, All\'{e}e du 6 Ao\^{u}t, B-4000 Li\`{e}ge, Belgium\relax                                                              \label{inst:0081}
\and ATG Europe for ESA/ESAC, Camino bajo del Castillo, s/n, Urbanizacion Villafranca del Castillo, Villanueva de la Ca\~{n}ada, 28692 Madrid, Spain\relax                                                   \label{inst:0084}
\and \'{A}rea de Lenguajes y Sistemas Inform\'{a}ticos, Universidad Pablo de Olavide, Ctra. de Utrera, km 1. 41013, Sevilla, Spain\relax                                                                     \label{inst:0088}
\and ETSE Telecomunicaci\'{o}n, Universidade de Vigo, Campus Lagoas-Marcosende, 36310 Vigo, Galicia, Spain\relax                                                                                             \label{inst:0090}
\and Large Synoptic Survey Telescope, 950 N. Cherry Avenue, Tucson, AZ 85719, USA\relax                                                                                                                      \label{inst:0095}
\and Observatoire Astronomique de Strasbourg, Universit\'{e} de Strasbourg, CNRS, UMR 7550, 11 rue de l'Universit\'{e}, 67000 Strasbourg, France\relax                                                       \label{inst:0096}
\and Kavli Institute for Cosmology, University of Cambridge, Madingley Road, Cambride CB3 0HA, United Kingdom\relax                                                                                          \label{inst:0099}
\and Aurora Technology for ESA/ESAC, Camino bajo del Castillo, s/n, Urbanizacion Villafranca del Castillo, Villanueva de la Ca\~{n}ada, 28692 Madrid, Spain\relax                                            \label{inst:0103}
\and Laboratoire Univers et Particules de Montpellier, Universit\'{e} Montpellier, Place Eug\`{e}ne Bataillon, CC72, 34095 Montpellier Cedex 05, France\relax                                                \label{inst:0104}
\and Department of Physics and Astronomy, Division of Astronomy and Space Physics, Uppsala University, Box 516, 75120 Uppsala, Sweden\relax                                                                  \label{inst:0107}
\and CENTRA, Universidade de Lisboa, FCUL, Campo Grande, Edif. C8, 1749-016 Lisboa, Portugal\relax                                                                                                           \label{inst:0108}
\and Universit\`{a} di Catania, Dipartimento di Fisica e Astronomia, Sezione Astrofisica, Via S. Sofia 78, 95123 Catania, Italy\relax                                                                        \label{inst:0109}
\and INAF - Osservatorio Astrofisico di Catania, via S. Sofia 78, 95123 Catania, Italy\relax                                                                                                                 \label{inst:0110}
\and University of Vienna, Department of Astrophysics, T\"{ u}rkenschanzstra{\ss}e 17, A1180 Vienna, Austria\relax                                                                                           \label{inst:0111}
\and CITIC – Department of Computer Science, University of A Coru\~{n}a, Campus de Elvi\~{n}a S/N, 15071, A Coru\~{n}a, Spain\relax                                                                        \label{inst:0113}
\and CITIC – Astronomy and Astrophysics, University of A Coru\~{n}a, Campus de Elvi\~{n}a S/N, 15071, A Coru\~{n}a, Spain\relax                                                                            \label{inst:0114}
\and INAF - Osservatorio Astronomico di Roma, Via di Frascati 33, 00078 Monte Porzio Catone (Roma), Italy\relax                                                                                              \label{inst:0115}
\and Space Science Data Center - ASI, Via del Politecnico SNC, 00133 Roma, Italy\relax                                                                                                                       \label{inst:0116}
\and University of Helsinki, Department of Physics, P.O. Box 64, 00014 Helsinki, Finland\relax                                                                                                               \label{inst:0120}
\and Finnish Geospatial Research Institute FGI, Geodeetinrinne 2, 02430 Masala, Finland\relax                                                                                                                \label{inst:0121}
\and Isdefe for ESA/ESAC, Camino bajo del Castillo, s/n, Urbanizacion Villafranca del Castillo, Villanueva de la Ca\~{n}ada, 28692 Madrid, Spain\relax                                                       \label{inst:0122}
\and Institut UTINAM UMR6213, CNRS, OSU THETA Franche-Comt\'{e} Bourgogne, Universit\'{e} Bourgogne Franche-Comt\'{e}, 25000 Besan\c{c}on, France\relax                                                      \label{inst:0126}
\and STFC, Rutherford Appleton Laboratory, Harwell, Didcot, OX11 0QX, United Kingdom\relax                                                                                                                   \label{inst:0128}
\and Dpto. de Inteligencia Artificial, UNED, c/ Juan del Rosal 16, 28040 Madrid, Spain\relax                                                                                                                 \label{inst:0131}
\and Elecnor Deimos Space for ESA/ESAC, Camino bajo del Castillo, s/n, Urbanizacion Villafranca del Castillo, Villanueva de la Ca\~{n}ada, 28692 Madrid, Spain\relax                                         \label{inst:0139}
\and Thales Services for CNES Centre Spatial de Toulouse, 18 avenue Edouard Belin, 31401 Toulouse Cedex 9, France\relax                                                                                      \label{inst:0140}
\and Department of Astrophysics/IMAPP, Radboud University, P.O.Box 9010, 6500 GL Nijmegen, The Netherlands\relax                                                                                             \label{inst:0142}
\and European Southern Observatory, Karl-Schwarzschild-Str. 2, 85748 Garching, Germany\relax                                                                                                                 \label{inst:0149}
\and ON/MCTI-BR, Rua Gal. Jos\'{e} Cristino 77, Rio de Janeiro, CEP 20921-400, RJ,  Brazil\relax                                                                                                             \label{inst:0151}
\and OV/UFRJ-BR, Ladeira Pedro Ant\^{o}nio 43, Rio de Janeiro, CEP 20080-090, RJ, Brazil\relax                                                                                                               \label{inst:0152}
\and Department of Terrestrial Magnetism, Carnegie Institution for Science, 5241 Broad Branch Road, NW, Washington, DC 20015-1305, USA\relax                                                                 \label{inst:0160}
\and Universit\`{a} di Torino, Dipartimento di Fisica, via Pietro Giuria 1, 10125 Torino, Italy\relax                                                                                                        \label{inst:0167}
\and Departamento de Astrof\'{i}sica, Centro de Astrobiolog\'{i}a (CSIC-INTA), ESA-ESAC. Camino Bajo del Castillo s/n. 28692 Villanueva de la Ca\~{n}ada, Madrid, Spain\relax                                \label{inst:0171}
\and Departamento de Estad\'{i}stica, Universidad de C\'{a}diz, Calle Rep\'{u}blica \'{A}rabe Saharawi s/n. 11510, Puerto Real, C\'{a}diz, Spain\relax                                                       \label{inst:0177}
\and Astronomical Institute Bern University, Sidlerstrasse 5, 3012 Bern, Switzerland (present address)\relax                                                                                                 \label{inst:0180}
\and EURIX S.r.l., Corso Vittorio Emanuele II 61, 10128, Torino, Italy\relax                                                                                                                                 \label{inst:0181}
\and Harvard-Smithsonian Center for Astrophysics, 60 Garden Street, Cambridge MA 02138, USA\relax                                                                                                            \label{inst:0185}
\and HE Space Operations BV for ESA/ESAC, Camino bajo del Castillo, s/n, Urbanizacion Villafranca del Castillo, Villanueva de la Ca\~{n}ada, 28692 Madrid, Spain\relax                                       \label{inst:0188}
\and Kapteyn Astronomical Institute, University of Groningen, Landleven 12, 9747 AD Groningen, The Netherlands\relax                                                                                         \label{inst:0194}
\and University of Turin, Department of Computer Sciences, Corso Svizzera 185, 10149 Torino, Italy\relax                                                                                                     \label{inst:0204}
\and SRON, Netherlands Institute for Space Research, Sorbonnelaan 2, 3584CA, Utrecht, The Netherlands\relax                                                                                                  \label{inst:0205}
\and Dpto. de Matem\'{a}tica Aplicada y Ciencias de la Computaci\'{o}n, Univ. de Cantabria, ETS Ingenieros de Caminos, Canales y Puertos, Avda. de los Castros s/n, 39005 Santander, Spain\relax             \label{inst:0207}
\and Unidad de Astronom\'ia, Universidad de Antofagasta, Avenida Angamos 601, Antofagasta 1270300, Chile\relax                                                                                               \label{inst:0214}
\and CRAAG - Centre de Recherche en Astronomie, Astrophysique et G\'{e}ophysique, Route de l'Observatoire Bp 63 Bouzareah 16340 Algiers, Algeria\relax                                                       \label{inst:0225}
\and University of Antwerp, Onderzoeksgroep Toegepaste Wiskunde, Middelheimlaan 1, 2020 Antwerp, Belgium\relax                                                                                               \label{inst:0228}
\and INAF - Osservatorio Astronomico d'Abruzzo, Via Mentore Maggini, 64100 Teramo, Italy\relax                                                                                                               \label{inst:0230}
\and INAF - Osservatorio Astronomico di Capodimonte, Via Moiariello 16, 80131, Napoli, Italy\relax                                                                                                           \label{inst:0232}
\and Instituto de Astronomia, Geof\`{i}sica e Ci\^{e}ncias Atmosf\'{e}ricas, Universidade de S\~{a}o Paulo, Rua do Mat\~{a}o, 1226, Cidade Universitaria, 05508-900 S\~{a}o Paulo, SP, Brazil\relax          \label{inst:0233}
\and Department of Astrophysics, Astronomy and Mechanics, National and Kapodistrian University of Athens, Panepistimiopolis, Zografos, 15783 Athens, Greece\relax                                            \label{inst:0244}
\and Leibniz Institute for Astrophysics Potsdam (AIP), An der Sternwarte 16, 14482 Potsdam, Germany\relax                                                                                                    \label{inst:0247}
\and RHEA for ESA/ESAC, Camino bajo del Castillo, s/n, Urbanizacion Villafranca del Castillo, Villanueva de la Ca\~{n}ada, 28692 Madrid, Spain\relax                                                         \label{inst:0249}
\and ATOS for CNES Centre Spatial de Toulouse, 18 avenue Edouard Belin, 31401 Toulouse Cedex 9, France\relax                                                                                                 \label{inst:0251}
\and School of Physics and Astronomy, Tel Aviv University, Tel Aviv 6997801, Israel\relax                                                                                                                    \label{inst:0254}
\and UNINOVA - CTS, Campus FCT-UNL, Monte da Caparica, 2829-516 Caparica, Portugal\relax                                                                                                                     \label{inst:0255}
\and School of Physics, O'Brien Centre for Science North, University College Dublin, Belfield, Dublin 4, Ireland\relax                                                                                       \label{inst:0266}
\and Dipartimento di Fisica e Astronomia, Universit\`{a} di Bologna, Via Piero Gobetti 93/2, 40129 Bologna, Italy\relax                                                                                      \label{inst:0272}
\and Barcelona Supercomputing Center - Centro Nacional de Supercomputaci\'{o}n, c/ Jordi Girona 29, Ed. Nexus II, 08034 Barcelona, Spain\relax                                                               \label{inst:0283}
\and Department of Computer Science, Electrical and Space Engineering, Lule\aa{} University of Technology, Box 848, S-981 28 Kiruna, Sweden\relax                                                            \label{inst:0289}
\and Max Planck Institute for Extraterrestrial Physics, High Energy Group, Gie{\ss}enbachstra{\ss}e, 85741 Garching, Germany\relax                                                                           \label{inst:0291}
\and Astronomical Observatory Institute, Faculty of Physics, Adam Mickiewicz University, S{\l}oneczna 36, 60-286 Pozna{\'n}, Poland\relax                                                                    \label{inst:0310}
\and Konkoly Observatory, Research Centre for Astronomy and Earth Sciences, Hungarian Academy of Sciences, Konkoly Thege Mikl\'{o}s \'{u}t 15-17, 1121 Budapest, Hungary\relax                               \label{inst:0316}
\and E\"{ o}tv\"{ o}s Lor\'and University, Egyetem t\'{e}r 1-3, H-1053 Budapest, Hungary\relax                                                                                                               \label{inst:0317}
\and American Community Schools of Athens, 129 Aghias Paraskevis Ave. \& Kazantzaki Street, Halandri, 15234 Athens, Greece\relax                                                                             \label{inst:0320}
\and Faculty of Mathematics and Physics, University of Ljubljana, Jadranska ulica 19, 1000 Ljubljana, Slovenia\relax                                                                                         \label{inst:0323}
\and Villanova University, Department of Astrophysics and Planetary Science, 800 E Lancaster Avenue, Villanova PA 19085, USA\relax                                                                           \label{inst:0324}
\and Physics Department, University of Antwerp, Groenenborgerlaan 171, 2020 Antwerp, Belgium\relax                                                                                                           \label{inst:0326}
\and McWilliams Center for Cosmology, Department of Physics, Carnegie Mellon University, 5000 Forbes Avenue, Pittsburgh, PA 15213, USA\relax                                                                 \label{inst:0332}
\and Astronomical Institute, Academy of Sciences of the Czech Republic, Fri\v{c}ova 298, 25165 Ond\v{r}ejov, Czech Republic\relax                                                                            \label{inst:0336}
\and Telespazio for CNES Centre Spatial de Toulouse, 18 avenue Edouard Belin, 31401 Toulouse Cedex 9, France\relax                                                                                           \label{inst:0341}
\and Institut de Physique de Rennes, Universit{\'e} de Rennes 1, 35042 Rennes, France\relax                                                                                                                  \label{inst:0344}
\and Shanghai Astronomical Observatory, Chinese Academy of Sciences, 80 Nandan Rd, 200030 Shanghai, China\relax                                                                                              \label{inst:0351}
\and School of Astronomy and Space Science, University of Chinese Academy of Sciences, Beijing 100049, China\relax                                                                                           \label{inst:0352}
\and Niels Bohr Institute, University of Copenhagen, Juliane Maries Vej 30, 2100 Copenhagen {\O}, Denmark\relax                                                                                              \label{inst:0354}
\and DXC Technology, Retortvej 8, 2500 Valby, Denmark\relax                                                                                                                                                  \label{inst:0355}
\and Las Cumbres Observatory, 6740 Cortona Drive Suite 102, Goleta, CA 93117, USA\relax                                                                                                                      \label{inst:0356}
\and Astrophysics Research Institute, Liverpool John Moores University, 146 Brownlow Hill, Liverpool L3 5RF, United Kingdom\relax                                                                            \label{inst:0364}
\and Baja Observatory of University of Szeged, Szegedi \'{u}t III/70, 6500 Baja, Hungary\relax                                                                                                               \label{inst:0369}
\and Laboratoire AIM, IRFU/Service d'Astrophysique - CEA/DSM - CNRS - Universit\'{e} Paris Diderot, B\^{a}t 709, CEA-Saclay, 91191 Gif-sur-Yvette Cedex, France\relax                                        \label{inst:0370}
\and Warsaw University Observatory, Al. Ujazdowskie 4, 00-478 Warszawa, Poland\relax                                                                                                                         \label{inst:0408}
\and Institute of Theoretical Physics, Faculty of Mathematics and Physics, Charles University in Prague, Czech Republic\relax                                                                                \label{inst:0409}
\and AKKA for CNES Centre Spatial de Toulouse, 18 avenue Edouard Belin, 31401 Toulouse Cedex 9, France\relax                                                                                                 \label{inst:0416}
\and Vitrociset Belgium for ESA/ESAC, Camino bajo del Castillo, s/n, Urbanizacion Villafranca del Castillo, Villanueva de la Ca\~{n}ada, 28692 Madrid, Spain\relax                                           \label{inst:0422}
\and HE Space Operations BV for ESA/ESTEC, Keplerlaan 1, 2201AZ, Noordwijk, The Netherlands\relax                                                                                                            \label{inst:0430}
\and Space Telescope Science Institute, 3700 San Martin Drive, Baltimore, MD 21218, USA\relax                                                                                                                \label{inst:0439}
\and QUASAR Science Resources for ESA/ESAC, Camino bajo del Castillo, s/n, Urbanizacion Villafranca del Castillo, Villanueva de la Ca\~{n}ada, 28692 Madrid, Spain\relax                                     \label{inst:0440}
\and Fork Research, Rua do Cruzado Osberno, Lt. 1, 9 esq., Lisboa, Portugal\relax                                                                                                                            \label{inst:0445}
\and APAVE SUDEUROPE SAS for CNES Centre Spatial de Toulouse, 18 avenue Edouard Belin, 31401 Toulouse Cedex 9, France\relax                                                                                  \label{inst:0448}
\and Nordic Optical Telescope, Rambla Jos\'{e} Ana Fern\'{a}ndez P\'{e}rez 7, 38711 Bre\~{n}a Baja, Spain\relax                                                                                              \label{inst:0453}
\and Spanish Virtual Observatory\relax                                                                                                                                                                       \label{inst:0458}
\and Fundaci\'{o}n Galileo Galilei - INAF, Rambla Jos\'{e} Ana Fern\'{a}ndez P\'{e}rez 7, 38712 Bre\~{n}a Baja, Santa Cruz de Tenerife, Spain\relax                                                          \label{inst:0470}
\and INSA for ESA/ESAC, Camino bajo del Castillo, s/n, Urbanizacion Villafranca del Castillo, Villanueva de la Ca\~{n}ada, 28692 Madrid, Spain\relax                                                         \label{inst:0475}
\and Dpto. Arquitectura de Computadores y Autom\'{a}tica, Facultad de Inform\'{a}tica, Universidad Complutense de Madrid, C/ Prof. Jos\'{e} Garc\'{i}a Santesmases s/n, 28040 Madrid, Spain\relax            \label{inst:0476}
\and H H Wills Physics Laboratory, University of Bristol, Tyndall Avenue, Bristol BS8 1TL, United Kingdom\relax                                                                                              \label{inst:0479}
\and Institut d'Estudis Espacials de Catalunya (IEEC), Gran Capita 2-4, 08034 Barcelona, Spain\relax                                                                                                         \label{inst:0485}
\and Applied Physics Department, Universidade de Vigo, 36310 Vigo, Spain\relax                                                                                                                               \label{inst:0487}
\and Stellar Astrophysics Centre, Aarhus University, Department of Physics and Astronomy, 120 Ny Munkegade, Building 1520, DK-8000 Aarhus C, Denmark\relax                                                   \label{inst:0501}
\and Argelander-Institut f\"{ ur} Astronomie, Universit\"{ a}t Bonn,  Auf dem H\"{ u}gel 71, 53121 Bonn, Germany\relax                                                                                       \label{inst:0507}
\and Research School of Astronomy and Astrophysics, Australian National University, Canberra, ACT 2611 Australia\relax                                                                                       \label{inst:0514}
\and Sorbonne Universit\'{e}s, UPMC Univ. Paris 6 et CNRS, UMR 7095, Institut d'Astrophysique de Paris, 98 bis bd. Arago, 75014 Paris, France\relax                                                          \label{inst:0516}
\and Department of Geosciences, Tel Aviv University, Tel Aviv 6997801, Israel\relax                                                                                                                          \label{inst:0518}
}